\documentclass[twocolumn,prd,floatfix,preprintnumbers,nofootinbib,superscriptaddress, aps]{revtex4-1}

\usepackage{subcaption}
\usepackage{ragged2e}
\DeclareCaptionJustification{justified}{\justifying}
\usepackage{amssymb,amsmath,verbatim,mathtools,needspace,enumitem,etoolbox,graphicx,physics,microtype,afterpage,bm,soul}
\usepackage{esint}
\usepackage[dvipsnames]{xcolor}
\definecolor{linkcolor}{rgb}{0.0,0.3,0.5}
\usepackage[unicode, colorlinks=true, linkcolor=linkcolor, citecolor=linkcolor, filecolor=linkcolor,urlcolor=linkcolor, pdfusetitle]{hyperref}
\usepackage{orcidlink}
\usepackage[all]{hypcap}
\usepackage{lipsum}
\usepackage[T1]{fontenc}
\usepackage[utf8]{inputenc}
\usepackage{tabularx}
\usepackage{cancel}
\usepackage{pifont}
\newcommand\underrel[3][]{\mathrel{\mathop{#3}\limits_{%
      \ifx c#1\relax\mathclap{#2}\else#2\fi}}}
\usepackage{soul}
\usepackage{lmodern}
\allowdisplaybreaks
\usepackage{tikz}
\usepackage{cleveref}
\usepackage{color}
\usepackage{framed}
\usepackage{hyperref}
\hypersetup{colorlinks, citecolor=bluscuro, linkcolor=black, urlcolor=bluscuro}
\definecolor{rossos}{cmyk}{0,1,1,0.55}
\definecolor{bluscuro}{rgb}{0.15, 0.2, .85}
\newcommand{\be}{\begin{equation}}
\newcommand{\ee}{\end{equation}}
\renewcommand{\d}{{\mathrm d}}

\def\NS{\text{\tiny NS}}

\newcommand{\Mp}{M_\mathrm{Pl}}
\newcommand{\e}{\mathrm{e}}
\newcommand{\p}{\partial}

\def\lsim{\mathrel{\rlap{\lower4pt\hbox{\hskip0.5pt$\sim$}}
    \raise1pt\hbox{$<$}}}         
\def\gsim{\mathrel{\rlap{\lower4pt\hbox{\hskip0.5pt$\sim$}}
    \raise1pt\hbox{$>$}}}         

\makeatletter
\newcommand{\subsetsim}{\mathrel{\mathpalette\subset@sim\relax}}
\newcommand{\subset@sim}[2]{%
  \vtop{\offinterlineskip\m@th
    \ialign{\hfil##\cr
     ~$#1\subset$\cr\noalign{\kern0.5pt}\scalebox{0.9}{$#1\sim$}\cr
    }%
  }%
}
\makeatother
\makeatletter
\def\l@subsubsection#1#2{}
\makeatother

\newcommand{\sapienza}{Dipartimento di Fisica, Sapienza Universit\`a di Roma, Piazzale Aldo Moro 5, 00185, Roma, Italy}
\newcommand{\infn}{INFN, Sezione di Roma, Piazzale Aldo Moro 2, 00185, Roma, Italy}

\begin{document}

\title{Dynamical tidal response of neutron stars:\\
From effective field theory to gravitational waveforms}

\author{Thomas Apostolidis\orcidlink{0009-0002-2189-2338}}
\email{apostolidis@apc.in2p3.fr}
\affiliation{Université Paris Cité, CNRS, Astroparticule et Cosmologie, 10 Rue Alice Domon et Léonie Duquet, F-75013 Paris, France}

\author{Valerio De Luca\orcidlink{0000-0002-1444-5372}}
\email{vdeluca2@jh.edu}
\affiliation{William H.\ Miller III Department of Physics and Astronomy, Johns Hopkins University, 3400 North Charles Street, Baltimore, Maryland, 21218, USA}

 \author{Leonardo Gualtieri\orcidlink{0000-0002-1097-3266}}
 \email{leonardo.gualtieri@unipi.it}
 \affiliation{Dipartimento di Fisica, Universit\`a di Pisa, Largo B. Pontecorvo 3, 56127 Pisa, Italy}
 \affiliation{INFN, Sezione di Pisa, Largo B. Pontecorvo 3, 56127 Pisa, Italy}

 \author{Takuya Katagiri\orcidlink{0000-0002-3755-3093}}
 \email{takuya.katagiri@uniroma1.it}
 \affiliation{\sapienza}
 \affiliation{\infn}

 \author{Paolo Pani\orcidlink{0000-0003-4443-1761}}
 \email{paolo.pani@uniroma1.it}
 \affiliation{\sapienza}
 \affiliation{\infn}

 \author{Luca Santoni\orcidlink{0000-0001-9319-6054}}
 \email{santoni@apc.in2p3.fr}
 \affiliation{Université Paris Cité, CNRS, Astroparticule et Cosmologie, 10 Rue Alice Domon et Léonie Duquet, F-75013 Paris, France}


\begin{abstract}
\noindent
We investigate the fully relativistic dynamical tidal response of neutron stars up to second order in the frequency. Combining the worldline effective field theory for extended gravitating bodies with perturbation theory of relativistic stellar models, we derive the tidal deformation induced by an external time-dependent field, including a universal logarithmic running term.
In the effective theory, we work in dimensional regularization and, through a consistent matching procedure, obtain for the first time the complete leading-order dynamical tidal corrections to both the conservative dynamics and the gravitational-wave signal of compact binaries, including the scheme-dependent finite terms in addition to the running. 
We show that, in the relativistic regime, dynamical effects cannot be fully captured by mode excitations alone. The magnitude of the additional contribution depends on the stellar compactness, the equation of state, and the running term.
Dynamical Love numbers are significantly enhanced with respect to their static counterparts for relatively small compactness. As a result, although they formally enter the gravitational-wave phase at eighth post-Newtonian order, dynamical tidal effects yield a nonnegligible contribution during the late inspiral. 
Using a Fisher-matrix analysis, we show that third-generation detectors such as the Einstein Telescope could measure dynamical Love numbers for a range of neutron-star masses and equations of state. Conversely, neglecting these effects can lead to significant biases in the inference of static Love numbers, and hence on the nuclear equation of state. 
Our results highlight the importance of dynamical tidal effects for high-precision gravitational-wave modeling with future detectors.
\end{abstract}

\preprint{ET-0313A-26}
\maketitle
\tableofcontents

\newpage

\section{Introduction}
\label{sec:intro}
\noindent
Neutron stars provide a unique arena to study matter under extreme conditions~\cite{Lattimer:2021emm}. Their cores can reach densities exceeding those of atomic nuclei, making them natural laboratories to investigate the behavior of strongly interacting matter and its connection to nuclear and particle physics. Understanding their internal structure is therefore a central goal in modern astrophysics.

The macroscopic properties of neutron stars are governed by the equation of state~(EoS), which encodes the relation between pressure, density, and temperature and depends sensitively on the underlying microphysics~\cite{Lattimer:2021emm,Burgio:2021vgk}. The EoS determines observable quantities such as the mass-radius relation, the maximum mass, and the tidal deformability. The advent of gravitational-wave (GW) astronomy has opened a direct observational window onto these properties: signals from coalescing compact binaries carry imprints of the internal structure of neutron stars, allowing us to probe fundamental physics in regimes inaccessible to terrestrial experiments (see Refs.~\cite{Baiotti:2019sew,Chatziioannou:2020pqz} for reviews).

In a binary system, each neutron star experiences the gravitational field generated by its companion, which induces multipolar deformations~\cite{Poisson_Will_2014,Hinderer:2007mb,Flanagan:2007ix,Damour:2009vw,Binnington:2009bb}. These deformations are commonly characterized by the tidal Love numbers~(TLNs), which relate the external tidal field to the induced multipole moments. Tidal interactions affect both the conservative dynamics of the binary and the emitted GW signal, leaving measurable imprints on the waveform~\cite{Hinderer:2009ca,Vines:2011ud}. Since the first binary neutron-star observation GW170817~\cite{LIGOScientific:2017vwq}, tidal effects have become a key probe of dense-matter physics and compact objects (see Refs.~\cite{Chakraborty:2026qru,Rodriguez:2026iot} for recent overviews).

Most studies to date have focused on the \emph{static} (or adiabatic) tidal response, in which the induced multipole moments instantaneously track the external tidal field. This approximation is valid when the orbital timescale is much longer than the characteristic internal timescales of the star. As the binary inspirals toward merger, however, the orbital frequency increases and the tidal response acquires a nontrivial frequency dependence, giving rise to genuinely \emph{dynamical} tidal effects. These include resonant excitations of stellar oscillation modes, dissipative phenomena, and phase lags between the tidal field and the induced deformation, all of which can modify the late-inspiral GW signal.

The physics of dynamical tides has a long history in Newtonian gravity, where the tidal response of a self-gravitating fluid body can be described in terms of the excitation of its normal modes~\cite{Lai:1993di,Lai:1997wh,Ho:1998hq,Chakrabarti:2013xza,Andersson:2019ahb,Andersson:2019dwg,Passamonti:2020fur,Passamonti:2022yqp,Pnigouras:2022zpx,Yu:2024uxt, Pnigouras:2025muo}. In this framework, the frequency-dependent tidal deformability admits a mode-sum representation in which the dominant contribution typically arises from the fundamental fluid ($f$-) mode because of its relatively low frequency and large overlap with the external tidal field~\cite{Lai:1993di,Ho:1998hq,Andersson:2019ahb}. Even away from resonance, the coupling to the $f$-mode can produce sizable corrections to the static Love number, motivating effective descriptions in which the static tidal deformability is supplemented by more resonant contributions.

Although Newtonian models provide valuable physical intuition, a quantitatively accurate description of tidal interactions in compact binaries ultimately requires a relativistic treatment. Considerable progress in this direction has been achieved in recent years. A fully relativistic analysis of the dynamical tidal response of nonrotating stars was recently presented in Refs.~\cite{Pitre:2023xsr,HegadeKR:2024agt,HegadeKR:2025qwj,HegadeKR:2026iou,Counsell:2024pua}. The relativistic mode structure underlying the dynamical response has also been investigated in detail in Refs.~\cite{Saes:2025jvr,HegadeKR:2026kku}, where it was  compared to the Newtonian overlap-integral formalism.

An important outcome of these studies is that the relativistic mode expansion differs qualitatively from its Newtonian counterpart. In particular, the operators defining the natural inner product for relativistic stellar perturbations are not positive definite, implying that the associated mode sums are not strictly convergent~\cite{HegadeKR:2026kku,Saes:2025jvr}. This feature reflects the fact that, in general relativity, the dynamical tidal response cannot be fully decomposed in terms of a sum over normal-mode excitations. Closely related issues emerge when comparing the exact relativistic tidal response with its low-frequency expansion~\cite{Saes:2025jvr}: while the expansion reproduces the weak-field regime, sizable deviations can arise for sufficiently compact stars, signaling the increasing importance of genuinely relativistic effects.

The relativistic formulation of dynamical tides has also been developed from the perspective of worldline effective field theory (EFT)~\cite{Goldberger:2004jt,Goldberger:2007hy,Porto:2005ac}. In this framework, the internal structure of a compact object is encoded in frequency-dependent response functions entering a point-particle effective action. Dynamical tidal effects in compact binaries have been investigated in EFT approaches in Refs.~\cite{Chakrabarti:2013lua,Steinhoff:2016rfi,Jakobsen:2023pvx,Mandal:2023hqa,Mandal:2023lgy,Saketh:2024juq,Kosmopoulos:2025rfj,Jarequi:2026cyp, Bhattacharyya:2026itm,Martinez-Rodriguez:2026omk}.
This formalism provides a coordinate-independent and systematically improvable description of finite-size effects and is particularly suited to connect stellar perturbation theory with post-Newtonian~(PN) and waveform calculations.
Furthermore, the EFT calculation clarifies the appearance of logarithmic terms characterizing the running of the dynamical Love numbers~\cite{Chakrabarti:2013lua,Charalambous:2021mea,Saketh:2023bul,Perry:2023wmm,Jakobsen:2023pvx,Mandal:2023hqa,Ivanov:2024sds,Katagiri:2024wbg,Glazer:2024eyi,Caron-Huot:2025tlq,Combaluzier--Szteinsznaider:2025eoc,Chakraborty:2025wvs,Ivanov:2026icp,Jarequi:2026cyp} and highlighting the necessity of a rigorous matching procedure. As we discuss below, a complete matching has not yet been  performed for the dynamical Love numbers of a neutron star and this would be one of our primary goals.
 
Dissipative aspects of the tidal response have also recently received considerable attention. Frequency-dependent dissipative tidal deformabilities, including the effects of viscosity and mode damping, were investigated in Refs.~\cite{Counsell:2024pua,HegadeKR:2024agt,HegadeKR:2025qwj,HegadeKR:2026iou}. These works showed that dissipative effects are generally much smaller than the conservative dynamical response for realistic neutron-star models, although they remain sensitive to the internal composition and transport properties of dense matter. At the same time, they highlighted that the conservative dynamical tidal response can vary significantly with the EoS and stellar compactness.

From a PN perspective, dynamical tidal effects formally enter the GW phase at high order, comparable to other subleading finite-size contributions~\cite{Chakrabarti:2013lua,Hinderer:2016eia,Steinhoff:2016rfi,Poisson:2020vap,Mandal:2023hqa,Pitre:2023xsr,Katagiri:2024wbg,HegadeKR:2024agt,Chakraborty:2025wvs,Combaluzier--Szteinsznaider:2025eoc,Kobayashi:2025vgl}, including nonlinear tidal effects beyond linear perturbation theory~\cite{Pani:2025qxs}. Nevertheless, these contributions can be enhanced by resonant mode excitations, inverse compactness scaling, and by relativistic strong-field effects, making them potentially observable with future detectors. Accurate modeling of dynamical tides is therefore essential both to avoid systematic biases in parameter estimation and to fully exploit the scientific potential of third-generation GW observatories~\cite{Gamba:2020wgg,JimenezForteza:2018rwr,Gupta:2024gun,ET:2025xjr}.

In this work we address the following question:
\emph{How does the frequency dependence of the tidal response of relativistic neutron stars affect the dynamics of compact binaries and their GW emission?}
While previous studies established the qualitative importance of dynamical tides, a complete and systematic treatment connecting relativistic stellar perturbation theory, EFT response functions, binary dynamics, all the way down to GW observables is still lacking. 
This is particularly relevant in the presence of running terms,\footnote{Other examples where static Love number couplings can run include black holes in higher dimensions or in theories beyond general relativity, see e.g.~\cite{Kol:2011vg,Hui:2020xxx,Rodriguez:2023xjd,Charalambous:2023jgq, Cardoso:2017cfl, DeLuca:2022tkm, Barbosa:2025uau}.} as is the case for dynamical tidal effects in compact objects, and to avoid ambiguities that inevitably plague perturbation-theory calculations, if the latter are not properly matched to GW observables (see Refs.~\cite{Rodriguez:2026iot,Chakraborty:2026qru} for recent reviews).

In this work, we complete this program. First, we compute the frequency-dependent tidal response of neutron stars by directly solving the relativistic perturbation equations, both perturbatively and nonperturbatively in frequency, without relying on a mode decomposition. 
Second, building on Ref.~\cite{Combaluzier--Szteinsznaider:2025eoc} (see also Ref.~\cite{Caron-Huot:2025tlq}), we match the resulting response to the worldline EFT at loop order, thereby obtaining for the first time the dynamical Love numbers including both the logarithmic running and the scheme-dependent finite contributions in dimensional regularization. 
Crucially, matching the EFT to the underlying general-relativistic neutron-star dynamics in order to obtain the complete Love-number couplings cannot be achieved at tree level within the EFT. Instead, it requires the evaluation of higher-loop worldline diagrams at the classical level. These diagrams are formally divergent and therefore require regularization and renormalization, including the introduction of a subtraction prescription and an associated renormalization scale---a structure that closely mirrors the renormalization procedure and renormalization-group running familiar from quantum field theory. Matching the results of stellar perturbation theory to the EFT couplings enables the construction of observable quantities that are independent of the choice of coordinates and perturbation variables.

We further compare the full relativistic response with resonant models based on the dominant $f$-mode, quantifying the regime of validity of these approximations in agreement with recent results~\cite{HegadeKR:2026kku}.
Finally, after matching the stellar-perturbation theory response with the EFT action, we derive the complete leading-order dynamical tidal corrections to the conservative dynamics and GW signal of compact binaries. Since the tidal contributions to the PN waveform can themselves be derived from the EFT action~\cite{Vines:2010ca,Vines:2011ud}, fixing the split between the tidal field and the response through matching to the EFT guarantees that the coefficients we compute are precisely those entering the waveform. We then assess their detectability with current and future GW detectors for realistic neutron-star EoS. Interestingly, we find that---due to its strong enhancement at moderate compactness---the 8PN contribution coming from the dynamical Love number is relevant for parameter estimation with third-generation interferometers such as the Einstein Telescope~(ET)~\cite{ET:2025xjr}.

\vspace{0.1in}
\noindent
\textit{Outline:}
The paper is organized as follows. In Sec.~\ref{sec:EFT} we introduce the worldline EFT
framework and define the static and dynamical Love numbers through the finite-size effective
action. In Sec.~\ref{sec:UV} we describe the relativistic stellar perturbation theory,
presenting both a perturbative expansion and a  nonperturbative treatment in the frequency. In Sec.~\ref{sec:NStidalresponse} we match the interior and exterior solutions,
extract the tidal Wilson coefficients  for a range of realistic EoS, and discuss
universal relations and the connection to the $f$-mode resonance model. In Sec.~\ref{sec:measurability}
we derive the complete 8PN tidal waveform phase and assess the observability of the
dynamical tide with LIGO-Virgo-KAGRA (LVK) O4 and ET~\cite{ET:2025xjr}. We conclude in
Sec.~\ref{sec:discussion}. Technical details are collected in four appendices: the EFT
computation of the graviton one-point function (Appendix~\ref{app:ppEFT}),  technical details of the stellar perturbation analysis (Appendix~\ref{app:StellarPerturbation}), the exterior Zerilli
matching (Appendix~\ref{app:matchingtoGR}), and a comparison with the tidal constants used in the
literature (Appendix~\ref{app:comparison}).

\vspace{0.1in}
\noindent
\textit{Conventions:} We adopt the mostly-plus metric signature, $(-,+,\cdots,+)$, and work in natural units with $\hbar = c = 1$. The reduced Planck mass is defined as $\Mp=  (8 \pi G)^{-\frac1{D-2}}$, with $G$ Newton's gravitational constant and $D$ the number of spacetime dimensions. Although our main interest is in $D=4$, we keep $D$ generic in certain sections where it is needed for dimensional regularization in the worldline effective theory. 
In Sec.~\ref{sec:UV}, we will also set $G=1$ to simplify the notation.
Our curvature conventions is ${R^\rho}_{\sigma\mu\nu}=\partial_\mu\Gamma^\rho_{\nu\sigma}+\dots$ and $R_{\mu\nu}={R^\rho}_{\mu\rho\nu}$.
Fields are decomposed in spherical harmonics and Fourier modes as $\psi(t,r,\theta,\varphi) = \sum_{\ell,m} \int \frac{\mathrm{d}\omega}{2\pi} \e^{-i\omega t} \, \psi^{\ell m}(\omega,r) \, Y_{\ell m}(\theta,\varphi)$ and, for brevity, we will often suppress the arguments of $\psi^{\ell m}$ when context suffices.   We use Greek indices, $\mu, \nu, \rho, \dots$, to label spacetime coordinates, while Latin indices, $i, j, k, \dots$, refer to spatial coordinates.

\section{Worldline Effective Field Theory}
\label{sec:EFT}
\noindent
We define the dynamical Love numbers of the neutron star within the point-particle EFT framework~\cite{Goldberger:2004jt,Goldberger:2007hy,Porto:2005ac}, which provides a robust and gauge-invariant description of an object's tidal response. In this work, we focus on nonrotating, perfect-fluid  stars, leaving the study of spin and dissipative effects to future work.

The EFT of a nonrotating particle takes the following form (see,  e.g., Refs.~\cite{Rothstein:2014sra,Porto:2016pyg,Levi:2018nxp,Goldberger:2022rqf} for some reviews):
\begin{equation}
    S=S_{\mathrm{pp}}+S_{\text {bulk }}+S_{\text {int}} \,.
\label{eq:ppEFT0}
\end{equation}
Here, $S_{\mathrm{pp}}$ is the point-particle action, given by the standard expression
\begin{align}
	S_{\rm pp}  = -M \int \d \tau = 
	-M \int \d \sigma 
	\sqrt{-g_{\mu\nu}(X)\frac{\d X^\mu}{\d \sigma}\frac{\d X^\nu}{\d \sigma}} 
	\label{eq:Spp}
\end{align}
where $\tau$ is the proper time measured along the particle's worldline, $M$ the particle's mass, $X^\mu(\sigma)$ its spacetime trajectory, and $\sigma$ an affine parameter describing the worldline.

The action $S_{\mathrm{bulk}}$ governs instead the dynamics of the gravitational field in the bulk spacetime and is given by the standard Einstein--Hilbert term,
\begin{equation}
    S_{\mathrm{bulk}} = \frac{M_{\mathrm{Pl}}^{2}}{2} \int \mathrm{d}^4 x \, \sqrt{-g} \, R \,.
\end{equation}
The object is a point particle only at leading order. To describe tidal responses, we introduce a series of higher-dimensional operators $S_{\mathrm{int}}$ that account for the object's finite size, evaluated along its worldline and coupled to the bulk fields~\cite{Goldberger:2004jt,Goldberger:2005cd,Chakrabarti:2013lua,Goldberger:2020fot,Saketh:2023bul,Combaluzier--Szteinsznaider:2025eoc}: 
\begin{equation}
\begin{split}
    S_{\mathrm{int}} & =  \int \mathrm{d}\tau \,
    Q_E^{ij}(\tau) \, E_{ij} 
\\
    & \quad + \text{magnetic} + \text{higher orders} \,,
\end{split}
\label{SintE}
\end{equation}
where $E_{ij}$ corresponds to the electric component of the Weyl tensor $C_{\mu\nu\rho\sigma}$,
\begin{equation}
    E_{ij} \equiv C_{0i0j} \,.
\end{equation}
The quantity $Q_E^{ij}(\tau)$ in Eq.~\eqref{SintE} represents a composite operator, constructed from the internal degrees of freedom, and encodes all information about the microscopic interior dynamics, including tidal deformations, absorption, internal hydrodynamic modes, resonances, and any other  finite-size effects.\footnote{This operator coincides with the quadrupolar tensor defined in Ref.~\cite{Flanagan:2007ix}, up to a normalization factor of $-1/4$.}

Note that in the second line of Eq.~\eqref{SintE} we have omitted operators involving the magnetic component of the Weyl tensor. Because of the relative velocity suppression between the magnetic and electric tidal fields, magnetic tidal effects arise at higher PN order compared to their electric counterparts. For the same reason, additional higher-derivative terms involving gradients of $E_{ij}$ have also been neglected in Eq.~\eqref{SintE}, as they are subleading in the PN expansion. Since this work focuses on the dominant dynamical quadrupolar relativistic response of neutron stars, we will restrict our attention to the leading operator \eqref{SintE}, leaving a more general analysis, including subdominant terms, for future work.

The process we aim to compute is the following: we probe the point particle with an external gravito-electric tidal field and then determine the induced response field.
Since a long-distance observer cannot resolve the detailed internal microdynamics of the object, we integrate out its internal degrees of freedom. Operationally, this amounts to solving for the composite operator $Q_E^{ij}(\tau)$ within  response theory. In the following, we will restrict ourselves to linear response,\footnote{See Refs.~\cite{Bern:2020uwk,DeLuca:2023mio,Riva:2023rcm,Iteanu:2024dvx,Combaluzier-Szteinsznaider:2024sgb,Pani:2025qxs} for a generalization to nonlinear response in the point-particle EFT.} but retain  frequency-dependent corrections. In addition, since our focus here is on perfect-fluid (nonrotating) neutron stars, for which viscous effects vanish, we will retain only the conservative part of Eq.~\eqref{SintE}.\footnote{Although we match the EFT to a nondissipative relativistic star in this work, Appendix~\ref{app:ppEFT} presents the general treatment using the Schwinger--Keldysh formalism (see also Ref.~\cite{Combaluzier--Szteinsznaider:2025eoc} for details). This is useful for discussing general properties of the response that hold for any object, independently of the details of the UV matching.}

In this limit, the most general quadratic finite-size action for the gravito-electric conservative sector, up to quadratic order in frequency, boils down to
\begin{equation}
    S_{\mathrm{int}}  = \frac{1}{2} \int \mathrm{d}\tau 
    \left( \frac{R_\star^5}{G}c_E E_{ij}E^{ij} + \frac{R_\star^8}{G^2M}c_{\dot{E}} \dot E_{ij}\dot E^{ij}  \right)\,,
\label{eq:Sintmain}
\end{equation}
where $R_\star$ is the radius of the object, and $c_{{E}}$ and $c_{\dot{E}}$ are the (dimensionless) static and dynamical quadrupolar Love number couplings, respectively.

The action \eqref{eq:ppEFT0} can be used to compute any long-distance observable involving the point particle and the gravitational field. In this work, we focus on an off-shell quantity, namely the graviton one-point function induced by an external quadrupolar tidal field $\bar{E}_{ij}$ coupled to the point particle. This response can be obtained by solving the equations of motion derived from the action \eqref{eq:ppEFT0} or, more generally, by using the Schwinger--Keldysh in-in formalism (which allows one to incorporate dissipative effects when present). The relevant formalism is reviewed in Appendix~\ref{app:ppEFT}; see also Ref.~\cite{Combaluzier--Szteinsznaider:2025eoc}.
The result for the graviton one-point function at tree-level (in $D = 4$) reads
\begin{multline}
\label{ErrD4}
 E_{rr} 
       = -2\e^{-i\omega t}c_{\mathrm{ext}} Y_{2m} \\
       \times\left[1  +  36 \, 
      \left(\frac{R_\star}{r} \right)^{5} \left(c_E+ \omega^2   \frac{R_\star^3}{GM}c_{\dot E}\right)\right]    \, ,
\end{multline}
where $\omega$ is the perturbation's frequency and $c_{\mathrm{ext}}$  characterizes the amplitude of the external (quadrupole) tidal field, located at a distance $r$ from the object. Eq.~\eqref{ErrD4} captures the leading distance behavior of the gravito-electric field in terms of the object's tidal couplings.

To match it to the full relativistic solution, however, we must include the coupling to gravity, which generates
subleading terms and induces the appearance of ultraviolet divergences, which must be regularized. As in standard quantum field theory, these divergences are absorbed by introducing suitable counterterms and adopting a renormalization scheme. Throughout this work, we employ dimensional regularization together with minimal subtraction~\cite{Combaluzier--Szteinsznaider:2025eoc}.
As an intermediate step toward obtaining the renormalized quadrupolar response, we compute the Zerilli field, which satisfies the  Zerilli equation [see Eq.~\eqref{eq:Z_EWE}]
\begin{equation}
    \left(\frac{\d^2}{\d r^2}-\frac{(2-\varepsilon)(3-\varepsilon)}{r^2} \right)\Psi_{\text{Z}}(r) = V_{\Psi_\text{Z}}(r)\Psi_\text{Z}(r)\,,
    \label{eq:Z_EWEmain}
\end{equation}
in the bulk, where $\varepsilon=2-D/2$ and $V_{\Psi_\text{Z}}$ is the effective potential capturing the nonlinearities of the Einstein--Hilbert theory.
After resumming the gravitational nonlinearities through a Born-series expansion~\cite{Caron-Huot:2025tlq} and canceling the ultraviolet divergences with the appropriate counterterms, we obtain the renormalized Zerilli field [see Eq.~\eqref{eq:psiZRapp} in Appendix~\ref{app:ppEFT}, and Ref.~\cite{Combaluzier--Szteinsznaider:2025eoc} for details]:
\begin{multline}
    \Psi_\text{Z}^{\mathrm{R}}=\, \bar{B}_\text{reg} \left\{r^3 \left[1-\bar{G}^2 \omega ^2 \left(\frac{214}{105} \log (\mu  r)+\frac{2731}{9800}\right)\right] \right. 
\\
\left.+\frac{\bar{G}^7 \omega ^2}{r^2}\left(\frac{3011}{80} \log (\mu  r)+\frac{1545209}{57600}\right)-\frac{189 \bar{G}^5}{32r^2}\right\}\\
    +\frac{\bar{B}_\text{irr}}{r^2} \left[\bar{G}^2 \omega ^2 \left(\frac{214}{525} \log (\mu  r)+\frac{111383}{441000}\right)+\frac{1}{5}\right]\,,
\label{eq:psiZRappmain}
\end{multline}
where $\bar G \equiv GM$, $\mu$ is the renormalization scale, and $\bar B_\text{reg}$ and $\bar B_\text{irr}$ are the (frequency-dependent) renormalized integration constants of the flat-space Zerilli equation. 
Equation~\eqref{eq:psiZRappmain} can then be used to construct the gravito-electric component $E_{ij}$ of the Weyl tensor [see Eq.~\eqref{eq:ErrevenZ} of Appendix~\ref{app:ppEFT}]. Matching this result to the one-point function computed from the EFT action in Eq.~\eqref{ErrD4} allows us to relate the renormalized tidal couplings $c_E$ and $c_{\dot E}$ to the coefficients $\bar B_{\rm reg}$ and $\bar B_{\rm irr}$ via [see Eqs.~\eqref{eq:KpmcEs} and \eqref{Bratio}]:\footnote{We stress that the couplings on the left-hand side of Eq.~\eqref{eq:ccBB} are the renormalized ones. For ease of notation, we drop the superscript R that distinguishes them from the bare couplings. The distinction should be clear from the context and should not be a source of confusion.}
\begin{equation}
    c_E + \omega^2   \frac{R_\star^3}{GM}c_{\dot E} = \frac{1}{180R_\star^5} \frac{\bar B_\text{irr}(\omega)}{ \bar B_\text{reg}(\omega)}\,,
\label{eq:ccBB}
\end{equation}
where, consistently with our perturbative framework, the right-hand side should be expanded to quadratic order in $\omega r_s$, where $r_s=2GM$ is the Schwarzschild radius of the star.
Once the Zerilli solution \eqref{eq:psiZRappmain} is matched to the relativistic solution for $\Psi_\text{Z}$, and the coefficients $\bar B_\text{reg}$ and $\bar B_\text{irr}$ are fixed in terms of the parameters of the full theory, the Love numbers $c_E$ and $c_{\dot E}$ are completely determined by the neutron star’s properties and EoS via Eq.~\eqref{eq:ccBB}. This matching will be discussed in Sec.~\ref{sec:Matching}.

\section{Stellar perturbation theory}
\label{sec:UV}
\noindent
In this section, we describe time-dependent, linear relativistic stellar perturbation theory. 
We employ two complementary approaches: one which is nonperturbative in the frequency and is valid to any order in $M\omega$ (Sec.~\ref{subsection:nonperturbative}) and another which is perturbative in the frequency and valid up to quadratic order (Sec.~\ref{subsection:perturbative}). We set $G=1$ throughout this section.

\subsection{Background configuration}
\noindent
We consider as a background a  static and spherically symmetric configuration. In Schwarzschild-like coordinates~$(t,r,\theta,\varphi)$, the line element takes the form
\begin{align}
    g_{\mu\nu} \d x^\mu \d x^\nu = -\e^\nu \d t^2+\e^\lambda \d r^2 + r^2\left(\d\theta^2+\sin^2\theta \,\d\varphi^2\right),\label{eq:BGmetric}
\end{align}
where $\nu=\nu(r)$ and $\lambda=\lambda(r)$. We describe the matter configuration by a perfect-fluid model. The energy-momentum tensor of the unperturbed fluid is given by
\begin{align}
    T_{\mu \nu}=\left(\epsilon+p\right)u_\mu u_\nu +pg_{\mu \nu}\,,\label{eq:BGfluid}
\end{align}
where $\epsilon$ and $p$ are the total energy density and pressure, respectively. The unperturbed fluid four-velocity~$u^\mu=(\e^{-\nu/2},0,0,0)$ is normalized by $g_{\mu \nu} u^\mu u^\nu=-1$.

Within the present setup, Einstein's equations reduce to the standard set of Tolman--Oppenheimer--Volkov differential equations:
\begin{align}
    m'&=4\pi r^2 \epsilon\,,\\
    \nu'&=\frac{2m+8\pi r^3 p}{r\left(r-2m\right)}\,,\\
    p'&=-\left(\epsilon+p\right)\frac{m+4\pi r^3p}{r\left(r-2m\right)}\,,
\end{align}
where the prime denotes a derivative with respect to $r$ and $m(r)=(r/2)(1-\e^{-\lambda})$ is the gravitational mass in a sphere of radius $r$. The system is closed by imposing the EoS of the fluid composing the star, which we assume to be barotropic, i.e.~$p=p(\epsilon)$. The radius of the star $r=R_\star$ is defined by $p(R_\star)=0$. In the exterior $r>R_\star$, $\epsilon=p=0$ and the geometry is described by the Schwarzschild metric, i.e.~$m=M$, $\e^\nu = \e^{-\lambda}=1-2M/r$.

\subsection{Dynamical tidal perturbations: Nonperturbative treatment in frequency} \label{subsection:nonperturbative}

\subsubsection{Lindblom-Detweiler equations}
\noindent
We consider nonradial, time-dependent perturbations of the background \eqref{eq:BGmetric}--\eqref{eq:BGfluid} with polar parity. Following Lindblowm and Detweiler (LD)\,\cite{Lindblom:1983ps,Detweiler:1985zz}, we choose the Regge--Wheeler gauge\,\cite{Regge:1957td} and expand the metric perturbations as: 
\begin{multline}
\delta g_{\mu\nu}\d x^\mu \d x^\nu=- r^\ell \left[\e^\nu H^{\ell m}_0 \d t^2-2i \omega r H^{\ell m}_1 \d t\d r\right.\\
\left.+\e^\lambda H_0^{\ell m} \d r^2 +r^2 K^{\ell m}(\d\theta^2+\sin^2\theta \d\varphi) \right]Y_{\ell m}\e^{-i\omega t}\,,\label{eq:PerturbedMetric}
\end{multline}
where $Y_{\ell m}=Y_{\ell m}(\theta,\varphi)$ are the spherical harmonics,  $H_0^{\ell m}=H_0^{\ell m}(r)$, $H^{\ell m}_1=H^{\ell m}_1(r)$, and $K^{\ell m}=K^{\ell m}(r)$ are radial functions describing the perturbations, and we leave implicit an integration in $\d\omega$. Note the opposite sign convention for the Fourier transform to that of LD~\cite{Lindblom:1983ps,Detweiler:1985zz}.

In the perturbed configuration, the fluid variables are displaced as $\epsilon\to \epsilon+\delta \epsilon$, $\rho\to \rho+\delta \rho$, $u^\mu \to u^\mu+\delta u^\mu$. Here, $\delta$ denotes the Eulerian perturbation and is related to the Lagrangian perturbation~$\Delta$ through $\Delta=\delta +{\cal L}_\xi$, where ${\cal L}_\xi$ denotes the Lie derivative along the displacement vector field $\xi^\mu$. The latter is parametrized by two radial functions $W^{\ell m}(r)$ and $V^{\ell m}(r)$ as~\cite{Lindblom:1983ps,Detweiler:1985zz}
\begin{align}
    \xi^r& =r^{\ell-1}\e^{-\lambda/2}W^{\ell m}Y_{\ell m}\e^{-i\omega t}\,,\label{eq:expW}
\\
    \xi_A& =-r^\ell V^{\ell m} {\cal  Y}_A^{\ell m} \e^{-i\omega t}\,,\label{eq:expV}
\end{align}
where ${\cal  Y}_A^{\ell m}=(\partial_\theta Y_{\ell m},\partial_\varphi Y_{\ell m})$ are the even-parity vector spherical harmonics. Finally, we expand the Lagrangian perturbation  of the pressure in terms of a further radial function, $X^{\ell m}$, as:
\begin{align}
\Delta p=-\e^{\nu/2}r^{-\ell}  X^{\ell m}Y_{\ell m} \e^{-i \omega t}\,.
\end{align}
Henceforth, we omit the superscript $\ell m$ of the perturbation radial variables.

For a static background, the space components of the Eulerian four-velocity perturbations are $\delta u^i=u^0 \partial_t \xi^i =-i\omega \e^{-\nu/2}\xi^i$ and can thus be expanded in terms of $W,V$ from Eqs.\,\eqref{eq:expW} and \eqref{eq:expV}. By enforcing the normalization  
\begin{equation}
(g_{\mu\nu}+\delta g_{\mu\nu})(u^\mu+\delta u^\mu)
(u^\nu+\delta u^\nu)=-1\,,
\end{equation}
we find $\delta u^0=-\frac{1}{2}\e^{-\nu/2}r^\ell H_0Y_{\ell m}\e^{-i\omega t}$.

Then, the conservation laws of the baryon number current and the energy-momentum tensor for the perturbed configuration allow us to express $\Delta \epsilon$ as
\begin{multline}
    \Delta \epsilon=\left(\epsilon+p\right)r^\ell Y_{\ell m} \e^{-i\omega t}\\
    \times \left[\frac{H_0}{2}+K-\frac{\ell\left(\ell+1\right)}{r^2}V-\frac{\left(\ell+1\right)W+r W'}{\e^{\lambda/2} r^2}\right].
\end{multline}
Assuming that the perturbed configuration remains barotropic, $\Delta p$ is related to $\Delta \epsilon$ via
\begin{equation}
    \Delta p =c_s^2 \Delta \epsilon\,, 
\end{equation}
where $c_s:=(\partial p/\partial \epsilon)^{1/2}$ is the adiabatic speed of sound. 

Then, the linearized Einstein equations and the energy-momentum conservation reduce to the {\it LD equations}~\cite{Lindblom:1983ps,Detweiler:1985zz}, a set of four first-order differential equations in the variables~$(H_1,K,W,X)$:
\begin{widetext}
\begin{align}
H_1'&=\frac{\e^\lambda}{r} H_0+\left[4\pi r \left(\epsilon-p\right)\e^\lambda -\frac{2m \e^\lambda}{r^2}-\frac{\ell+1}{r}\right]H_1+\frac{\e^\lambda}{r} K-16 \pi\frac{\epsilon+p}{r}\e^\lambda V\,,\label{eq:eqforpH1}\\
K'&=\frac{H_0}{r}+\frac{\ell\left(\ell+1\right)}{2r}H_1+\left(\frac{\nu'}{2}-\frac{\ell+1}{r}\right)K-8\pi\frac{\epsilon+p}{r}\e^{\lambda/2}W\,,\label{eq:eqforpK}\\
W'&=\frac{r}{2}\e^{\lambda/2}H_0 +r \e^{\lambda/2}K+\frac{r \e^{(\lambda-\nu)/2}}{c_s^2\left(\epsilon+p\right)}X-\frac{\ell+1}{r}W-\frac{\ell\left(\ell+1\right)}{r}\e^{\lambda/2}V\,,\label{eq:eqforpW}\\
X'&=\frac{\epsilon+p}{2}\e^{\nu/2}\left(\frac{1}{r}-\frac{\nu'}{2}\right)H_0+\frac{\epsilon+p}{2}\e^{\nu/2}\left[\omega^2 r \e^{-\nu}+\frac{\ell\left(\ell+1\right)}{2r}\right]H_1+\frac{\epsilon+p}{2}\e^{\nu/2}\left(\frac{3\nu'}{2}-\frac{1}{r}\right)K  \nonumber \\ \label{eq:eqforpX}
&\quad-\frac{\ell}{r}X-\frac{\epsilon+p}{r}\e^{(\lambda+\nu)/2}\left[\omega^2 \e^{-\nu}+\frac{\e^{-\lambda}}{4r^2}\left\{\e^{2\lambda}\left(1+8\pi r^2 p \right)^2+2\e^\lambda\left(3+8\pi r^2 p \right)-7\right\}\right]W+\frac{\ell\left(\ell+1\right)}{r^2}\e^{\nu/2}p' V\,,
\end{align}
supplemented by two algebraic equations which give the remaining functions $(H_0,V)$ in terms of $(H_1,K,W,X)$:
\begin{multline}
\left[3m+\frac{\left(\ell+2\right)\left(\ell-1\right)}{2}r+4\pi r^3 p \right]H_0=-\left[\frac{\ell\left(\ell+1\right)}{2}\left(m+4\pi r^3 p\right) -\omega^2 r^3 \e^{-\lambda-\nu}\right]H_1\\ +\left[\frac{\left(\ell+2\right)\left(\ell-1\right)}{2}r-\omega^2 r^3 \e^{-\nu}-\frac{e^\lambda}{r}\left(m+4\pi r^3 p \right)\left(3m-r+4\pi r^3 p \right)\right]K+8 \pi r^3\e^{-\nu/2}X\,,\label{eq:eqforV}
\end{multline}
\begin{equation}
    X=\frac{\epsilon+p}{2}\e^{\nu/2}H_0-\frac{p'}{r}\e^{(\nu-\lambda)/2}W+\omega^2 \left(\epsilon+p\right)\e^{-\nu/2}V\,.\label{eq:eqforX}
\end{equation}
\end{widetext}

To obtain physical acceptable solutions, we integrate Eqs.~\eqref{eq:eqforpH1}--\eqref{eq:eqforpX} 
imposing regularity  at the center of the star.
Given the two central amplitudes~$(K,W)|_{r=0}$, one can determine the central values of the other variables by~\cite{Detweiler:1985zz}
\begin{align}
    H_1\big|_{r=0}&=\frac{1}{\ell(\ell+1)}\left[2\ell K+16\pi(\epsilon+p)W\right]\bigg|_{r=0}\,,\\
    X\big|_{r=0}&=(\epsilon+p) \e^{\nu/2} \nonumber\\
    & \times \left[\left(\frac{4\pi (\epsilon+3p)}{3}-\frac{\omega^2\e^{-\nu}}{\ell}\right)W+\frac{K}{2}\right]\bigg|_{r=0}\,.
\end{align}
Moreover, we impose that the Lagrangian pressure perturbation vanishes at the surface of the star, i.e.~$X(R_\star)=0$. This fixes the ratio $W/K|_{r=0}$, and we are left with one free constant at the center of the star, which 
controls the overall normalization of the perturbations and hence does not affect the extraction of the TLNs. We set it to unity without loss of generality, thereby constructing a unique regular perturbed configuration for a given stellar model.  

In Sec.~\ref{sec:NStidalresponse}, we will match the interior solution obtained above with an exterior solution at $r=R_\star$. To this end, we construct the Zerilli function~$\Psi_\mathrm{Z}$ and its derivative at $r=R_\star$ through~\cite{Sotani:2001bb}
\begin{align}
  \Psi_\mathrm{Z}(r)=&\, \frac{r^{\ell+2}}{n r+3M}\left(K-\e^\nu H_1\right)\,,\nonumber\\
\frac{\d\Psi_\mathrm{Z}}{\d r_*}(r)=&\, \frac{r^\ell}{\left(n r+3M\right)^2}\left[-(nr^2-3M(n r+M))K\right.\label{eq:LogDeriveOfZerilli}\\
&\left.+\e^\nu\left(n (n +1)r^2+3M(n r+2M)\right)H_1\right]\,,\nonumber
\end{align}
where $r_*=r+2M\log (r/2M-1)$ is the Schwarzschild tortoise coordinate, and $n\equiv(\ell+2)(\ell-1)/2$. The matching to the exterior solution, outlined in Sec.~\ref{sec:NStidalresponse}, will be performed by requiring the logarithmic derivative of $\Psi_\mathrm{Z}$ to be continuous at the stellar surface. To this end, we define the interior logarithmic derivative as
\begin{align}
\left.y^{\rm int}(\omega)\equiv \frac{2M}{\Psi_\mathrm{Z}}\frac{\d\Psi_\mathrm{Z}}{\d r_*}\right|_{r=R_\star}\,,\label{eq:yint}
\end{align}
where the quantities on the right-hand side are computed from Eq.~\eqref{eq:LogDeriveOfZerilli}.

We compute $y^{\rm int}(\omega)$ in the 
low-frequency regime, $\omega r_s\ll1$. Although the equations become numerically unstable  
as $\omega\to0$, they 
yield reliable results for $\omega r_s\gtrsim0.01$, where Eq.~\eqref{eq:yint}
is well approximated by a quadratic fit $y^{\rm int}\simeq y^{\rm int}_0+(\omega r_s)^2y^{\rm int}_2$. 
The leading-order coefficient agrees with the result for static perturbations (see e.g.~Ref.~\cite{Hinderer:2007mb}) within percent accuracy.

\subsubsection{Alternative formulation}
\noindent
As an independent analysis, we solve the stellar perturbation problem within a different formulation, equivalent to that of LD. The LD equations are reformulated as three differential equations~\cite{Katagiri:2025qze}:
\begin{align}
    {\cal H}''& =\alpha_{\cal H, H'}{\cal H}'+\alpha_{\cal H,H}H+\alpha_{{\cal H},W} W+\alpha_{{\cal H},V}V\,,~\label{eq:eqforppH}\\
    W'& =\alpha_{W,{\cal H}'}{\cal H}'+\alpha_{W,{\cal H}}{\cal H}+\alpha_{W,W}W+\alpha_{W,V}V\,,\label{eq:eqforpWp}\\
    V'& =\alpha_{V,{\cal H}}{\cal H}'+\alpha_{V,{\cal H}}{\cal H}+\alpha_{V,W}W+\alpha_{V,V}V\,,\label{eq:eqforpV}
\end{align}
where the radial function~${\cal H}$ is defined through ${\cal H}\equiv -r^\ell H_0$ in Eq.~\eqref{eq:PerturbedMetric}. The coefficients depend on the background quantities and $\omega$. The explicit forms are provided in Ref.~\cite{gitTK}. As in the LD equations, there are two unconstrained parameters~$({\cal H},W)|_{r=0}$. Their ratio is fixed by imposing $\Delta p=0$ at $r=R_\star$ and the remaining central amplitude is set to unity without loss of generality. We obtain a regular interior solution for any given stellar model.

To match the interior solution with the exterior at linear order in the perturbation, we compute $y^{\rm int}$ given in Eq.~\eqref{eq:yint} by using Eq.~\eqref{eq:LogDeriveOfZerilli}. The functions~$H_1$ and $K$ are obtained from ${\cal H}$, $W$, and $V$ by using the following relations:
\begin{align}
-i \omega r^{\ell+1}H_1& =\alpha_{H_1,{\cal H}'} {\cal H}'+\alpha_{H_1,{\cal H}} {\cal H}+\alpha_{H_1,V} V+\alpha_{H_1,W}W\,,\label{eq:H1}\\
- r^\ell K& =\alpha_{K,{\cal H}'} {\cal H}'+\alpha_{K,{\cal H}} {\cal H}+\alpha_{K,V} V+\alpha_{K,W} W\,,\label{eq:K}
\end{align}
where the coefficients can be found in Ref.~\cite{gitTK}. We have verified that the values computed in this approach agree  with those in the LD approach.

\subsection{Dynamical tidal perturbations: perturbative treatment in frequency}\label{subsection:perturbative}
\noindent
To match the perturbation theory result with the EFT expansion \eqref{eq:Sintmain}, it is useful to obtain the perturbative solutions order-by-order in $\omega r_s\ll1$, up to quadratic order.\footnote{Note that, although   the action \eqref{eq:Sintmain} is formally expressed as an expansion in the number of time derivatives, one could instead keep the EFT nonperturbative in frequency (while still expanding in spatial gradients) and match the relativistic result directly to the kernel $K_{+-}^{(E)}\left(\omega\right)$ defined in Appendix~\ref{app:ppEFT}.}

First, we expand ${\cal H}, V, W$ as
\begin{align}
    {\cal H}& ={\cal H}_{0}+\left(\omega r_s\right)^2 {\cal H}_{2}\,,\nonumber\\
    V& =V_{0}+\left(\omega r_s\right)^2 V_{2}\,,\label{eq:PerturbativeHVW}\\
    W& =W_{0}+\left(\omega r_s\right)^2 W_{2}\,.\nonumber
\end{align}
Then, Eqs.~\eqref{eq:eqforppH}--\eqref{eq:eqforpV} reduce to 
\begin{align}
    {\cal L}_{\cal H}\left[ {\cal H}_0\right]& =0\,,\label{eq:eqforH0}\\
    {\cal L}_V\left[V_0,W_0,{\cal H}_0\right]& =0\,,\label{eq:eqforV0}\\
    {\cal L}_W\left[W_0,V_0,{\cal H}_0\right]& =0\label{eq:eqforW0}\,,
\end{align}
at the leading order, and
\begin{align}
    {\cal L}_{\cal H}\left[{\cal H}_2\right]&={\cal S}_{\cal H}\,,\label{eq:eqforH2}\\\
     {\cal L}_V\left[V_2,W_2, {\cal H}_2\right]&={\cal S}_V\,,\\
      {\cal L}_W\left[W_2,V_2,{\cal H}_2\right]&={\cal S}_W\,,\label{eq:eqforW2}
\end{align}
at the quadratic order. Here, we introduce the differential operators
\begin{multline}
{\cal L}_{\cal H}=\frac{\d^2}{\d r^2}+\left[\frac{2}{r}+\e^\lambda\left(\frac{2m}{r^2}+ 4 \pi r \left(p-\epsilon\right) \right)\right]\frac{\d}{\d r}\\
-\e^\lambda \left[ \frac{\ell \left(\ell+1\right)}{r^2}-4 \pi \frac{p+\epsilon}{c_s^2} - 4\pi \left(5 \epsilon+9p\right) \right]-\nu'^2\,,
\end{multline}
and 
\begin{align}
{\cal L}_V\left[V_i,W_i,{\cal H}_i\right]=& \, V_i'+\beta_{V,{\cal H'}}{\cal H}_i'+ \beta_{V,{\cal H}}{\cal H}_i \label{eq:OpeLV}\\
&+\beta_{V,V}V_i+\beta_{V,W}W_i\,,\nonumber\\
{\cal L}_W\left[W_i,V_i,{\cal H}_i\right]=& \, W_i'+\beta_{W,{\cal H'}}{\cal H}_i'+ \beta_{W,{\cal H}}{\cal H}_i\label{eq:OpeLW}\\
&+\beta_{W,V}V_i+\beta_{W,W}W_i\,,\nonumber
\end{align}
where $i=0,2$, and the source terms 
\begin{align}
 {\cal S}_{\cal H}  & = S_{\cal H, H'} {\cal H}_0'+ S_{\cal H, H} {\cal H}_0 + S_{{\cal H},V} V_0 + S_{{\cal H},W} W_0\,,\label{eq:SH}\\
  {\cal S}_{V} & = S_{V, \cal  H'} {\cal H}_0'+ S_{V, \cal H} {\cal H}_0 + S_{V,V} V_0 + S_{V,W} W_0\,,\label{eq:SV}\\
   {\cal S}_W & = S_{W, \cal H'} {H}_0'+ S_{W, \cal H} {\cal H}_0 + S_{W,V} V_0 + S_{W,W} W_0\,,\label{eq:SW}
\end{align}
where the explicit forms of the coefficients are provided in Appendix~\ref{app:coefficients}.

We now briefly describe the perturbative construction of the interior solution. At zeroth order in $\omega$, Eq.~\eqref{eq:eqforH0} reduces to the standard static tidal perturbation equation, coinciding, for $\ell=2$, with Eq.~(15) of Ref.~\cite{Hinderer:2007mb}. After obtaining the regular solution for ${\cal H}_0$ at the stellar center, we solve Eqs.~\eqref{eq:eqforV0} and~\eqref{eq:eqforW0} simultaneously to determine the corresponding regular solutions for $V_0$ and $W_0$. 

At quadratic order in $\omega$, the reduced equations~\eqref{eq:eqforH2}--\eqref{eq:eqforW2} are sourced by the leading-order solutions. Solving them with the same regularity conditions yields ${\cal H}_2$, $V_2$, and $W_2$. As in the nonperturbative analysis, we impose $\Delta p=0$ at $r=R_\star$ up to ${\cal O}(\omega^2r_s^2)$, thereby selecting the physical perturbative solution.

Substituting the perturbative solutions for ${\cal H}$, $V$, and $W$ into the low-frequency expansions of Eqs.~\eqref{eq:H1} and~\eqref{eq:K}, we compute $H_1$ and $K$ perturbatively and construct the logarithmic derivative of the Zerilli function through Eq.~\eqref{eq:LogDeriveOfZerilli}. As shown in Appendix~\ref{app:compNP}, the resulting perturbative expression for $y^{\rm int}$ in Eq.~\eqref{eq:yint}, truncated at ${\cal O}((\omega r_s)^2 )$, follows the fully numerical result throughout most of the frequency range relevant to the GW data analysis of Sec.~\ref{sec-dataanalysis}.

Note that the inhomogeneous equations~\eqref{eq:eqforH2}--\eqref{eq:eqforW2} are defined
modulo the addition of the homogeneous solution which is regular at the center (with an arbitrary amplitude).
However, as expected, $y^{\rm int}$ is independent of the particular choice of the homogeneous regular solution for ${\cal H}_2$, $V_2$, and $W_2$, since this ambiguity only affects the internal decomposition of the regular solution, whose physical content is already fixed (up to normalization) by regularity and the boundary condition $\Delta p=0$.

\vspace{0.5cm}

\section{Dynamical tidal response of a neutron star}
\label{sec:NStidalresponse}
\noindent
In this section, we first describe the procedure to extract the dynamical tidal response of a neutron star. We then present our main results and discuss the qualitative and quantitative behavior of the dynamical TLN~$c_{\dot E}$ computed for several stellar models.

\subsection{Extraction of the tidal response}\label{sec:Matching}
\noindent
The tidal perturbation in the exterior is described by the Zerilli function~$\Psi_\mathrm{Z}$, which is obtained by solving the Zerilli equation
\begin{align}
    &f(r)\partial_r\left[f(r) \partial_r \Psi_{\mathrm{Z}}\right]\\
    &+\left[\omega^2-f(r)\left(\frac{r_s}{r^3}+\frac{2 n}{3 r^2}+\frac{8 n^2(2 n+3)}{3\left(2 n r+3 r_s\right)^2}\right)\right] \Psi_{\mathrm{Z}}=0\,,\nonumber
\end{align}
where $f(r)= 1-r_s/r$.
In the low-frequency expansion~$\omega r_s\ll 1$, the general solution for $\Psi_\mathrm{Z}$ can be written schematically, up to quadratic order, as 
\begin{align}
    \Psi_\mathrm{Z}(r)=\Psi_{\mathrm{Z},0}(r) +\left(\omega r_s\right)^2 \Psi_{\mathrm{Z},2}(r)\,.
\end{align}
Following the procedure outlined in Appendix~\ref{app:matchingtoGR}, we obtain the general solution for $\Psi_{\mathrm{Z},0}$ and $\Psi_{\mathrm{Z},2}$. The large-distance expansion of $\Psi_\mathrm{Z}$ for the quadrupole mode, $\ell=2$, then reads:\footnote{Since we will connect the exterior solution to the interior solution obtained from a nondissipative stellar model, we have set $a_1=b_1=0$ in Eq.~\eqref{Zer_sol_inf_GR}.}
\begin{widetext}
    \begin{align}
     \Psi_{\mathrm{Z}}^{\ell=2}(r) \xrightarrow{r\rightarrow\infty}
      &\,\, a_0\frac{r^3}{r_s^3}-\left(\frac{189 }{1024}a_0-\frac{1 }{5}b_0\right)\frac{r_s^2 }{r^2}\nonumber\\
     &+\omega^2r_s^2\left[\frac{r^3 }{ r_s^3}\left( a_2+\frac{107}{210}a_0 \log \left(\frac{r_s}{r}\right) \right) +
     \left(b_2 -\left(\frac{107}{1050} b_0 +\frac{3011}{10240}a_0\right)\log \left(\frac{r_s}{r}\right)\right)\frac{r_s^2}{r^2}\right]+\cdots,
\label{eq:psiZ2txt}
\end{align}
\end{widetext}
where $a_0,b_0,a_2,b_2$ are integration constants. The full solution, including subleading terms in the $1/r$ expansion, is provided in the Mathematica code available at the GitHub repository~\cite{gitTA}. We match this expression to the renormalized Zerilli function~$\Psi_\mathrm{Z}^{\rm R}$ given in Eq.~\eqref{eq:psiZRappmain}, thereby relating $a_0,b_0,a_2$, and $b_2$ to the renormalized integration constants of $\Psi_\mathrm{Z}^{\rm R}$, namely $\bar{B}_{\rm reg}$ and $\bar{B}_{\rm irr}$, via\footnote{See Eqs.~\eqref{eq:Breg}--\eqref{eq:Birr} with $a_1=b_1=0$.}
\begin{widetext}
\begin{align}
r_s^3\bar{B}_\text{reg}&  = a_0 +  \omega^2 r_s^2\left[a_2+  a_0 \left(\frac{2731}{39200}+\frac{107}{210}\log (\mu  r_s)\right) \right],
\\
r_s^{-2}\bar{B}_\text{irr} & = b_0 + \omega^2r_s^2 \left[\frac{945 }{1024}a_2 +5 b_2 -\frac{1015283 }{1032192}a_0-\frac{111383}{352800}b_0 -  
\left(a_0+\frac{107}{210}b_0\right) \log (\mu   r_s)
\right].
\end{align}
\end{widetext}
Thus, once $a_0$, $b_0$, $a_2$, and $b_2$ are determined by matching to the interior solution, the static and dynamical TLNs, $c_E$ and $c_{\dot E}$, can be obtained from Eq.~\eqref{eq:ccBB}, as shown in Appendix~\ref{app:matchingtoGR}:
\begin{align}
    c_E & = \frac{8G^5M^5}{45 R_\star^5} \frac{b_0}{a_0}\,,\label{eq:CE}
\\
    c_{\dot E} & = \frac{32G^8M^8}{45 R_\star^8} \bigg[\!- \left(1+\frac{107}{105}\frac{b_0}{a_0}\right)\log (\mu r_s)
\nonumber \\
&\qquad\qquad\quad\,\,    -\frac{67981 }{176400}\frac{ b_0}{ a_0}
-\frac{b_0a_2}{a_0^2} \label{eq:cEdot} \\
&\qquad\qquad\quad\,\, + \frac{945 a_2}{1024 a_0}+\frac{5 b_2}{a_0}-\frac{1015283}{1032192}
    \bigg].\nonumber
\end{align}
Before proceeding, it is worth commenting on the structure of the result \eqref{eq:cEdot}. Note that the coefficient of $\log(\mu r_s)$ in $c_{\dot E}$ contains two types of contributions. One is a universal term, independent of the integration constants, which arises solely from the gravitational nonlinearities of the Einstein--Hilbert action and is therefore common to any object, in particular black holes (see, e.g., Refs.~\cite{Chakrabarti:2013lua,Saketh:2023bul,Chakraborty:2025wvs,Combaluzier--Szteinsznaider:2025eoc,Kobayashi:2025vgl}). The other contribution is $107 b_0/(105 a_0)$, which is proportional to the static Love number $c_E$ and is thus unambiguously fixed by the ratio $b_0/a_0$.
This should be contrasted with the remaining (scheme-dependent) finite terms on the second and third lines of Eq.~\eqref{eq:cEdot}, which depend explicitly on the free parameters $a_2$ and $b_2$, and therefore require a full matching at second order in $\omega$ to be determined.

The exterior solution for $\Psi_\mathrm{Z}$ is then matched to the interior solution, obtained in the previous section, by requiring continuity of the logarithmic derivative of the Zerilli function at the stellar surface. For the exterior solution, we introduce
\begin{align}
 y^{\rm ext}(\omega) \equiv &\left.\frac{2M}{\Psi_\mathrm{Z}}\frac{\d\Psi_\mathrm{Z}}{\d r_*}\right|_{r=R_\star}.\label{eq:yext}
\end{align}
By equating the exterior logarithmic derivative $y^{\rm ext}$ to the interior one, $y^{\rm int}$, given in Eq.~\eqref{eq:yint}, we match the interior and exterior solutions for $\Psi_Z$ to second order in $\omega r_s$. This determines the ratios between $b_0$, $a_2$, and $b_2$ with respect to $a_0$ for a given stellar model, from which the TLNs $c_E$ and $c_{\dot E}$ are obtained.

\subsection{Dynamical Love numbers}
\label{sec:results}
\noindent

\begin{figure}[t]
    \centering
    \includegraphics[width=1\linewidth]{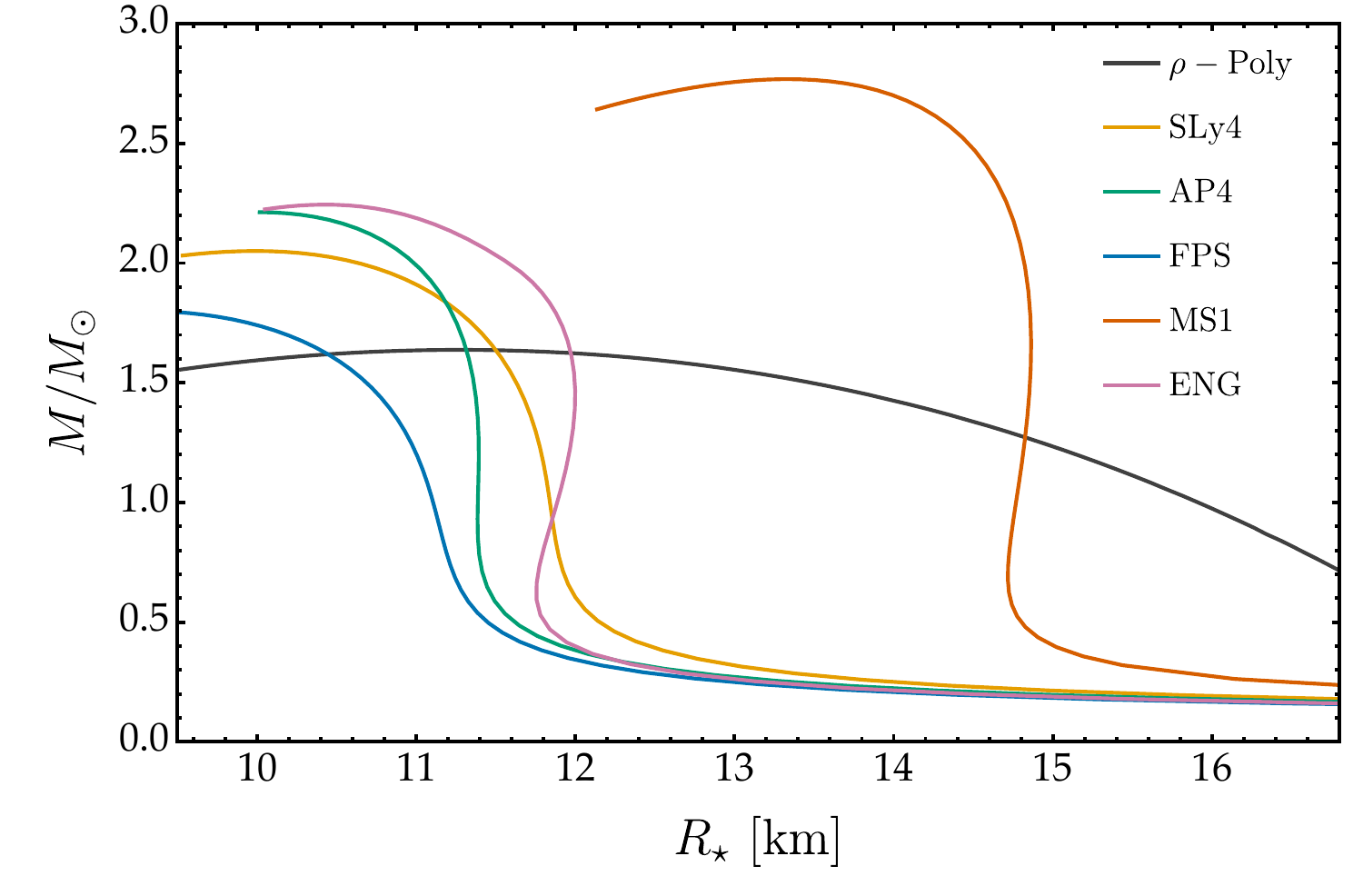}
\caption{Mass-radius relations for neutron stars with several nuclear-physics tabulated EoS considered in this work, together with the rest-mass polytrope~($\rho$-Poly), defined by $\epsilon=(p/K_P)^{1/2}+p$ with $K_P=100 M_\odot^2$, as a reference.}
\label{fig:MassRadiusRelation}
\end{figure}

\begin{figure*}[t]
    \centering
    \includegraphics[width=0.49\linewidth]{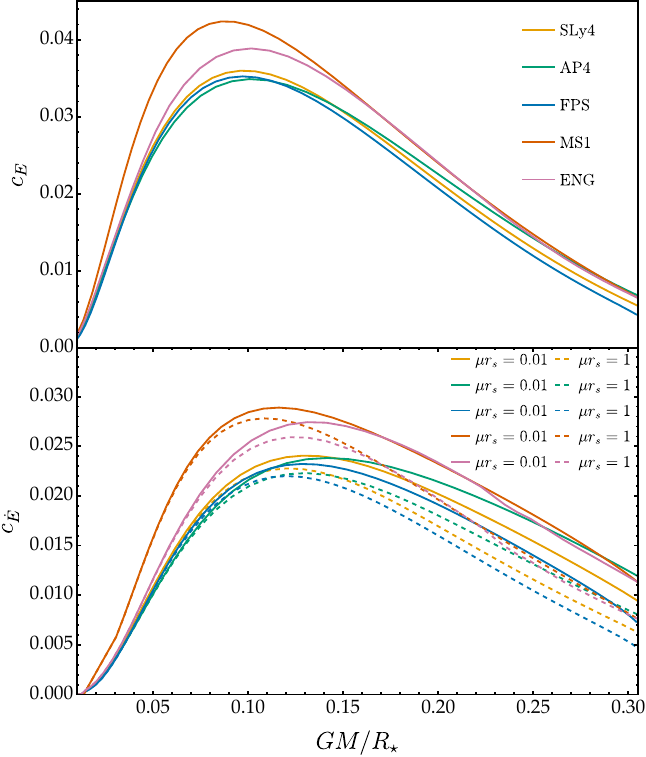}
        \includegraphics[width=0.49\linewidth]{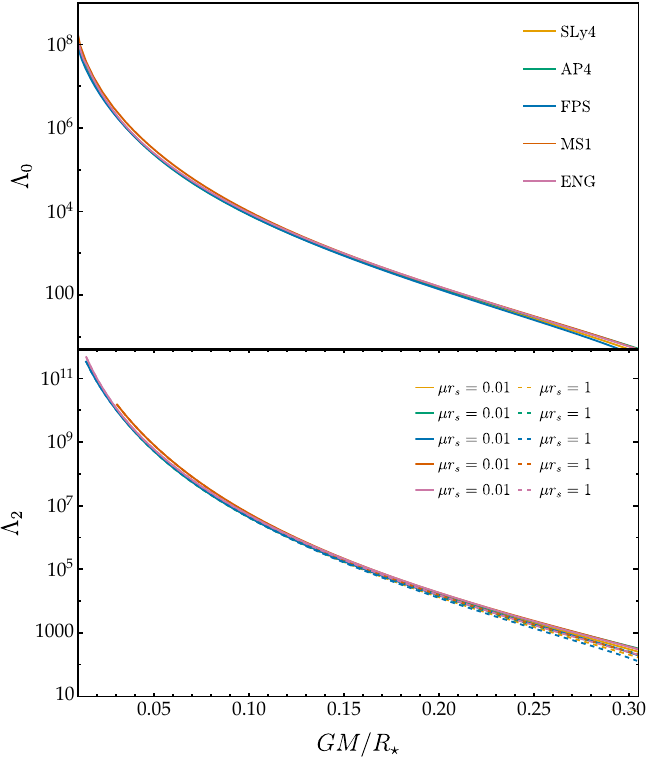}
\caption{Left: The static TLNs~$c_E$ (top panel) and the dynamical TLNs~$c_{\dot E}$ (bottom panel) for the realistic EoS as functions of compactness. Here, $\mu$ is normalized by the Schwarzschild radius~$r_s=2GM$. Right: The corresponding $\Lambda_0$ (top) and $\Lambda_2$ (bottom) as functions of compactness. } 
\label{fig:TLNs}
\end{figure*}

\noindent
In what follows, we present several properties of the dynamical tidal response of neutron stars. We emphasize that the TLNs computed through Eqs.~\eqref{eq:CE} and~\eqref{eq:cEdot} are defined within the point-particle EFT framework and admit a direct connection to the gravitational waveform discussed in Sec.~\ref{sec:waveform}. For the reader's convenience, in Appendix~\ref{app:comparison} we summarize the connections with the tidal constants introduced in alternative approaches based on the PN framework~\cite{Poisson:2020vap,Pitre:2023xsr,HegadeKR:2024agt}.

First of all, we construct several stellar configurations using realistic, tabulated EoS models released online~\cite{lalsuite}. Specifically, we consider SLy4~\cite{Douchin:2001sv}, AP4~\cite{Akmal:1998cf}, FPS~\cite{1981NuPhA.361..502F}, MS1~\cite{Mueller:1996pm}, and ENG~\cite{Engvik:1994tj}. Figure~\ref{fig:MassRadiusRelation} presents the mass-radius relations for the stellar models studied in this work.

The results of the matching are shown in Fig.~\ref{fig:TLNs}. The upper-left and lower-left panels display the static ($c_E$) and dynamical ($c_{\dot E}$) TLNs, respectively, as functions of the neutron-star compactness for various EoS. Accounting for the different conventions adopted in the 
literature---namely $c_E=k_0/3$, where $k_0$ is the Love number defined in Refs.~\cite{Hinderer:2007mb,Binnington:2009bb}---the upper panel reproduces  the results for the static TLN reported in previous works, see e.g.~\cite{Hinderer:2009ca,
Postnikov:2010yn}.

The bottom panel shows that, for the EoS considered in this work, $c_{\dot{E}}$ exhibits trends qualitatively similar to those of $c_E$: (i)~it reaches a maximum around $GM/R_\star \simeq 0.12$; (ii) it decreases toward both the low- and high-compactness regimes; and (iii) stiffer EoS generally yield larger peak values. A distinctive feature of $c_{\dot E}$ is its explicit scale dependence through the $\log (\mu r_s)$ term in Eq.~\eqref{eq:cEdot}. We find that this dependence is relatively mild at low compactness, whereas smaller values of $\mu$ enhance the peak of $c_{\dot{E}}$ and produce a slower decay in the high-compactness regime.
This is consistent with the fact that the running term is a relativistic effect.

\begin{figure}[t]
    \centering
    \includegraphics[width=1.01\linewidth]{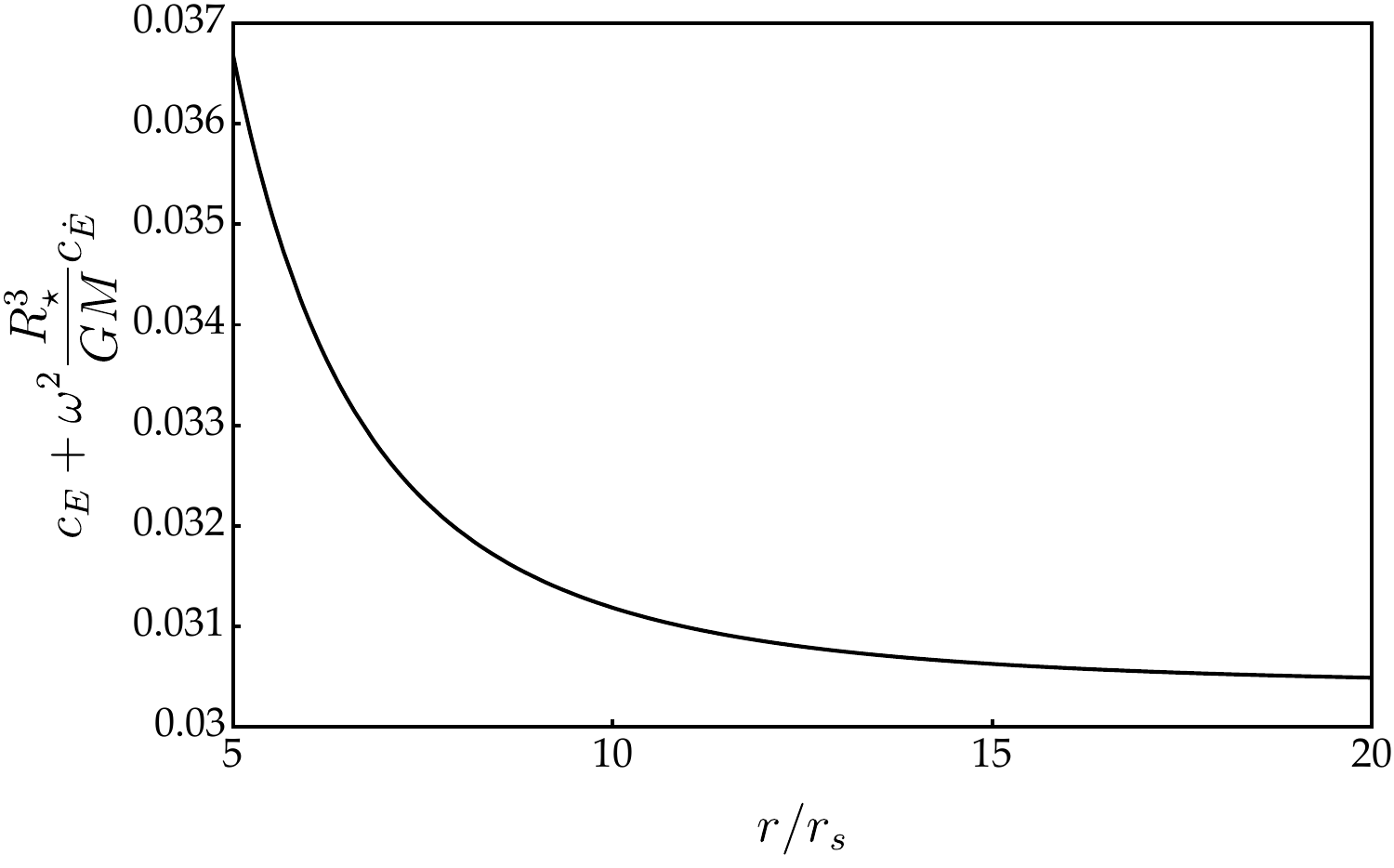}
\caption{Estimated evolution of $c_E+\omega^2 R_\star^3/(GM) c_{\dot E}$ as a function of the orbital length scale~$r$. Here, $c_E$ and $c_{\dot E}$ are computed for the AP4 EoS with $GM/R_\star\simeq 0.153$. We set $\omega=(G M/r^3)^{1/2}$ and $\mu=1/r$, so that the logarithmic scale dependence of $c_{\dot E}$ is evaluated at the orbital scale.}
\label{fig:Evolution_tides}
\end{figure}

It is worth noting that $\mu$ is associated with the characteristic length scale of the binary system and is therefore expected to scale as $\mu\sim 1/r$, where $r$ is the orbital separation. This implies that $\mu r_s$ is much smaller than unity at the early stage of the late-inspiral phase and grows to ${\cal O}(10^{-1})$ as the binary evolves. 
Equation~\eqref{eq:cEdot} then indicates that, because the logarithmic term enters with a minus sign, $c_{\dot E}$ may decrease during the binary evolution. In this respect, the behavior is reminiscent of an asymptotically free coupling: as the binary shrinks, the renormalization scale $\mu \sim 1/r$ grows and the dynamical Love number coupling decreases logarithmically, much like the running of the strong coupling constant in quantum chromodynamics toward higher energies. In both cases, the running is driven by the self-interactions of the mediator: in quantum chromodynamics, gluons carry color charge and interact among themselves, antishielding the source and yielding a negative beta function; in gravity, gravitons carry energy-momentum and self-interact through the nonlinearities of general relativity, generating the analogous logarithmic flow of the tidal coupling. However, the decrease in $c_{\dot E}$ does not necessarily imply that the tidal response is suppressed in the late inspiral. Indeed, Fig.~\ref{fig:Evolution_tides} shows that the combination~$c_E+\omega^2 R_\star^3/(G M) c_{\dot E}$ of Eq.~\eqref{eq:ccBB} increases as the orbital length scale decreases. This behavior results from the competition between the power-law growth in $\omega^2 (r)$ and the logarithmic running. In the range shown, the former dominates over the latter.

In the right panels of Fig.~\ref{fig:TLNs}, we show the dimensionless tidal coefficients $\Lambda_0$ and $\Lambda_2$ as functions of the stellar compactness, $C \equiv GM/R_\star$, defined by
\begin{equation}
\Lambda_0 = 2\left(\frac{GM}{R_\star}\right)^{-5} c_E,
\qquad
\Lambda_2 =2 \left(\frac{GM}{R_\star}\right)^{-8} c_{\dot E}\,.
\label{Lambdas}
\end{equation}
The top panel reproduces the well-known behavior of the static tidal deformability; see, e.g., Ref.~\cite{Yagi:2013awa}. The bottom panel shows that $\Lambda_2$ is typically about three orders of magnitude larger than $\Lambda_0$. This difference is primarily driven by the stronger compactness dependence of $\Lambda_2$, which scales as $C^{-8}$ compared to the $C^{-5}$ scaling of $\Lambda_0$. As we discuss in Sec.~\ref{sec:measurability}, $\Lambda_0$ and $\Lambda_2$ are precisely the combinations that enter the gravitational waveform and therefore provide the relevant parameters for describing dynamical tidal effects in GW observations.

\subsection{Approximate universal relations}
\label{sec:UR}
\begin{figure}[ht]
    \centering
    \includegraphics[width=1\linewidth]{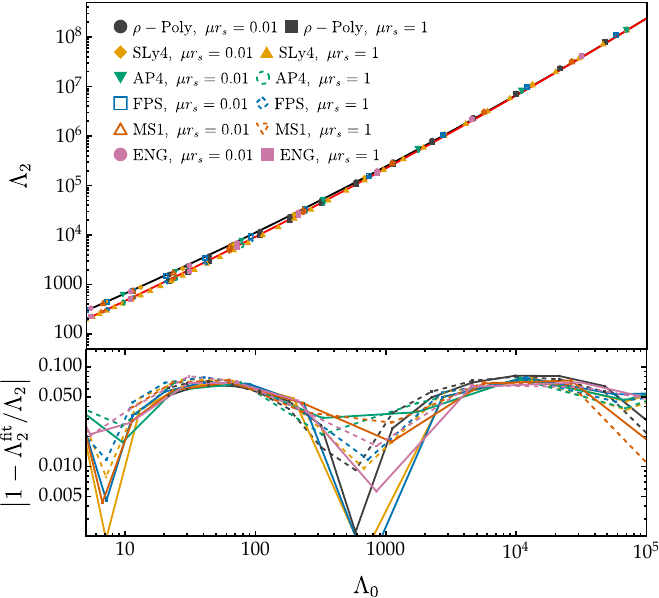}
\caption{Top: approximate universal relations between $\Lambda_0$ and $\Lambda_2$ with various EoS, demonstrating approximate universality at fixed $\mu r_s$. The black and red solid lines denote the fitting formulas~\eqref{eq:UniversalityFit} and~\eqref{eq:universalrel_murs1} with $\mu r_s=0.01$ and $\mu r_s=1$, respectively. Bottom: relative difference between the empirical fit and each data of $(\Lambda_0,\Lambda_2)$ for a given stellar model and given $\mu r_s$.}
\label{fig:UniversalRelation}
\end{figure}

\noindent
Neutron-star observables are known to exhibit approximately EoS-insensitive relations~\cite{Yagi:2013bca,Yagi:2013awa,Yagi:2016bkt}. It is therefore reasonable to expect that an approximate universal relation holds between $\Lambda_0$ and $\Lambda_2$. Indeed, such a relation was found in Ref.~\cite{Saes:2025jvr} for the dynamical Love numbers defined in that work. Here, we establish the approximate universality for our TLNs, which are defined unambiguously and directly connected to the gravitational waveform. We stress that this approximate universality becomes manifest only when using the waveform couplings $\Lambda_0$ and $\Lambda_2$, while it remains hidden for other combinations of
the tidal parameters~\cite{Majumder:2015kfa}. 
Figure~\ref{fig:UniversalRelation} shows that an approximate universal relation emerges only for fixed values of $\mu r_s$. We fit the numerical data using the following fitting formulas:
\begin{align}
    Y|_{\mu r_s=0.01}&= 3.9171 +1.0707 X+0.023100 X^2\,,\label{eq:UniversalityFit}\\
     Y|_{\mu r_s=1}&= 3.3840 +1.1621 X+0.019180 X^2\,,\label{eq:universalrel_murs1}
\end{align}
where $(X,Y)=(\log \Lambda_0, \log \Lambda_2)$. The typical relative errors between the data and these empirical fits are smaller than $10\%$, consistent with the results of Ref.~\cite{Saes:2025jvr}. 
Unlike the standard scale-independent I-Love-Q relations~\cite{Yagi:2013bca,Yagi:2013awa}, the relation between $\Lambda_0$ and $\Lambda_2$ inherits the explicit scale dependence of the dynamical Love number through $\mu$, so that the fit coefficients in Eqs.~\eqref{eq:UniversalityFit} and \eqref{eq:universalrel_murs1} change with $\mu r_s$. 

As we shall discuss in Sec.~\ref{sec:measurability}, the running of the dynamical TLNs has only a subdominant impact on GW observations, with the overall tidal effect being primarily controlled by the scale-independent component. For this reason, it will be convenient to adopt the choice $\mu r_s=1$, which allows using the approximate universal relation in Eq.~\eqref{eq:universalrel_murs1}.

\subsection{Comparison with mode-sum representation}
\noindent
As in the Newtonian case~\cite{Lai:1993di,Lai:1997wh,Ho:1998hq,Chakrabarti:2013xza,Andersson:2019ahb,Andersson:2019dwg,Passamonti:2020fur,Passamonti:2022yqp,Pnigouras:2022zpx,Yu:2024uxt,Pitre:2023xsr}, one may expect the dynamical tidal response to be closely related to the characteristic oscillation spectrum of the star, namely its quasi-normal modes. These are the complex eigenfrequencies for which the homogeneous perturbation equations discussed in Sec.~\ref{subsection:nonperturbative} admit solutions that are regular at the stellar center and purely outgoing at infinity~\cite{Kokkotas:1999bd}.

For neutron stars, the oscillation spectrum is typically dominated by the fundamental fluid ($f$-) mode, whose frequency is roughly given by~\cite{Kokkotas:1999bd}
\begin{equation}
\omega_f \simeq c_f\sqrt{\frac{GM}{R_\star^3}} \approx \mathcal{O}(1)\,\mathrm{kHz}\,,
\end{equation}
where $c_f$ is an $\mathcal{O}(1)$ coefficient that depends on the EoS (see, for example, Ref.~\cite{Andersson:1997eq} for an explicit evaluation for polytropic EoS).
At the Newtonian level, the dynamical
tidal deformability admits a mode-sum representation in which the $f$-mode contribution plays a dominant role
because of its relatively low frequency~\cite{Lai:1993di,Ho:1998hq,Andersson:2019ahb,Pitre:2023xsr}. Consistent with this picture, 
$\omega_f$ typically agrees with an effective frequency~$\omega_*$ within $0.1\%$, where $\omega_*$ is defined by~\cite{Pitre:2023xsr}\footnote{Notice that the definition of $\omega_*$ in Eq.~\eqref{eq:effectiveomega} differs slightly from the quantity introduced in Eq.~(1.14) of Ref.~\cite{Pitre:2023xsr}, as the latter is defined in terms of the tidal constants adopted therein.  }
\begin{align}
    \omega_* \equiv \sqrt{\frac{GM}{R_\star^3}\frac{c_E}{c_{\dot E}}}\,.\label{eq:effectiveomega}
\end{align}
Therefore, in the Newtonian regime the dynamical TLN can be very well approximated in terms of the $f$-mode frequency and the static TLN,
\begin{align}
    c_{\dot E}\approx\frac{GM}{R_\star}\frac{c_{E}}{\omega_f^2}\,.
\end{align}
In the relativistic regime, however, the relation between tidal response and quasi-normal modes becomes more subtle. In particular, a mode-based description may not be complete, and additional contributions can cause $\omega_*$ to deviate from $\omega_f$~\cite{Pitre:2023xsr,Saes:2025jvr,HegadeKR:2026kku}, and hence the dynamical TLN cannot be simply obtained from the quasi-normal modes.

\begin{figure}
      \includegraphics[width=0.5\textwidth]{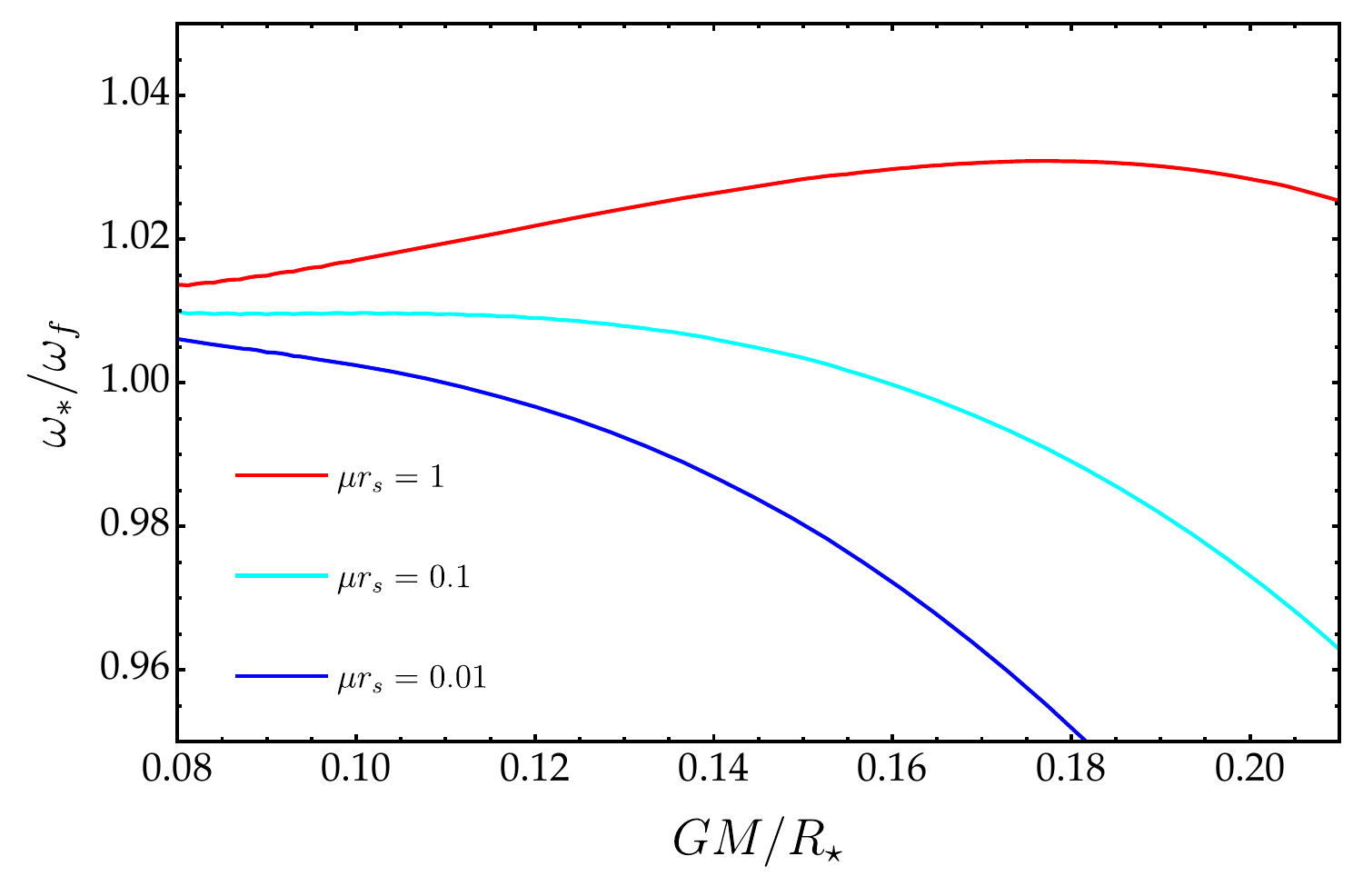}
    \caption{The ratio $\omega_*/\omega_f$ as a function of $GM/R_\star$ for the rest-mass polytrope~$\epsilon=(p/K_P)^{1/2}+p$ with $K_P=100 M_\odot^2$. Deviations from unity indicate the extent to which the dynamical tidal response is captured by the $f$-mode contribution.} 
\label{fig:fmodeapproximation}
\end{figure}

To quantify this discrepancy, in
Fig.~\ref{fig:fmodeapproximation} we show the ratio $\omega_*/\omega_f$ as a function of compactness for several values of $\mu$. For sufficiently small values of $\mu$, the deviation from unity decreases with compactness, systematically lowering the ratio $\omega_*/\omega_f$. Remarkably, throughout the most interesting interval $0.1\lesssim \mu r_s\lesssim 1$, the ratio remains within approximately $5\%$ of unity over the entire compactness range considered.
This suggests that, despite relativistic corrections, the dynamical tidal response of realistic neutron stars during the late inspiral is still largely controlled by the $f$-mode frequency. Furthermore, we stress that the ratio is compatible with the corresponding prediction obtained from the tidal constants introduced in Refs.~\cite{Poisson:2020vap,Pitre:2023xsr}, once the convention difference in Eq.~\eqref{eq:cEandPP} is accounted for (see Appendix~\ref{app:comparison} for their definition and a comparison with our dynamical TLN, as well as with other approaches~\cite{HegadeKR:2024agt,Saes:2025jvr}).

Our findings are qualitatively consistent with those of Ref.~\cite{HegadeKR:2026kku}, while providing an independent validation of this behavior within the EFT framework developed here, including the effects of renormalization-group running.

\section{Measurability with GW detections}
\label{sec:measurability}
\noindent
Having computed the dynamical TLNs of neutron stars, we now
assess their impact on the GW signal emitted during the
inspiral of a compact binary and their measurability with current and
future detectors.  In Sec.~\ref{sec:waveform} we derive the PN
waveform phase, including both the static and dynamical tidal contributions
and the distinctive logarithmic running of the latter, following Ref.~\cite{Chakraborty:2025wvs}. As shown in that work, the dynamical Love number enters the waveform at the 8PN order.  In
Sec.~\ref{sec-dataanalysis} we quantify the observability of these effects
through a Fisher-matrix analysis and a mismatch calculation, considering the
LVK O4 network and ET at its nominal sensitivity.
Together, these two analyses provide complementary views of
the same physics: the Fisher matrix quantifies the statistical precision
with which the dynamical Love number can be extracted, while the mismatch
measures the systematic error incurred by neglecting it.

\subsection{PN waveform derivation}
\label{sec:waveform}
\noindent
In this section, we use the EFT Lagrangian \eqref{eq:Sintmain}, with the renormalized induced dynamical tidal response couplings obtained in the previous sections, to study a binary inspiral. Although the renormalized couplings were derived for an isolated neutron star, no additional divergences are expected when the same Lagrangian is applied to the two-body problem. As in any EFT, once all counterterms consistent with the symmetries have been included up to a given order in the derivative expansion, the divergences are removed once and for all. The theory  becomes fully predictive at that order and can be used to compute any process consistently, with finite results. In particular, the renormalization scale $\mu$ drops out of physical observables---such as cross sections, scattering angles, or waveform phases---which depend only on the renormalized couplings evaluated at a chosen scale together with the corresponding logarithmic dependence on the relevant physical scale (e.g., energy or distance).

The procedure for deriving the interaction Hamiltonian, including the counterterms, is discussed in detail in e.g.~Ref.~\cite{Mandal:2023hqa}.\footnote{Although the renormalization scheme adopted in Ref.~\cite{Mandal:2023hqa} differs from ours, a nontrivial consistency check is that both approaches yield the same counterterms and renormalization-group running of the couplings.} For the purposes of this work, however, we follow a simpler approach~\cite{Chakraborty:2025wvs}. Since the renormalized dynamical Love number $c_{\dot E}$ depends on the renormalization scale $\mu$, and the natural scale in the binary problem is set by the inverse orbital separation, we identify $\mu \sim \gamma/r$, with $\gamma$  some constant number. Because our goal is  to estimate the potential observability of dynamical Love number effects, the precise relation between $\mu$ and $r$ will not be important.
As the inspiral proceeds and the orbital separation decreases, the tidal response evolves logarithmically, as illustrated for example by Eq.~\eqref{eq:psiZ2txt}. Importantly, this running cannot in general be absorbed into a redefinition of the scale-independent part of $c_{\dot E}$. It therefore leaves a characteristic imprint on the GW phase in the form of a frequency-dependent 8PN correction, qualitatively distinct from the standard frequency-independent tidal contributions. In practice, however, this effect is suppressed in the small-compactness regime and, as we shall see explicitly below, does not significantly affect our estimates.

To connect with the waveform calculation, recall that the dynamical couplings $c_E$ and $c_{\dot E}$ enter the effective two-body Lagrangian through Eq.~\eqref{eq:Sintmain}. Their effects on the GW phase are conveniently encoded in the dimensionless tidal parameters $\Lambda_0$ and $\Lambda_2$, defined in Eq.~\eqref{Lambdas}. It is therefore useful to express the renormalization-group running directly in terms of these quantities. Following Ref.~\cite{Chakraborty:2025wvs}, we write the running of the dynamical tidal parameter as
\begin{equation}
  \Lambda_2(\mu) \;\to\; \Lambda_2(r)
    \equiv \bar{\Lambda}_2
      + \beta_2\,\log\left(\frac{\gamma\,r_s}{r}\right),
\label{eq:Lambda2_running}
\end{equation}
where $\bar{\Lambda}_2$ captures the value of the dynamical TLN at the UV
matching scale $r_s$, given by the neutron-star Schwarzschild radius, while $\beta_2$ encodes the logarithmic running associated with the renormalization procedure. A similar approach has been followed in Ref.~\cite{Mandal:2023hqa}, where an effective radial-dependent Love number has been included in the binary Lagrangian.  The relation between
$(\bar{\Lambda}_2,\,\beta_2)$ and the EFT coupling $c_{\dot{E}}$ of Sec.~\ref{sec:EFT} is given by 
\begin{align}
  \bar{\Lambda}_2 &= 2\left(\frac{GM}{R_\star}\right)^{-8} c_{\dot{E}} \, (\mu = r_s^{-1})\,, \nonumber \\
  \quad
  \beta_2 &= 2\left(\frac{GM}{R_\star}\right)^{-8}
            c_{\dot{E}}^{(\log)} = - \frac{64}{45} \left(1 + \frac{321}{112} \Lambda_0 \right)\,,
\label{eq:Lambda2_dict}
\end{align}
where $c_{\dot{E}}^{(\log)}$ is the coefficient of $\log(\mu r_s)$ in
Eq.~\eqref{eq:cEdot}, while the static counterpart is
$\Lambda_0 = 2(GM/R_\star)^{-5}c_E$ as in Eq.~\eqref{Lambdas}.
The key observation is that, since $\Lambda_2\propto C^{-8}$ while
$\Lambda_0\propto C^{-5}$, the dynamical parameter is enhanced by three
additional inverse powers of the compactness $C = GM/R_\star$ relative to
the static one, an amplification that, as shown in Fig.~\ref{fig:TLNs},
reaches several orders of magnitude at typical neutron-star compactnesses
and is the principal reason why the 8PN dynamical tide may be observable
despite its formally high PN order (see Ref.~\cite{Pani:2025qxs} for a similar enhancement for the quadratic TLNs).

We now derive the GW phase following Ref.~\cite{Chakraborty:2025wvs}.
Consider a binary of two nonrotating neutron stars with masses $m_1$,
$m_2$, total mass $M_\text{\tiny tot} = m_1+m_2$,  symmetric
mass ratio $\eta = m_1 m_2/M_\text{\tiny tot}^2$, and orbital angular frequency $\omega$.  The
effective binary Lagrangian for circular orbits, including the leading tidal
interactions of both bodies described in Eq.~\eqref{eq:Sintmain}, reads
\begin{align}
  L &= \frac{1}{2}\eta M_\text{\tiny tot} \dot{r}^2
    + \frac{1}{2}\eta M_\text{\tiny tot} r^2\dot{\varphi}^2
    + \frac{G \eta  M_\text{\tiny tot}^2}{r} \nonumber \\
    & + \frac{m_2 (Gm_2)^4}{4}
      \left[\Lambda^{(2)}_0\,E_{ij}E^{ij}
           + (G m_2)^2\,\Lambda^{(2)}_2(r)\,\dot{E}_{ij}\dot{E}^{ij}\right]  \nonumber \\
    & + (1 \leftrightarrow 2)\,,
\label{eq:Lagrangian}
\end{align}
where ${E}_{ij}$ describes the tidal field sourced by the companion neutron star~\cite{Vines:2011ud}. Time derivatives are denoted by a dot, and the superscript $(i)$ labels the tidal response of body $i = 1,2$. 
Evaluating the Euler-Lagrange
equations on circular orbits, 
one obtains the expression for the binding energy~\cite{Chakraborty:2025wvs}
\begin{align}
  \mathcal{E}(\omega) & = -\frac{1}{2}\eta M_\text{\tiny tot} (G M_\text{\tiny tot} \omega)^{2/3}
  \bigg\{1
    - \frac{9\,(G m_1 \omega)^4}{(GM_\text{\tiny tot} \omega)^{2/3}}
     \frac{m_2}{M_\text{\tiny tot}} \Lambda^{(1)}_0  \nonumber \\ 
     & - \frac{9\, (G m_1 \omega)^6}{(GM_\text{\tiny tot} \omega)^{2/3}} \frac{m_2}{M_\text{\tiny tot}}
    \left[ 
       5 \bar{\Lambda}^{(1)}_2 + \frac{2}{3} \beta^{(1)}_2 \right. \nonumber \\
       &  \left.  - 5 \beta^{(1)}_2\log\left(\frac{(GM_\text{\tiny tot} \omega)^{1/3}}{\omega \, \gamma \,  r_s^{(1)}} 
            \right)
    \right] + (1\!\leftrightarrow\!2)
  \bigg\}\,,
\label{eq:Ebind}
\end{align}
and the GW flux
\begin{align}
  \mathcal{F}(\omega)
  &= \frac{32}{5} \eta^2 (GM_\text{\tiny tot} \omega)^{10/3}
  \bigg\{1
     \nonumber \\
    & + \frac{ 6 \, (Gm_1 \omega)^4 }{(GM_\text{\tiny tot} \omega)^{2/3}}
    \left(2\frac{m_2}{M_\text{\tiny tot}}+1\right)\Lambda^{(1)}_0
          \nonumber \\
  & 
    + \frac{6 \, (Gm_1 \omega)^6}{(GM_\text{\tiny tot} \omega)^{2/3}}
    \left[ \frac{m_2}{M_\text{\tiny tot}}\beta^{(1)}_2
         + 2\left(3\frac{m_2}{M_\text{\tiny tot}}+2\right) \right.  \nonumber \\
         & \left. \times \left(\bar{\Lambda}^{(1)}_2 - \beta^{(1)}_2
         \log\left(\frac{ (GM_\text{\tiny tot} \omega)^{1/3}}{\omega \, \gamma \, r_s^{(1)}}\right) \right)  \right] 
          +  (1\!\leftrightarrow\!2)
  \bigg\}\,.
\label{eq:flux}
\end{align}
Note that, in deriving these expressions, the orbital separation $r$ is eliminated in favor of the orbital frequency $\omega$ via the condition for circular orbits, $\partial L/\partial r = 0$; crucially, the resulting relation $r(\omega)$ receives leading-order dynamical tidal corrections~\cite{Chakraborty:2025wvs}, which must be consistently included when expanding the energy and flux in powers of the frequency. 

In the stationary-phase approximation, $\d^2\psi/\d\omega^2 = -(2/\mathcal{F})\,\d\mathcal{E}/\d\omega$~\cite{Vines:2011ud},
and integrating twice in frequency yields the waveform phase as a function of the
dimensionless PN parameter $x = (GM_\text{\tiny tot}\omega)^{2/3}$~\cite{Chakraborty:2025wvs}\footnote{Notice that the expression for the 8PN phase differs from the one of Ref.~\cite{Chakraborty:2025wvs} by a typo in their last derivation.}
\begin{align}
  \psi_\text{\tiny tidal}(x) & = \frac{3}{128\,\eta\,x^{5/2}}
  \left[
    1 - \frac{39}{2}\,\Lambda_\text{\tiny 5PN}\,x^5
    - \frac{15}{11} \Lambda_\text{\tiny 8PN}\,x^8 \right. \nonumber \\
    & \left. - \frac{15}{11}\sum_i B^{(i)}_\text{\tiny 8PN}\,x^8\log\left(\frac{ GM_\text{\tiny tot}}{x\, \gamma \, r_s^{(i)}} \right)
  \right]\,,
\label{eq:phase}
\end{align}
where we have defined the combinations
\begin{align}
\Lambda_\text{\tiny 5PN}
  &= \frac{16}{13} \sum_i \frac{m_i^3(m_i+11\eta M_\text{\tiny tot})}{M_\text{\tiny tot}^4} \Lambda^{(i)}_0\,, \nonumber \\
  \Lambda_\text{\tiny 8PN}
  &= \sum_i \frac{m_i^5}{M_\text{\tiny tot}^6} 
    \left[(8m_i+147\eta M_\text{\tiny tot})\bar{\Lambda}^{(i)}_2  \right. \nonumber \\
         & \left. \quad - \left(\frac{38}{11}m_i+\frac{1253}{44}\eta M_\text{\tiny tot}\right)\beta^{(i)}_2\right]\,, \nonumber \\
  B^{(i)}_\text{\tiny 8PN}
  &= - \frac{m_i^5(8m_i+147\eta M_\text{\tiny tot})}{M_\text{\tiny tot}^6}\,\beta^{(i)}_2\,.
\label{eq:B2_def}
\end{align}
Equation~\eqref{eq:phase} is the central result of this section. It contains three qualitatively distinct
contributions.  The first is the well-known 5PN static tidal term~\cite{Flanagan:2007ix,Hinderer:2007mb},
proportional to $\Lambda_0$.  The second and third arise from the dynamical tidal
coupling $c_{\dot{E}}$ computed in Sec.~\ref{sec:NStidalresponse}, and enter the waveform at 8PN
order.  The term proportional to $\Lambda_\text{\tiny 8PN}$ captures the finite, EoS-dependent
contribution of the dynamical TLN, while the logarithmic terms $B^{(i)}_\text{\tiny 8PN}$ encode its
renormalization-group running. 
Crucially, because $\bar{\Lambda}_2 \propto C^{-8}$, the 8PN dynamical tide is enhanced by three
additional inverse powers of the compactness compared to the 5PN static term $\Lambda_0 \propto C^{-5}$. This makes the dynamical tide potentially
observable despite its formally high PN order and also makes it much larger than (currently unknown) 8PN point-particle terms~\cite{Pani:2025qxs}.
Furthermore, the logarithmic term in Eq.~\eqref{eq:phase} provides a
qualitative handle to, in principle, disentangle the dynamical contribution from
higher-order point-particle corrections entering at the same PN order,
since the latter do not carry the same frequency dependence. However, as we will comment below, this
logarithmic running appears to be hardly measurable even with next-generation GW detectors: it
enters at subleading order relative to the bare $\Lambda_\text{\tiny 8PN}$ term, and
the numerical coefficient $\gamma$ that relates the renormalization scale to the orbital separation
does not significantly affect the waveform.
The dominant and observationally relevant quantity is therefore the EoS-dependent constant
$\Lambda_\text{\tiny 8PN}$.

\begin{figure}[t!] 
    \includegraphics[width=0.49\textwidth]{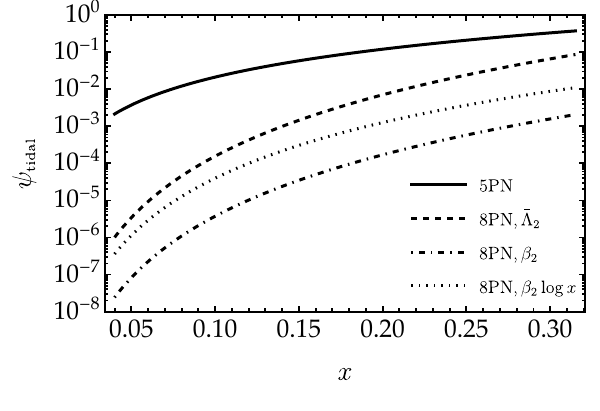}
    \caption{Absolute contributions to the tidal phase  as a
function of the dimensionless binary frequency $x$, for a representative
neutron-star binary with equal masses $m_\NS = 2.2 M_\odot$ and AP4 EoS. The four contributions correspond to the leading static (5PN) tidal
term $\propto \Lambda_0$ (solid line), the two finite dynamical (8PN) terms $\propto
\bar{\Lambda}_2$ (dashed line) and $\propto
\beta_2$ (dot-dashed line), and the logarithmic running term $\propto \beta_2 \log x$
(dotted line), as defined in Eqs.~\eqref{eq:phase} and~\eqref{eq:B2_def}. The static Love
number contribution dominates across the entire inspiral band, while the finite 8PN
dynamical term exceeds the logarithmic one, the latter being suppressed by additional factors of compactness, $C^{-3}$.}
\label{fig:phase}
\end{figure}

This hierarchy is manifest when looking at Fig.~\ref{fig:phase}, which shows the absolute
contributions to the tidal GW phase $\psi_\text{\tiny tidal}$ from the static (solid),
finite dynamical (dashed and dot-dashed), and logarithmic (dotted) TLNs, computed from
Eqs.~\eqref{eq:phase} and~\eqref{eq:B2_def}, as a function of the dimensionless binary
frequency $x$, taken to span the frequency range between the ET minimum frequency and the neutron star contact frequency (see Eq.~\eqref{eq:contact} below). The hierarchy among these contributions is immediately
apparent: the static Love numbers, entering at 5PN order, overwhelmingly dominate the
tidal phase throughout the inspiral. At 8PN order, the finite dynamical contribution
$\bar{\Lambda}_2$ exceeds the logarithmic term $\beta_2$ associated with the
renormalization-group running of the Love numbers. Despite the logarithmic coefficient
being proportional to the static Love number, see Eq.~\eqref{eq:Lambda2_dict}, it is
suppressed relative to the finite term due to the additional factors of compactness $C^{-3}$. This clear separation of scales motivates
our choice, in Sec.~\ref{sec-dataanalysis}, to focus the data-analysis study on the bare 8PN
coefficient $\Lambda_\text{\tiny 8PN}$. Furthermore, as discussed in Sec.~\ref{sec:UR},
this hierarchy suggests that the approximate universal relation between the 8PN and 5PN
tidal parameters (neglecting the running) could be leveraged to reduce the waveform parameter space, potentially
tightening constraints on the latter; we leave a detailed exploration of this avenue to
future work.

\begin{figure*}[t!] 
    \includegraphics[width=0.49\textwidth]{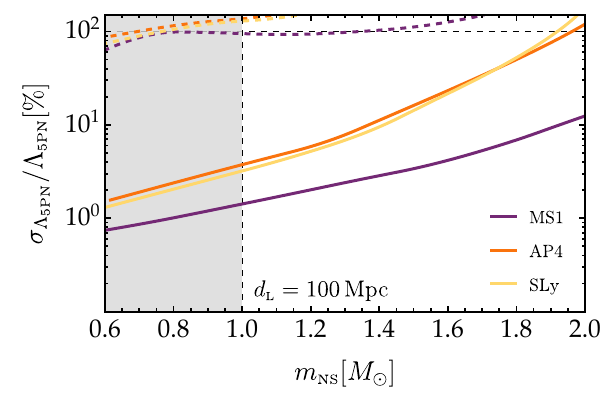}
        \includegraphics[width=0.49\textwidth]{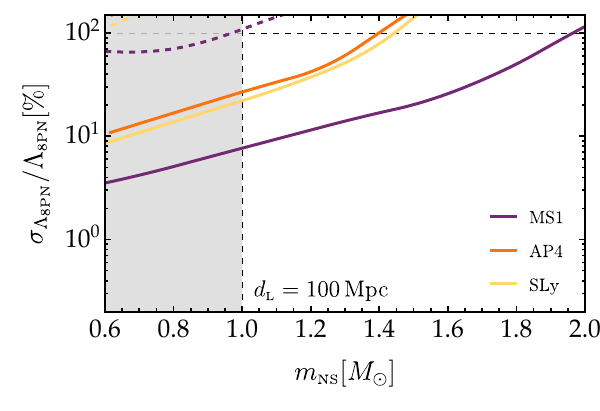}
    \caption{Relative $1\sigma$ uncertainty on the static $\Lambda_\text{\tiny 5PN}$ (left panel) and dynamical $\Lambda_\text{\tiny 8PN}$ (right panel) TLN parameters, for an equal-mass, nonspinning binary neutron star system with individual masses $m_\NS$, observed by LVK O4 (dashed) and ET (solid), assuming a luminosity distance of $d_\text{\tiny L} = 100\,{\rm Mpc}$. The colored lines correspond to different choices of realistic EoS. A measurement with $100\%$ error corresponds to the horizontal dashed line. }
\label{fig:Fisher}
\end{figure*}

\subsection{Data analysis}
\label{sec-dataanalysis}
\noindent
We now assess the observability of the 8PN dynamical tidal corrections of
Eq.~\eqref{eq:phase}, following the approach of Ref.~\cite{Chakraborty:2025wvs}. We consider two complementary diagnostics:
a Fisher-matrix analysis, which quantifies statistical measurement
precision, and a mismatch calculation, which directly probes the systematic
bias incurred when the dynamical tide is omitted from the template.

We model the GW signal using the \textsc{IMRPhenomD} waveform
family~\cite{Husa:2015iqa,Khan:2015jqa}, which captures the full inspiral
evolution. The tidal phase of Eq.~\eqref{eq:phase} is appended to the point-particle
phase, including the 5PN static term $\Lambda_\text{\tiny 5PN}$ and the 8PN dynamical term
$\Lambda_\text{\tiny 8PN}$.  In the frequency
domain the template reads
\begin{equation}
  \tilde{h}(f;\,\boldsymbol{\theta})
    = \mathcal{C}_\Omega\,\mathcal{A}(f;\,\boldsymbol{\theta})\,
      \e^{\,i\left[\psi_\text{\tiny  PP}(f;\,\boldsymbol{\theta})
              +\psi_\text{\tiny tidal}(f;\,\boldsymbol{\theta})\right]}\,,
\label{eq:waveform}
\end{equation}
where $\psi_\text{\tiny PP}$ describes the standard point-particle phase, while the leading-order amplitude is
\begin{equation}
  \mathcal{A}(f;\,\boldsymbol{\theta})
    = \sqrt{\frac{5}{24}}\,
      \frac{G^{5/6} \mathcal{M}_c^{5/6}}{\pi^{2/3}\,d_\text{\tiny L}\,f^{7/6}}\,,
\label{eq:amplitude}
\end{equation}
in terms of the chirp mass  $\mathcal{M}_c = \eta^{3/5}M_\text{\tiny tot}$ and luminosity distance $d_\text{\tiny L}$. The geometric factor $\mathcal{C}_\Omega$ captures the detector’s
response and depends on the binary’s inclination angle and antenna pattern functions (which in turn vary
with the source’s sky location and polarization angle). For simplicity, we assume optimally oriented
binaries, thereby neglecting these dependencies in our
analysis.  We consider
equal-mass ($m_1 = m_2 = m_\NS$), nonspinning binary neutron-star systems at
$d_\text{\tiny L} = 100\,\text{Mpc}$ and three representative EoS: SLy4~\cite{Douchin:2001sv},
AP4~\cite{Akmal:1998cf}, and MS1~\cite{Mueller:1996pm}.

To forecast measurement uncertainties, we use the Fisher information matrix
formalism~\cite{Vallisneri:2007ev}, which is reliable in the limit of large
signal-to-noise ratio (SNR).  The Fisher matrix is
\begin{equation}
  \Gamma_{ij}
    = \left\langle\frac{\partial h}{\partial\theta^i}
      \,\bigg|\,
      \frac{\partial h}{\partial\theta^j}\right\rangle
      \bigg|_{\boldsymbol{\theta}=\hat{\boldsymbol{\theta}}}\,,
\label{eq:fisher}
\end{equation}
where the noise-weighted inner product is
\begin{equation}
  \langle h_1 | h_2 \rangle
    = 4\,\operatorname{Re}\int_{f_\text{\tiny min}}^{f_\text{\tiny max}}
      \frac{\tilde{h}_1^*(f)\,\tilde{h}_2(f)}{S_n(f)}\,{\rm d}f\,,
\label{eq:inner_product}
\end{equation}
with $S_n(f)$ the one-sided noise power spectral density, and $f_\text{\tiny min}$,
$f_\text{\tiny max}$ the lower and upper limits of the integration band.  The $1\sigma$ uncertainty on the
$i$-th parameter is $\sigma_i = \sqrt{(\Gamma^{-1})_{ii}}$ and the SNR is
$\rho = \sqrt{\langle h|h\rangle}$.
The full parameter vector is
$\boldsymbol{\theta} = \{\mathcal{M}_c,\,\eta,\,\chi_s,\,\chi_a,\,t_c,\,\phi_c,\,\Lambda_\text{\tiny 5PN},\,\Lambda_\text{\tiny 8PN}\}$
with injected coalescence time and phase $(t_c,\phi_c)=(0,0)$, and vanishing injected spin terms, $\chi_s = \chi_a = 0$.  We consider two
detector configurations: the LVK network
at O4 sensitivity with $(f_\text{\tiny det,min},f_\text{\tiny det,max}) = (10, 4 \cdot 10^3)\,{\rm Hz}$, and ET in its ET-D configuration~\cite{ET:2025xjr,Branchesi:2023mws}
with $(f_\text{\tiny det,min},f_\text{\tiny det,max}) = (2, 10^4)\,{\rm Hz}$.  The lower limit is set by the detector, $f_\text{\tiny min} = f_\text{\tiny det,min}$.
The inspiral sequence proceeds until the neutron stars come into
contact, which for two stars of radius $R_*$ occurs at orbital separation $r = 2R_*$,
corresponding to the frequency
\begin{equation}
\label{eq:contact}
f_\text{\tiny contact} = 6 \sqrt{6} \, C^{3/2} f_\text{\tiny ISCO}\,,
\end{equation}
in terms of their compactness and frequency at the innermost
circular orbit (ISCO)
\begin{equation}
\label{eq:fISCO}
f_\text{\tiny ISCO} = 4.4 \, {\rm kHz} \left(\frac{M_\odot}{M_\text{\tiny tot}} \right) \,.
\end{equation}
Accordingly, we take the upper end of the integration band to be
the minimum of the contact, ISCO, and detector cutoff frequencies,
\begin{equation}
\label{eq:fmax}
f_\text{\tiny max} = \min \left(f_\text{\tiny contact},\, f_\text{\tiny ISCO},\, f_\text{\tiny det,max}\right)\,,
\end{equation}
so that the adiabatic PN phasing is never integrated beyond the ISCO, where it
ceases to be valid.

As a complementary and more conservative diagnostic, we compute the mismatch between a
signal template that includes the dynamical tide and one that neglects it:
\begin{equation}
  \mathcal{M}
    = 1 - \underset{\{\phi_c, t_c\}}{\rm max} \frac{\langle h_1 | h_2 \rangle}
               {\sqrt{\langle h_1|h_1\rangle\,\langle h_2|h_2\rangle}}\,,
\label{eq:mismatch}
\end{equation}
where $h_1$ includes $\Lambda_\text{\tiny 8PN}\neq 0$ and $h_2$ is computed with $\Lambda_\text{\tiny 8PN}=0$.
A mismatch $\mathcal{M} \lesssim 1/(2\rho^2)$ guarantees that two waveforms are equivalent within statistical errors at a given SNR, while exceeding this limit might indicate a systematic bias in parameter
estimation~\cite{Flanagan:1997sx}.

Figure~\ref{fig:Fisher} shows the relative $1\sigma$ uncertainties on $\Lambda_\text{\tiny 5PN}$
(left panel) and $\Lambda_\text{\tiny 8PN}$ (right panel) as functions of the individual neutron star mass
$m_\NS$, for the three EoS at $d_\text{\tiny L}=100\,\text{Mpc}$; the horizontal dashed line marks a relative error of 100\%.
For the static Love
number, ET (solid lines) lies well below this threshold across essentially the entire mass range, reaching $\sigma_{\Lambda_\text{\tiny 5PN}}/\Lambda_\text{\tiny 5PN} \lesssim 10\%$ for
$m_{\NS} \lesssim 1.4\,M_\odot$, and approaching the 100\% line only for the most massive, softest configurations. By contrast, the LVK O4 curves (dashed lines) sit close to or above the 100\% line, so that $\Lambda_\text{\tiny 5PN}$ is at best marginally constrained at current sensitivity.

Two features of the curves are noteworthy. First, the
relative error grows steeply with mass. This happens because more massive neutron stars are more compact, so
that both tidal deformabilities fall rapidly with compactness, shrinking the tidal signal; in addition, the
cutoff of the inspiral scales inversely with the total mass
through $f_\text{\tiny ISCO} \propto M_\text{\tiny tot}^{-1}$, so that the contact
frequency of Eq.~\eqref{eq:contact} moves to lower values and fewer wave
cycles accumulate in the tidal-sensitive band. Both effects reduce the measurable tidal imprint and inflate the statistical uncertainty. 
Second, at fixed mass the relative error is smallest for the stiffest EoS
(MS1) and largest for the softer ones (AP4, SLy). Since the tidal terms enter the phase linearly,
the absolute Fisher uncertainty $\sigma_{\Lambda_\text{\tiny 5PN}}$ is set essentially by the detector, the masses, and
the available frequency band, and depends only weakly on the EoS; the relative error
$\sigma_{\Lambda_\text{\tiny 5PN}}/\Lambda_\text{\tiny 5PN}$ is therefore controlled primarily by the magnitude of $\Lambda_\text{\tiny 5PN}$ itself.
Stiffer EoS produce larger radii and hence lower compactness (see Fig.~\ref{fig:MassRadiusRelation}),
which through the steep inverse-compactness scaling translates into substantially larger $\Lambda_0$ and $\Lambda_2$,  and thus a stronger and more easily measured
tidal imprint. The lower compactness of stiffer stars also
lowers their contact
frequency, an effect
that by itself would reduce the in-band tidal information.

For the dynamical Love number the picture is more
nuanced: ET reaches $\sigma_{\Lambda_\text{\tiny 8PN}}/\Lambda_\text{\tiny 8PN} \lesssim 10\%$ only for $m_{\NS} \lesssim 1\,M_\odot$ and the stiffest EoS (MS1), the only case that drops appreciably below the 100\% line; for the softer EoS (SLy4, AP4) the relative error stays
below $100\%$ for masses $m_{\NS} \lesssim 1.4\,M_\odot$, after which $\Lambda_\text{\tiny 8PN}$ is not individually measurable on its own. LVK O4 yields uncertainties 
orders of magnitude above the $100\%$ for all configurations in the supersolar regime.
Clearly, the presented relative errors have a simple linear scaling with the source distance.

The key
factor enabling ET to approach the dynamical tide is the large
enhancement of $\Lambda_2$ relative to $\Lambda_0$ visible in Fig.~\ref{fig:TLNs}: since
$\Lambda_2 \propto C^{-8}$ while $\Lambda_0 \propto C^{-5}$, the dynamical parameter is
larger by roughly three orders of magnitude at typical neutron-star compactnesses,
compensating the formal $x^3$ suppression of the 8PN term relative to the 5PN one.  The strong EoS dependence of $\Lambda_\text{\tiny 8PN}$ implies that
a detection would simultaneously constrain the stiffness of dense matter at
supranuclear densities.
This
situation is qualitatively different from the black-hole case studied in
Ref.~\cite{Chakraborty:2025wvs}, where the dynamical tidal correction is unobservable even
with future-generation detectors because the analogous $(r_s/M)^8$ amplification
factor reduces to $\mathcal{O}(1)$. 
We also note that our Fisher analysis treats $\Lambda_\text{\tiny 8PN}$ as the
primary free parameter; the logarithmic correction $B^{(i)}_\text{\tiny 8PN}$ is not
independently varied, as it is numerically subleading relative to $\Lambda_\text{\tiny 8PN}$
for the orbital frequencies relevant to the late inspiral, and the sensitivity
to the free coefficient $\gamma$ is correspondingly negligible.
Finally,
employing the approximate universality relation between $\Lambda_0$
and $\Lambda_2$ established above would allow one to express $\Lambda_\text{\tiny 8PN}$
in terms of $\Lambda_\text{\tiny 5PN}$, thereby reducing the dimensionality of the
parameter space and improving the measurability of the
static coefficient.

\begin{figure}[t!] 
    \includegraphics[width=0.49\textwidth]{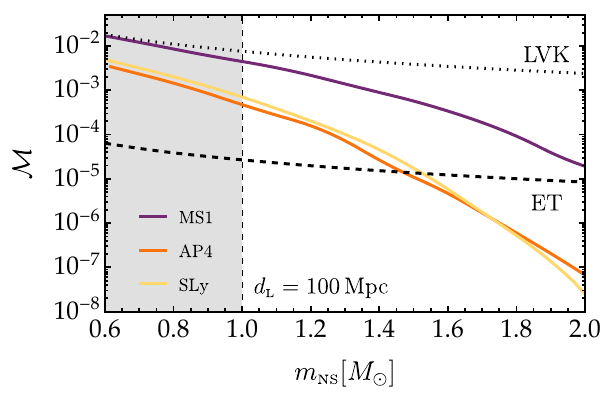}
    \caption{Mismatch $\mathcal{M}$ between a template that includes the
    dynamical tidal phase ($\Lambda_\text{\tiny 8PN}\neq 0$) and one that
    neglects it ($\Lambda_\text{\tiny 8PN}=0$), as a function of the individual neutron star
    mass $m_\NS$, for an equal-mass, nonspinning binary at
    $d_\text{\tiny L}=100\,{\rm Mpc}$.  Horizontal dashed lines indicate the
    systematic-bias threshold $1/(2\rho^2)$ for LVK O4 and ET, respectively.
    Colored lines correspond to different EoS.  Systems above the ET threshold
    are those for which omitting the dynamical tide induces a statistically
    significant bias in the inference of the static Love number.}
\label{fig:Mismatch}
\end{figure}

Figure~\ref{fig:Mismatch} shows the mismatch $\mathcal{M}$ as a function of $m_{\NS}$
for the same configurations, with the horizontal dashed lines marking the systematic-bias threshold
$1/(2\rho^2)$ for LVK O4 (upper line) and ET (lower line). A curve lying above its detector
threshold signals that the dephasing induced by omitting the
dynamical tide is larger than the statistical error, so that
neglecting $\Lambda_\text{\tiny 8PN}$ may bias the recovery of the remaining
parameters. Because the dynamical tide is strongest for
less compact stars, the mismatch is largest at low masses
and for the stiffer EoS, and decreases steeply as $m_\NS$ grows.

For ET and a binary neutron star at $d_\text{\tiny L}=100\,\text{Mpc}$, the
mismatch exceeds the threshold for
$m_{\NS} \lesssim 1.4\,M_\odot$ and the softest EoS (AP4, SLy), while being always above it for the stiffer EoS (MS1), whereas for LVK O4 it stays below
the (much higher) O4 threshold for all configurations, in
agreement with the Fisher-matrix result that the dynamical
tide is inaccessible at current sensitivity.

The two diagnostics are complementary, and their joint
reading is the central message of this section. The Fisher
analysis quantifies whether $\Lambda_\text{\tiny 8PN}$ can be measured as a parameter, whereas the mismatch quantifies the bias incurred
by leaving it out of the template altogether. The two need
not coincide: AP4, for instance, crosses the ET mismatch threshold even though its dynamical TLN is
not individually measurable at the $10 \%$ level according to Fig.~\ref{fig:Fisher}. The reason is that the mismatch is
sensitive to the total dephasing accumulated across
the inspiral band, and can flag a systematic
bias even when the corresponding parameter uncertainty is large. In practice this means that, for ET sources
at $d_\text{\tiny L} \sim 100 \, {\rm Mpc}$ (${\rm SNR} \gtrsim 100$), omitting the dynamical tidal
phase will bias the inference of the static Love number---and hence the extracted nuclear EoS---even in
regimes where the dynamical tide itself cannot be pinned
down. Including the full phase of Eq.~\eqref{eq:phase} in the template
is therefore necessary to avoid such biases, irrespective of
whether $\Lambda_\text{\tiny 8PN}$ is independently resolvable.

In summary, our analysis shows that the dynamical Love number is
potentially measurable with third-generation detectors for a range of neutron star masses and
EoS, and that neglecting it can introduce systematic biases in the inference
of nuclear-matter properties. The degree of observability depends sensitively on the EoS, with stiffer
EoS yielding both a larger $\Lambda_\text{\tiny 8PN}$ and a larger mismatch.  These
findings motivate including the full dynamical tidal phase of Eq.~\eqref{eq:phase} in
next-generation GW data-analysis pipelines, and suggest that joint
inference of $\Lambda_\text{\tiny 5PN}$ and $\Lambda_\text{\tiny 8PN}$ with ET could provide
a new window into the frequency-dependent tidal response of dense nuclear matter.

\section{Discussion}
\label{sec:discussion}
\noindent
In this work we have carried out a systematic and fully relativistic study of the dynamical
tidal response of neutron stars by combining worldline EFT with stellar
perturbation theory. We computed the dynamical quadrupolar Love number $c_{\dot{E}}$ for a
broad set of realistic EoS. The corresponding dimensionless
parameter $\Lambda_2 \propto C^{-8}$ is enhanced by three additional inverse powers of the
compactness relative to the static counterpart $\Lambda_0 \propto C^{-5}$, being significantly larger than its static counterpart for typical neutron-star compactnesses. 
The dynamical response coupling also exhibits a universal logarithmic running, in addition to a set of scheme-dependent finite terms, which are unambiguously fixed by the EFT matching and which we determine here for the first time using dimensional regularization.
We have shown that, for a fixed running parameter, approximate EoS-insensitive relations exist between $\Lambda_0$ and $\Lambda_2$, analogous to the I-Love-Q universality~\cite{Yagi:2013bca,Yagi:2013awa}, and that, taking motivated values of the running into account, the $f$-mode still provides an accurate
single-mode approximation to the dynamical response throughout the late inspiral, as in the Newtonian case.
Using these results, we derived the complete leading-order 8PN tidal phase correction to
the gravitational waveform of a binary neutron-star inspiral, including both the
EoS-dependent constant term  and the logarithmic running. While the logarithmic term is qualitatively distinctive---in
principle allowing one to disentangle the dynamical tide from unknown point-particle
contributions at the same PN order---it is numerically subleading during the late inspiral.

The waveform contribution associated with the dynamical Love number is also quantitatively comparable to that arising from the quadratic Love numbers~\cite{Pani:2025qxs,Pitre:2025qdf}. Both effects enter the GW phase at 8PN order and are enhanced in the low-compactness regime, scaling as $\sim C^{-8}$. Comparing our results with those of Ref.~\cite{Pani:2025qxs}, we find that the two contributions are generally of the same order of magnitude. This suggests that a consistent description of the leading 8PN tidal correction requires both dynamical and quadratic Love numbers to be included in the waveform model.

Our Fisher-matrix analysis shows that ET could measure the 8PN tidal contribution below the
$100\%$ level for $m_\NS \lesssim 1.4\,M_\odot$ and $m_\NS \lesssim 2\,M_\odot$ for the softest and stiffest EoS, respectively. For $m_\NS \lesssim 1\,M_\odot$, the accuracy improves to $10\%$ and percent level, respectively.
In contrast, LVK O4 cannot
constrain the dynamical tide for any of the configurations considered. The mismatch
analysis reveals a complementary and practically important finding: omitting the 8PN dynamical tide from the waveform
template introduces systematic biases in the inference of the static Love number for $m_\NS \lesssim 1.4\,M_\odot$ and soft EoS, and larger masses for the stiffer ones, at ET
sensitivity. Since the latter is the primary probe of the nuclear EoS in
current and planned GW observations, this finding motivates including the full dynamical
tidal phase in next-generation waveform models.

Our work can be extended along various directions.
Incorporating stellar rotation would introduce
gravito-magnetic dynamical tides and mode-mixing relevant for spinning neutron stars.
Extending the EFT matching to include tidal corrections to the GW amplitude,
beyond the phase computed here, would complete the leading-order dynamical tidal waveform.
Finally, combining the dynamical TLN measurement with independent nuclear-physics
constraints could significantly sharpen EoS inference with third-generation detectors,
making the framework developed here a concrete ingredient for high-precision GW science.

\vspace{0.3cm}
\noindent
\textbf{Note added:} While this work was being completed, Ref.~\cite{Saketh:2026trm} appeared, containing an independent discussion of the calculation of dynamical TLNs of neutron stars through EFT matching. We note that, differently from the approach of Ref.~\cite{Saketh:2026trm}, we work at the level of off-shell quantities rather than on-shell ones, and perform the matching at loop level, resumming gravitational nonlinearities within the EFT.

\section*{Acknowledgments}
\noindent
We acknowledge interesting discussions with Nils Andersson, Abhishek Hegade, Tanja Hinderer, Soumodeep Mitra, Raj Patil, Hector O.~Silva, and Nicolas Yunes. 
V.D.L. and L.S. are grateful to  the Institute for Fundamental Physics of the Universe (IFPU) in Trieste for hosting the workshop ``Perspectives on Gravity: From Theory to Observation'', where this work was completed.
V.D.L.~is supported by NSF Grants No.~AST-2307146, No.~PHY-2513337, No.~PHY-090003, and No.~PHY-20043, by NASA Grant No.~21-ATP21-0010, by John Templeton Foundation Grant No.~62840, by the Simons Foundation [MPS-SIP-00001698, Emanuele Berti], by the Simons Foundation International [SFI-MPS-BH-00012593-02], and by Italian Ministry of Foreign Affairs and International Cooperation Grant No.~PGR01167.
This work was carried out at the Advanced Research Computing at Hopkins (ARCH) core facility (\url{https://www.arch.jhu.edu/}), which is supported by the NSF Grant No.~OAC-1920103.
We acknowledge support by the MUR FIS2 Advanced Grant ET-NOW (CUP:~B53C25001080001), and by the INFN TEONGRAV initiative. 
This project has received funding from the European Union’s Horizon Europe research and
innovation programme under the Marie Skłodowska-Curie Staff Exchanges grant agreement
No. 101299389 (STRONG).
The research of L.S.~has been funded, in part, by the French National Research Agency (ANR) under Project No. ANR-24-CE31-1097-01. This work has received support under the program ``\textit{Investissement d'Avenir}'' launched by the French Government and implemented by ANR, with the reference No. ANR-18-IdEx-0001 as part of its program ``\textit{Emergence}''.

\begin{widetext}
\appendix
\section{Gravito-electric one-point function and renormalized couplings from point-particle EFT}
\label{app:ppEFT}
\noindent
In this Appendix, we illustrate the computation of the graviton one-point function within the point-particle EFT \eqref{eq:ppEFT0}. For generality, we work in the in-in Schwinger--Keldysh formalism and incorporate gravitational nonlinearities using the Born series developed in Ref.~\cite{Caron-Huot:2025tlq}  for the scalar field case (see also Ref.~\cite{Correia:2024jgr}), later generalized to the gravitational response in Ref.~\cite{Combaluzier--Szteinsznaider:2025eoc}. For further details and notation, we refer the reader to Ref.~\cite{Combaluzier--Szteinsznaider:2025eoc}.

We begin by doubling the fields along a closed-time contour, denoting the graviton fields on the forward and backward branches by $h_1^{\mu\nu}$ and $h_2^{\mu\nu}$, respectively. In the Keldysh basis, we define
\begin{equation}
h_+ \equiv \frac{1}{2}\left(h_1 + h_2\right), \qquad
h_- \equiv h_1 - h_2 \,.
\label{eq:keldyshbasispm}
\end{equation}
The effective action for $h_\pm$ is then obtained by integrating out the degrees of freedom on which the composite operators $Q_E$ [see Eq.~\eqref{SintE}] depend upon, which we collectively denote by $\mathcal{X}$:
\begin{equation}
\label{eq:PIXX}
    \e^{i\Gamma^\text{in-in}_\text{int}[h_\pm]} = \int\mathcal{D}\mathcal{X}_+\mathcal{D}\mathcal{X}_- \, \e^{iS\left[h_\pm,\mathcal{X}_\pm\right]} \,,
\end{equation}
where $S$ is the action \eqref{eq:ppEFT0}.
At leading order in the external tidal field amplitude, $Q_E$ can be expressed perturbatively in terms of its linear response as
\begin{equation}
    \langle Q_{E,I}^{ij}(\tau)\rangle=\int \d \tau'\,K_{IJ}^{(E)}{}^{ij\vert i' j'}(\tau-\tau'){E}^J_{i' j'}(\tau')\,,
\end{equation}
where $I,J = \{+,-\}$, and $K_{IJ}^{(E)ij \vert i' j'}$ is the linear response kernel. This kernel is related to the two-point function of $Q_E^{ij}$ in the Keldysh basis~\cite{Goldberger:2020fot,Saketh:2022xjb,Saketh:2023bul,Glazer:2024eyi} via
\begin{equation}
    \langle Q_{E,I}^{ij}(\tau)Q_{E,J}^{i j'}(\tau')\rangle=-iK_{IJ}^{(E)}{}^{i j\vert i' j'}(\tau-\tau')\,.
\end{equation}
Following standard EFT logic, we remain agnostic about the precise form of the response kernel and parametrize it in the most general way consistent with the symmetries of the problem. We simply assume it admits a low-frequency expansion---justified when the object's characteristic internal timescale (in the present case, the hydrodynamical timescale of the star) is much shorter than the timescales of the external perturbation.
For nonspinning objects, the spatial tensor structure of the kernel reduces to a product of Kronecker deltas, yielding the in-in effective action
\begin{equation}
    \Gamma^\text{in-in}_\text{int}[h_\pm] 
    = \int \mathrm{d}\tau_1 \, \mathrm{d}\tau_2  
    K_{{IJ}}^{(E)}(\tau_2-\tau_1) \, E^I_{i j}(\tau_2) \, E^J{}^{i j}(\tau_1) +\dots \,,
\label{eq:gammainineft}
\end{equation}
where the ellipsis indicates higher-order terms in the gradient expansion (i.e., operators corresponding to multipoles beyond the quadrupole).

To match with neutron-star perturbation theory, we need a gauge-invariant quantity independent of the coordinate system used in both the EFT and full-theory calculations. A natural choice is the (linearized) Weyl tensor---which is also the quantity appearing in the effective action~\eqref{eq:gammainineft}. Computing its one-point function amounts to evaluating the in-in path integral:
\begin{equation}
    \langle E_{+ab}(t,\vec{x})\rangle_\text{in-in}  =\int \mathcal{D}h_+\mathcal{D}h_- E_{+ab}(t,\vec{x})\e^{i\Gamma^\text{in-in}_\text{int}[h_\pm]}  \,,
    \label{eq:1-pt_fct_def}
\end{equation}
which yields
\begin{equation}
\label{E_opf}
    \left\langle E_{+a b}(t, \vec{x})\right\rangle_{\mathrm{in}-\mathrm{in}}=i  \int \mathrm{d} \tau_1 \mathrm{d} \tau_2 \, K_{+-}^{(E)}\left(\tau_2-\tau_1\right)\left\langle E_{+a b }(t, \vec{x}) E_{-ik}\left(\tau_2\right)\right\rangle \bar{E}_{+}^{ik}\left(\tau_1\right)\,,
\end{equation}
where $\bar{E}_{+}^{ik}$ denotes the gravito-electric component of the external (quadrupolar) tidal field, and where $K^{({E})}_{+-}(\tau)$ is proportional to the retarded Green's function of the $Q_E$ operators (see, e.g., Refs.~\cite{Saketh:2023bul,Glazer:2024eyi,Combaluzier--Szteinsznaider:2025eoc}).

The (quadrupolar) response function $K_{+-}^{(E)}\left(\tau_2-\tau_1\right)$ can be expressed in frequency space as 
\begin{equation}
    \begin{aligned} K_{+-}^{(E)}\left(\tau_2-\tau_1\right) & =\int \frac{\mathrm{d} \omega}{2 \pi} \mathrm{e}^{-i \omega\left(\tau_2-\tau_1\right)} K_{\ell}^{(E)}(\omega) \\ & = \int \frac{\mathrm{d} \omega}{2 \pi} \mathrm{e}^{-i \omega\left(\tau_2-\tau_1\right)}\left[\frac{R_\star^5}{G}c_E+i \omega M R_\star^5 \nu_E+ \omega^2   \frac{R_\star^8}{G^2M}c_{\dot E}+\cdots\right]\,,
    \end{aligned}
\label{eq:KpmcEs}
\end{equation}
where in the second line we stripped off some dimensionful quantities (with $R_\star$ and $M$ the radius and mass of the object, respectively), and where $c_E$, $\nu_E$, and $c_{\dot E}$ are dimensionless.

In the following, we compute Eq.~\eqref{E_opf} following Ref.~\cite{Combaluzier--Szteinsznaider:2025eoc}. A key difference with respect to Ref.~\cite{Combaluzier--Szteinsznaider:2025eoc}, which focused on matching to a black hole response, is that the static coefficient $c_E$ is nonzero for neutron stars. As a consequence, computing $\langle E_{+ab}(t, \vec{x}) \rangle$ up to the next-to-leading conservative order, $\mathcal{O}(\omega^2)$, requires evaluating the graviton propagator appearing in the integrand of \eqref{E_opf} to the same order.

\subsection{Gravito-electric one-point function}
\noindent
On the EFT side, we work in the de Donder gauge, defined by 
\begin{equation}
    \partial^{\mu} h_{\mu\nu} = \frac{1}{2} \partial_{\nu} h \,,
\end{equation}
with $h \equiv {h^\mu}_\mu$. 
Hereafter, $h_{\mu\nu}$ denotes the canonically normalized graviton field, defined via 
$g_{\mu\nu} = \eta_{\mu\nu} + 2h_{\mu\nu}/{M_{\mathrm{Pl}}^{(D-2)/2}}$ in $D$ spacetime dimensions. 
Indices are raised and lowered using the Minkowski metric $\eta_{\mu\nu}$. 
In this gauge, the linearized equations of motion reduce to $\Box h_{\mu\nu} = 0$.  
Some useful power counting to keep in mind for the rest is  $\partial^0\partial_0h_{\mu\nu}\sim\partial^i\partial_ih_{\mu\nu}   \sim\mathcal{O}(\omega^2) $ and also
$\partial^{i}h_{i0}\sim\partial_0h_{00}+\partial_0h\sim\mathcal{O}(\omega)$.

The electric component of the Weyl tensor on shell is given by 
\begin{equation}
\begin{split}
    E_{ab}=C_{0a0b}&=\frac{2}{\Mp^{\frac{D-2}{2}}}\left(\partial_a\partial_{[0}h_{b]0}-\partial_0\partial_{[0}h_{b]a}\right)
\\
    & =\frac{1}{\Mp^{\frac{D-2}{2}}}\left(-\partial_a\partial_bh_{00}+2\partial_0\partial_{(a}h_{b)0}-\partial_0\partial_0h_{ab}\right)\,,
\end{split}
\end{equation}
which is, by construction, traceless.

Let us start by computing the product $E_{-ik} \bar{E}_{+}^{ik}$ appearing in Eq.~\eqref{E_opf}. In what follows, we suppress the $+,-$ indices for simplicity and will reintroduce them at the end. We find
\begin{equation}
 \begin{aligned} 
 \label{El2_exp}E_{ik}\bar{E}^{ik}= 
 \frac{1}{\Mp^{D-2}}\Big[& \p_i\p_kh_{00}\p^i\p^k\bar{h}_{00}
-2\p_i\p_kh_{00}\p_0\p^i\bar{h}^{k}{}_0+\p_i\p_kh_{00}\p_0\p_0\bar{h}^{ik}\\
& 
-2\p_0\p_{i}h_{k0}\p^{i}\p^{k}\bar{h}_{00}
+2\p_0\p_ih_{k0}\p_0\p^i\bar{h}^{k}{}_0 +2\p_0\p_ih_{k0}\p_0\p^k\bar{h}^{i}{}_0
+\p_0\p_0h_{ik}\p^i\p^k\bar{h}_{00}
+\mathcal{O}(\omega^3)\Big]\,.
 \end{aligned} 
\end{equation}

We are now interested in the product $\langle E_{+ab} E_{-ik} \rangle \, \bar{E}_{+}^{ik}$ up to $\mathcal{O}(\omega^2)$. This involves various combinations of derivatives of the two-point function $\langle h^{+}_{\mu\nu} h^{-}_{\rho\sigma} \rangle$. Taking into account its tensor structure leads to significant simplifications. In particular, in the de Donder gauge we have 
\begin{equation}
\label{de_doner_prop}
\left\langle h_{+\,\mu \nu}(t, \vec{x}) h_{-\, \rho \sigma}\left(\tau_2, 0\right)\right\rangle=i \mathcal{P}_{\mu \nu \rho \sigma}^{\mathrm{dD}} \int \frac{\d^{D} p}{(2 \pi)^{D}} \frac{\e^{-i\omega(t-\tau_2)+ i\vec p \cdot \vec x}}{\vec{p}^2-\omega^2}\,,
\end{equation}
where
\begin{equation}
\mathcal{P}_{\mu \nu \rho \sigma}^{\mathrm{dD}} \equiv-\frac{1}{2}\left(\eta_{\mu \sigma} \eta_{\nu \rho}+\eta_{\mu \rho} \eta_{\nu \sigma}-\frac{2}{D-2} \eta_{\mu \nu} \eta_{\rho \sigma}\right)\,.
\end{equation}
This implies that $\left\langle h_{+\,ik} h_{-\, 0j}\right\rangle=\left\langle h_{+\,i0} h_{-\, 00}\right\rangle=0$.
Therefore, many of the terms vanish, leaving us with
\begin{multline}
\label{EEELS}
    \langle E_{a b } E_{ik}\rangle \bar{E}^{ik}=-\frac{1}{\Mp^{\frac{3}{2}(D-2)}}\Big[\langle\p_a\p_bh_{00}\p_i\p_kh_{00}\rangle\left(\p^i\p^k\bar{h}_{00}-2\p_0\p^i{\bar{h}^{k}}_0+\p_0\p_0\bar{h}^{ik}\right)
\\
+\left(\langle\p_a\p_bh_{00}\p_0\p_0h_{ik}\rangle+4\langle\p_0\p_{(a}h_{b)0}\p_0\p_{(i}h_{k)0}\rangle+\langle\p_0\p_0h_{ab}\p_i\p_kh_{00}\rangle\right)\p^i\p^k\bar{h}_{00}+\mathcal{O}(\omega^3)\Big]\,.
\end{multline}
Putting everything together, we obtain
\begin{equation}
\label{E_opfb}
\begin{split}
    \left\langle E_{+a b}(t, \vec{x})\right\rangle_{\mathrm{in}-\mathrm{in}} 
       =  \frac{\e^{-i\omega t}}{\Mp^{\frac{3}{2}(D-2)}} K_{+-}^{(E)}\left(\omega\right) \int\frac{\d^{D-1}\vec p}{(2\pi)^{D-1}}
     \frac{\e^{i\vec{p}\cdot\vec{x}}}{\vec{p}^2-\omega^2}\bigg[& \frac{3-D}{D-2}  
    p_ap_bp_ip_k \left(\p^i\p^k\bar{h}_{00}-2\p_0\p^i{\bar{h}^{k}}_0+\p_0\p_0\bar{h}^{ik}\right)
\\
    &   +2    \omega^2 p_i p_{(a} \p_{b)}\p^i\bar{h}_{00} -\frac{\omega^2}{D-2}\eta_{ab}p_ip_k\p^i\p^k \bar{h}_{00}\bigg]\,,
\end{split}
\end{equation}
where $\omega$ is the frequency of the (monochromatic) external probe, which we took to be  $\bar h_{\mu\nu}(\tau_1) = \e^{-i\omega \tau_1} \bar{h}_{\mu\nu}(\omega)$, with $\bar{h}_{\mu\nu}(\omega)$ solving the flat-space Einstein equations in de Donder gauge,  and expanded in powers of $\vec{x}$ around the location of the point particle. 

Since we are interested in performing the matching in a near zone $\vert \vec x \vert\ll1/\omega$~\cite{Combaluzier--Szteinsznaider:2025eoc},\footnote{Note that what we refer to as the ``near zone'' here corresponds to the ``intermediate zone'' of Ref.~\cite{Combaluzier--Szteinsznaider:2025eoc}. In that work, a finer distinction between near and intermediate zones was necessary because $r$ extends down to $r_s$. Here, since $R_\star$ is never parametrically close to $r_s$---namely $R_\star \gtrsim \mathcal{O}(1)r_s$---this distinction is not required. We also stress that, despite its name, the near zone covers a wide range of distances, reaching distances parametrically far from $R_\star$ ($r\gg R_\star$), while remaining smaller than $1/\omega$.} we can expand the graviton propagator in Eq.~\eqref{E_opfb} in powers of $\omega$.
After some tedious but straightforward algebra, Eq.~\eqref{E_opfb} boils down to
\begin{equation}
\label{E_opfc}
    \left\langle E_{+a b}(t, \vec{x})\right\rangle_{\mathrm{in}-\mathrm{in}} 
       =  -\frac{\e^{-i\omega t}}{\Mp^{D-2}} K_{+-}^{(E)}\left(\omega\right) \left[
       \frac{2\Gamma\left(2+\frac{D-3}{2}\right)}{\pi^{\frac{D-1}{2}}}
       \frac{D-3}{D-2} c_{ik}  \p_a\p_b
     \frac{x^ix^k}{\vert\vec{x}\vert^{1+D}} 
     +\mathcal{O}(\omega r)\right]     \,,
\end{equation}
where $r\equiv \vert \vec{x}\vert$, and  $c_{ij}$ is a symmetric and traceless tensor characterizing the amplitude of the $t$--$t$ component of the external (quadrupole) tidal field,  $\bar h_{00}(\omega)=\Mp^{(D-2)/2}  c_{ij}x^i x^j(1+\mathcal{O}(\omega^2 r^2))$. 

Some further comments on the result \eqref{E_opfc} are in order. First, we note that all corrections of order $\omega r$ and higher have been neglected. To determine the effective couplings $K_{+-}^{(E)}(\omega)$ from the bulk solution, it is sufficient to match the leading falloffs $r^\ell$ and $r^{-\ell-1}$. Subleading terms in $\omega r$ correspond to large-distance corrections to the intermediate zone solution and match automatically once $K_{+-}^{(E)}(\omega)$ is fixed at leading order. 
In particular, note that the leading term displayed in Eq.~\eqref{E_opfc} arises from the first contribution in parentheses in the first line of Eq.~\eqref{E_opfb}, with the graviton propagator taken to be instantaneous---the remaining terms are all proportional to the frequency and contribute to the subleading corrections omitted in Eq.~\eqref{E_opfc}---and  reproduces the result of Ref.~\cite{Combaluzier--Szteinsznaider:2025eoc} (see Eq.~(3.96) therein).

Specializing to the $r$--$r$ component of $E_{ab}$ and combining the one-point function \eqref{E_opfc} with the same quantity evaluated on the tidal-field solution, we obtain
\begin{equation}
\label{E_opfd}
 \bar{E}_{rr} +   \left\langle E_{+\, rr}(t, \vec{x})\right\rangle_{\mathrm{in}-\mathrm{in}} 
       = -2\e^{-i\omega t}c_{\mathrm{ext}} Y_{2m}\left[ 1+ \mu^{2\varepsilon} \frac{K_{+-}^{(E)}\left(\omega\right)}{\Mp^{D-2}}  
       \frac{\Gamma\left(2+\frac{D-3}{2}\right)}{\pi^{\frac{D-1}{2}}}
       D(D-1)\frac{D-3}{D-2} r^{-1-D} 
     +\mathcal{O}(\omega r)\right]    \, ,
\end{equation}
where we introduced $c_{\mathrm{ext}}$, defined by $c_{ij} x^i x^j = c_{\mathrm{ext}} r^2 Y_{2m}$, with $Y_{2m}$ the $\ell=2$ spherical harmonic, and replaced $K_{+-}^{(E)}$ by $\mu^{2\varepsilon} K_{+-}^{(E)}$, where $\mu$ is an arbitrary energy scale and $\varepsilon \equiv 2 - D/2$, to ensure that the response function has the same dimensions for $D \neq 4$~\cite{Combaluzier--Szteinsznaider:2025eoc}.

\subsection{Including gravitational nonlinearities}
\noindent
The solution \eqref{E_opfd} was obtained from a tree-level calculation in the EFT and captures the leading-$r$ behavior of the $E$ field. To match it to the full relativistic solution, however, we must include the coupling to gravity, which generates subleading terms. For this purpose, we closely follow the procedure described in Ref.~\cite{Combaluzier--Szteinsznaider:2025eoc}, based on earlier results of Refs.~\cite{Caron-Huot:2025tlq,Correia:2024jgr}. In particular, we work in dimensional regularization---this is why we kept $D$ generic in all previous expressions---and account for nonlinear $r_s$ corrections to the Minkowski-space solution via a Born-series expansion. 
The calculation is most naturally performed in Regge--Wheeler gauge, whereas the graviton one-point function was derived in de Donder gauge. To bridge the two, we express the coefficient $K_{+-}^{(E)}$ in terms of the integration constants of the homogeneous Zerilli equation in the bulk. This guarantees that the bulk solution satisfies the correct boundary conditions at the worldline source, with the delta-function fixing the behavior of the solution at infinity.

We start from the $D$-dimensional Zerilli equation:
\begin{equation}
\frac{\d^2\Psi_{\rm Z}}{\d r_\star^2} + \Big(\omega^2 - V_{\rm Z}(r)\Big)\Psi_{\rm Z} = 0\,,
\label{eq:ddimZeq}
\end{equation}
where
\begin{equation}
    \d r_\star \equiv \frac{\d r}{f(r)}\,,
    \qquad
    f(r)\equiv 1-\left(\frac{r_s}{r}\right)^{D-3} = 1-\frac{2 GM n_D \mu^{2\varepsilon}}{r^{D-3}}  \,,
\qquad {\rm with}~~n_D\equiv \frac{4\pi^{\frac{3-D}{2}}\Gamma\left(\frac{D-1}{2}\right)}{D-2}\,,
\end{equation}
and where the potential is (see Refs.~\cite{Kodama:2003jz,Hui:2020xxx} for details)
\begin{align}
\nonumber
V_\text{Z}(r) = \bigg[& 4(D-4)(D-2)^4 f^3-8(D-2)^2(D-2)(D-6)\ell(\ell+D-3)f^2\\\nonumber
&+4(D-2)(D-2)(D-12)\ell^2(\ell+D-3)^2f\\[2pt]
&+2(D-2)^3(D+2)r^3 f'{}^3-4(D-2)^2 (D-6)\ell(\ell+D-3)r^2f'{}^2\\[2pt]\nonumber
&-8(D-2)^2\ell^2(\ell+D-3)^2rf'+12(D-2)^5 rf^2f'+(D-2)^3(D(D+10)-32)r^2 ff'{}^2\\[2pt]\nonumber
&-4(D-2)^2(D-2)(3D-8)\ell(\ell+D-3)rff'\\[2pt]\nonumber
&+16\ell^2(\ell+D-3)^2(D-2)\ell(\ell+D-3)
\bigg]
\frac{f [2\ell(\ell+D-3)+(D-2)(rf'-2f)]^{-2}}{4(D-2)r^2 } \,,
\end{align}
which reduces to the well-known Zerilli potential in $D=4$~\cite{Regge:1957td,Zerilli:1970wzz}.\footnote{We refer to Ref.~\cite{Hui:2020xxx} for the precise relation between the  Zerilli field $\Psi_{\text{Z}}$ and the fluctuation of the metric tensor in standard perturbation theory of Sec.~\ref{sec:UV}.}
In order to connect with the EFT, we treat both $\omega$ and $r_s$ corrections in Eq.~\eqref{eq:ddimZeq} perturbatively. In this spirit, we move all these terms to the right-hand side of Eq.~\eqref{eq:ddimZeq} and solve the equation order by order in the frequency and background nonlinearities. 

We begin with the homogeneous equation, which reads
\begin{equation}
    \left(\frac{\d^2}{\d r^2}-\frac{(\ell-\varepsilon)(\ell-\varepsilon+1)}{r^2}\right)\Psi_{\mathrm{Z}}(r)=0\,,
\end{equation}
with $\varepsilon=2-D/2$. This equation admits two solutions, one decaying and one growing, given by
\begin{equation}
\label{homogen_psi}
    \Psi_{\mathrm{Z}}(r)=\mu^{-\varepsilon} B_{\mathrm{reg}} r^{\ell+1-\varepsilon}+\frac{\mu^{\varepsilon} B_{\mathrm{irr}}}{2 \ell+1-2 \varepsilon} r^{-\ell+\varepsilon}\,,
\end{equation}
where the factors of $\mu^{\mp\varepsilon}$ ensure that the field has the correct dimensions for $D \neq 4$, and the relative normalization is just a matter of convenience. The factors $B_{\mathrm{reg}}$ and $B_{\mathrm{irr}}$ are integration constants, which will later be related to the relativistic vacuum solution in the exterior of the star.

The solution \eqref{homogen_psi} can be used to compute the $E_{rr}$ component of the Weyl tensor via~\cite{Hui:2020xxx}
\begin{equation}
    E_{rr}=-\frac{1}{\Mp}\left(\partial_r^2 H_0 +2i\omega\partial_r  H_1\right)\,,
 \label{eq:ErrevenZ}
\end{equation}
where $H_0$ and $H_1$ (in the notation of Refs.~\cite{Hui:2020xxx,Combaluzier--Szteinsznaider:2025eoc}) are determined by the constraint equations obtained from the quadratic action for the perturbations in $D$ dimensions [see Eq.~(3.54) of Ref.~\cite{Hui:2020xxx}]:
\begin{align}
H_0 &=-\frac{\mu^\varepsilon r^{\frac{2-D}{2}}}{2[(D-2)(D-3)\ell(\ell-1)(\ell+D-3)(\ell+D-2)]^{1/2}}\bigg[2(D-3)(D-2)r\partial_r\Psi_\text{Z}  \nonumber \\
       &\qquad\qquad\qquad\quad + (D-3) \left(\left(D^2+2 D (\ell-3)+2\ell (\ell-3) +8\right)-2(D-2)r^2\omega^2\right)\Psi_\text{Z} \bigg] \,,
        \\
    H_1 &= \frac{i\omega \mu^\varepsilon r^{\frac{4-D}{2}}}{4[(D-2)(D-3)\ell(\ell-1)(\ell+D-3)(\ell+D-2)]^{1/2}}\bigg[2(D-2)r\partial_r\Psi_\text{Z} \nonumber  \\
       &\qquad\qquad\qquad\qquad\qquad\qquad\qquad\qquad\qquad\qquad
        + \left(D^2-2\ell(\ell-3)-2D(\ell+1)\right)\Psi_\text{Z}\bigg] \,.
\end{align}
Since the linearized Weyl tensor around flat space is gauge invariant, Eq.~\eqref{eq:ErrevenZ} can be readily matched onto the result \eqref{E_opfd}, yielding the following relation between the worldline effective couplings $K_{+-}^{(E)}$ and the integration constants $B_{\mathrm{reg}}$ and $B_{\mathrm{irr}}$ of the bulk solution~\cite{Combaluzier--Szteinsznaider:2025eoc}: 
\begin{equation}
\label{Bratio}
    \frac{K_{+-}^{(E)}(\omega)}{\Mp^{D-2}}\frac{D-3}{D-2}\frac{\Gamma(\frac{D+1}{2})}{\pi^{\frac{D-1}{2}}}D(D-1)(D+1)
    = \frac{B_\text{irr}}{ B_\text{reg}}\,.
\end{equation}
Given the homogeneous solution \eqref{homogen_psi} and the condition \eqref{Bratio}, we can now solve for gravitational nonlinearities by considering the inhomogeneous equation
\begin{equation}
    \left(\frac{\d^2}{\d r^2}-\frac{(\ell-\varepsilon)(\ell-\varepsilon+1)}{r^2} \right)\Psi_{\text{Z}}(r) = V_{\Psi_\text{Z}}(r)\Psi_\text{Z}(r)\,,
    \label{eq:Z_EWE}
\end{equation}
where the potential $V_{\Psi_\text{Z}}$ on the right-hand side encodes all corrections in $\omega$ and $r_s$ (see Ref.~\cite{Combaluzier--Szteinsznaider:2025eoc} for an explicit expression).

Using the Born series~\cite{Caron-Huot:2025tlq,Correia:2024jgr},  we obtain the following expression for $\Psi_\text{Z}^{\ell=2}$ in the $\varepsilon\to0$ limit~\cite{Combaluzier--Szteinsznaider:2025eoc}:
\begin{equation}
\label{eq:psibarel2}
\begin{aligned}
    \Psi_\text{Z}^{\ell=2}(r) = \, \,  & r^3B_\text{reg}\left[1+\omega^2 \bar{G}_D^2\left(-\frac{107}{210\varepsilon} -\frac{2731}{9800} -\frac{107}{70}\log(\mu r)\right)\right] \\
    &+ \frac{B_\text{reg}}{r^2}\left[ -\frac{189 \bar{G}_D^5}{32}+\omega^2\bar{G}_D^7\left(\frac{4937}{960\varepsilon}+  \frac{2802329}{403200} +\frac{64181}{960}\log(\mu r)\right)\right] \\
    &+ \frac{B_\text{irr}}{r^2}\left[\frac{1}{5}+\omega^2\bar{G}_D^2\left(\frac{107}{1050\varepsilon}+ \frac{129359}{441000} +\frac{107}{210}\log(\mu r)\right) \right]+\mathcal{O}(\varepsilon)+\dots\,,
\end{aligned}
\end{equation}
where $\bar{G}_D\equiv G M n_D\xrightarrow{D\rightarrow4}\bar{G}\equiv GM$, and the ellipsis denotes subleading terms in the $1/r$ expansion.
As expected, the bare solution \eqref{eq:psibarel2} exhibits divergences in the $\varepsilon \to 0$ limit, which must be regulated. This can be achieved by introducing a renormalization scheme and appropriate counterterms in the worldline action to absorb the infinities and enforce the renormalization conditions. Ultimately, this procedure amounts to defining the renormalized coefficients $\bar{B}_\text{reg}$ and $\bar{B}_\text{irr}$ as
\begin{equation}
\begin{pmatrix}
    {B}_{\text{reg}} \\
    {B}_{\text{irr}}
\end{pmatrix}
    = \begin{pmatrix}
    \bar{B}_\text{reg} \\
    \bar{B}_\text{irr}
\end{pmatrix}+
    \omega^2\begin{pmatrix}
       \frac{107 \bar{G}^2}{210\varepsilon} +\frac{107\bar{G}^2}{105}  (\gamma_\text{E} -1+\log (4\pi))  & 0 \\
       -\frac{32 \bar{G}^7}{3\varepsilon} -\frac{224\bar{G}^7}{3}   (\gamma_\text{E} -1+\log (4\pi))
        &-\frac{107 \bar{G}^2}{210\varepsilon} -\frac{107\bar{G}^2}{105}  (\gamma_\text{E} -1+\log (4\pi))
    \end{pmatrix}\begin{pmatrix}
    \bar{B}_\text{reg} \\
    \bar{B}_\text{irr}
\end{pmatrix},
\label{eq:subscheme}
\end{equation}
where $\gamma_\text{E}$ is the Euler--Mascheroni constant. Using the (non-minimal) subtraction scheme defined in Eq.~\eqref{eq:subscheme}, we obtain the following renormalized Zerilli solution~\cite{Combaluzier--Szteinsznaider:2025eoc}:\footnote{Note that the shift \eqref{eq:subscheme} should be performed at the level of Eq.~\eqref{eq:psibarel2} including the order-$\varepsilon$ terms, which we have omitted for simplicity. When multiplied by the poles in the shift \eqref{eq:subscheme}, these terms introduce additional finite contributions, which in turn lead to the result \eqref{eq:psiZRapp}.}
\begin{equation}
\begin{aligned}
    \Psi_\text{Z}^{\mathrm{R} ,\ell=2}(r)=\, &\bar{B}_\text{reg} \left[r^3 \left(1-\bar{G}^2 \omega ^2 \left(\frac{214}{105} \log (\mu  r)+\frac{2731}{9800}\right)\right)+\frac{\bar{G}^7 \omega ^2}{r^2}\left(\frac{3011}{80} \log (\mu  r)+\frac{1545209}{57600}\right)-\frac{189 \bar{G}^5}{32r^2}\right]\\
    &+\frac{\bar{B}_\text{irr}}{r^2} \left(\bar{G}^2 \omega ^2 \left(\frac{214}{525} \log (\mu  r)+\frac{111383}{441000}\right)+\frac{1}{5}\right)+\dots\,,
\end{aligned}
\label{eq:psiZRapp}
\end{equation}
which shows the characteristic running behavior in terms of the energy scale $\mu$. The full solution, including subleading terms in the $1/r$ expansion, is provided in the Mathematica code available at the GitHub repository~\cite{gitTA}.

\section{Technical details of stellar perturbation analysis}
\label{app:StellarPerturbation}

\subsection{Expressions for the coefficients of stellar perturbation theory}
\label{app:coefficients}
\noindent
Here, we provide the expressions for the coefficients of the perturbation equations in Sec.~\ref{sec:UV} (setting $G = 1$). First, the coefficients of Eqs.~\eqref{eq:OpeLV} and~\eqref{eq:OpeLW} are given by
\begin{align}
   \beta_{V,{\cal H}'}=&-\frac{
\e^{-\lambda} r^{2-\ell}}{
(\ell+2)(\ell+1)\ell(\ell-1)
}\left[
3+\e^{2\lambda}\left(1+8\pi r^2 p\right)^2
-2\e^{\lambda}\left(\ell(\ell+1)+16\pi r^2 p\right)
\right],\\
   \beta_{V,{\cal H}}=&-\frac{
\e^{-\lambda} r^{1-\ell}}{
(\ell+2)(\ell+1)\ell(\ell-1)
}\\
&\times \left\{
3+
\e^{3\lambda}\left(1+8\pi r^2 p\right)^3
-\e^{2\lambda}\left(1+8\pi r^2 p\right)
\left[
1+\ell(\ell+1)+8\pi r^2\left(\epsilon+6p\right)
\right]
-\e^{\lambda}
\left[
3-\ell(\ell+1)-8\pi r^2\left(3\epsilon+8p\right)
\right]
\right\}\,,\nonumber\\
   \beta_{V,V}=&\, \frac{
1+\ell-\e^{\lambda}\left(1+8\pi r^2 p\right)
}{r}\,,\\
   \beta_{V,W}=&\, \frac{
\e^{\lambda/2}}{
\ell(\ell+1)r
}\left[
\ell(\ell+1)-16\pi r^2(\epsilon+p)
\right],
\end{align}
and
\begin{align}
   \beta_{W,{\cal H}'}=&\frac{
\e^{-\lambda/2} r^{2-\ell}
}{
(\ell+2)(\ell-1)
}\left[
\e^{\lambda}\left(1+8\pi r^2 p\right)-1
\right]\,,\\
   \beta_{W,{\cal H}}=&-\frac{
\e^{-\lambda/2} r^{1-\ell}}{
2(\ell-1)(\ell+2)
\left[
1-\e^{\lambda}\left(1+8\pi r^2p\right)
\right]
(\epsilon+p)
}\nonumber\\
&\times \Big\{
2(\epsilon+p)
+2\e^{3\lambda}\left(1+8\pi r^2p\right)^3(\epsilon+p)
-\e^{2\lambda}\left(1+8\pi r^2p\right)(\epsilon+p)
\left[
8-3\ell(\ell+1)+16\pi r^2(\epsilon+4p)
\right]
\\
&\quad \quad +\e^{\lambda}
\left[
(\epsilon+p)
\left(
4-3\ell(\ell+1)+16\pi r^2(\epsilon+2p)
\right)
-2(\ell-1)(\ell+2)r\epsilon'
\right]
\Big\}\,,\nonumber\\
   \beta_{W,V}=&\,\frac{\e^{\lambda/2}}{r}\ell(\ell+1)\,,\\
   \beta_{W,W}=&\,\frac{\ell+1}{r}+\frac{\epsilon'}{\epsilon+p}\,.
\end{align}
The coefficients of Eqs.~\eqref{eq:SH}--\eqref{eq:SW} are given by
\begin{align}
    S_{\cal H, H'}=&
-\frac{
2\e^{-\lambda-\nu}r
}{
(\ell+2)(\ell+1)\ell(\ell-1)
}\left[
\e^{\lambda}\left(1+8\pi r^2p\right)-3
\right]
\left\{
3+\e^{2\lambda}\left(1+8\pi r^2p\right)^2
-2\e^{\lambda}\left[\ell(\ell+1)+16\pi r^2p\right]
\right\}\,,\\
 S_{\cal H, H}=&\, 
\frac{
\e^{-\lambda-\nu}
}{
(\ell+2)(\ell+1)\ell(\ell-1)
}\\
&\times
\Big\{
18
-2\e^{4\lambda}\left(1+8\pi r^2p\right)^4
+2\e^{3\lambda}\left(1+8\pi r^2p\right)^2
\left[
4+\ell(\ell+1)+8\pi r^2(\epsilon+9p)
\right]\nonumber\\
&
\quad \quad  +6\e^\lambda
\left[
\ell(\ell+1)-4+8\pi r^2(3\epsilon+7p)
\right]
-\e^{2\lambda}
\Big[
\ell(\ell+1)\left(\ell(\ell+1)+6\right)\nonumber\\
&
\quad \quad +32\pi r^2
\left[
3\epsilon
+p\left(
13+2\ell(\ell+1)+8\pi r^2(3\epsilon+13p)
\right)
\right]
\Big]
\Big\},\nonumber\\
S_{{\cal H},V}=&
-8\pi \e^{\lambda-\nu}r^\ell
\left[
\epsilon+p
-
\frac{2r\epsilon'}
{1-\e^\lambda\left(1+8\pi r^2p\right)}
\right]\,,\\
S_{\mathcal{H},W}=&\,
\frac{
32\pi \e^{\lambda/2-\nu} r^\ell
(\epsilon+p)
}{
\ell(\ell+1)
}\left[
3-\e^\lambda\left(1+8\pi r^2p\right)
\right]\,,
\end{align}
and
\begin{align}
    S_{V,{\cal H}'}=& -\frac{
\e^{-2\lambda-\nu}r^{4-\ell}
}{
(\ell+2)^2 (\ell+1)^2 \ell^2 (\ell-1)^2
}\nonumber \\
&\times
\Big\{
27
+\e^{4\lambda}\left(1+8\pi r^2p\right)^4
-2\e^{3\lambda}\left(1+8\pi r^2p\right)^2
\left[
1+3\ell(\ell+1)+40\pi r^2p
\right]\\
&\quad
-6\e^\lambda
\left[
1+5\ell(\ell+1)+72\pi r^2p
\right]
+4\e^{2\lambda}
\left[
3+\ell(\ell+1)\left(\ell^2+\ell+3\right)
+8\pi r^2p
\left(
8+7\ell(\ell+1)+72\pi r^2p
\right)
\right]
\Big\}\,,\nonumber\\
S_{V,{\cal H}}=&- \frac{
\e^{-2\lambda-\nu}r^{3-\ell}
}{
(\ell+2)^2 (\ell+1)^2 \ell^2 (\ell-1)^2
}\nonumber\\
& \times\Big\{
27
+\e^{5\lambda}\left(1+8\pi r^2p\right)^5
-3\e^\lambda
\left[
11+\ell(\ell+1)
-72\pi r^2(\epsilon+2p)
\right]
\nonumber\\
&-\e^{4\lambda}\left(1+8\pi r^2p\right)^3
\left[
3+5\ell(\ell+1)
+8\pi r^2(\epsilon+12p)
\right]\\
&+\e^{3\lambda}\left(1+8\pi r^2p\right)
\left[
8+15\ell(\ell+1)
+8\pi r^2
\left[
\epsilon\left(5+4\ell(\ell+1)+72\pi r^2p\right)
+p\left(41+31\ell(\ell+1)+424\pi r^2p\right)
\right]
\right]\nonumber
\\
&-\e^{2\lambda}
\left[
-\ell(\ell+1)(2\ell-3)(2\ell+5)
+8\pi r^2
\left[
3\epsilon\left(5+4\ell(\ell+1)+72\pi r^2p\right)
+p\left(55+47\ell(\ell+1)+792\pi r^2p\right)
\right]
\right]
\Big\}\,,\nonumber\\
S_{V,V}=& -\frac{
16\pi \e^{-\nu}r^3
(\epsilon+p)
}{
(\ell+2)(\ell+1)\ell(\ell-1)
}\left[
3-\e^\lambda\left(1+8\pi r^2p\right)
\right]\,,\\
S_{V,W}=&
-\frac{
16\pi \e^{-\lambda/2-\nu}r^3
(\epsilon+p)
}{
(\ell+2)(\ell+1)^2\ell^2(\ell-1)
}\left[
9
+\e^{2\lambda}\left(1+8\pi r^2p\right)^2
-2\e^\lambda\left(1+\ell(\ell+1)+24\pi r^2p\right)
\right]\,,
\end{align}
and
\begin{align}
    S_{W,{\cal H}'}=& -\frac{
\e^{-3\lambda/2-\nu}r^{4-\ell}
}{
(\ell+2)^2(\ell+1)\ell(\ell-1)^2
}\\
& \times \left[
9
-\e^{3\lambda}\left(1+8\pi r^2p\right)^3
+\e^{2\lambda}\left(1+8\pi r^2p\right)
\left(
3+4\ell(\ell+1)+56\pi r^2p
\right)
-\e^\lambda
\left(
3+8\ell(\ell+1)+120\pi r^2p
\right)
\right],\nonumber\\
S_{W,{\cal H}}=& - \frac{
\e^{-3\lambda/2-\nu}r^{3-\ell}
}{
(\ell+2)^2 (\ell-1)^2 (\ell+1) \ell
}\\
& \times \Big\{
9
-\e^{4\lambda}\left(1+8\pi r^2p\right)^4
+\e^{3\lambda}\left(1+8\pi r^2p\right)^2
\left(
4+3\ell(\ell+1)+8\pi r^2(\epsilon+9p)
\right)\nonumber
\\
&+\e^\lambda
\left(
(\ell-3)(\ell+4)+24\pi r^2(3\epsilon+7p)
\right)\nonumber
\\
&-2\e^{2\lambda}
\left[
-\ell(\ell+1)\left[\ell(\ell+1)-4\right]
+8\pi r^2
\left[
\epsilon\left(3+\ell(\ell+1)+24\pi r^2p\right)
+p\left(13+5\ell(\ell+1)+104\pi r^2p\right)
\right]
\right]
\Big\}\nonumber\,,\\
S_{W,V}=& -2\e^{\lambda/2-\nu}r^2
\left[
\frac{8\pi r(\epsilon+p)}{(\ell+2)(\ell-1)}
-\frac{\epsilon'}
{\left[1-\e^\lambda\left(1+8\pi r^2p\right)\right](\epsilon+p)}
\right]\,,\\
S_{W,W}=& -\frac{
16\pi \e^{-\nu}r^3
(\epsilon+p)
}{
(\ell+2)(\ell+1)\ell(\ell-1)
}\left[
3-\e^\lambda\left(1+8\pi r^2p\right)
\right]\,.
\end{align}

\subsection{Comparison between the nonperturbative and perturbative computations} \label{app:compNP}
\noindent
Here, we assess the validity of the perturbative analysis of Sec.~\ref{sec:UV} by comparing the resulting expression for $y^{\rm int}$ in Eq.~\eqref{eq:yint} with the corresponding nonperturbative result. Assessing the accuracy of $y^{\rm int}$ computed perturbatively provides a direct estimate of the impact of the low-frequency truncation on the parameter $\Lambda_2$ entering the waveform, derived in Sec.~\ref{sec:waveform}.

Figure~\ref{fig:PerturbativeVSNonperturbative} shows the nonperturbative and perturbative results for $y^{\rm int}$ as functions of frequency. The blue and red dashed vertical lines indicate the contact frequency in Eq.~\eqref{eq:contact} and the ISCO frequency in Eq.~\eqref{eq:fISCO}, respectively. In the Fisher analysis of Sec.~\ref{sec:waveform}, the upper cutoff of the frequency integration range is chosen as the minimum among the contact frequency, the ISCO frequency, and the detector maximum cutoff frequency. The dashed curves, corresponding to the perturbative results, follow the solid curves, corresponding to the nonperturbative results, throughout most of the frequency range below the chosen upper cutoff, although the relative difference between the nonperturbative and perturbative results increases toward the cutoff, reaching about $26\%~(C=0.1)$ and $31\%~(C=0.2)$. These values can be regarded as conservative endpoint estimates of the uncertainty in the $\Lambda_2$ contribution associated with the finite-frequency truncation at second order in the low-frequency expansion. The band-averaged impact in the Fisher analysis of Sec.~\ref{sec:waveform} is expected to be smaller because the discrepancy is largest only near the upper end of the frequency-integration range: lowering the upper
integration frequency to $\sim\!90\%$ of $f_\text{\tiny max}$, where the
perturbative and nonperturbative $y_{\rm int}$ agree well, increases the
forecast uncertainty on the 8PN term by $\sim\!30\%$. We have verified that the similar behavior is found for other choices of the EoS or stellar compactness.

\begin{figure}[t!] 
    \includegraphics[width=0.65\textwidth]{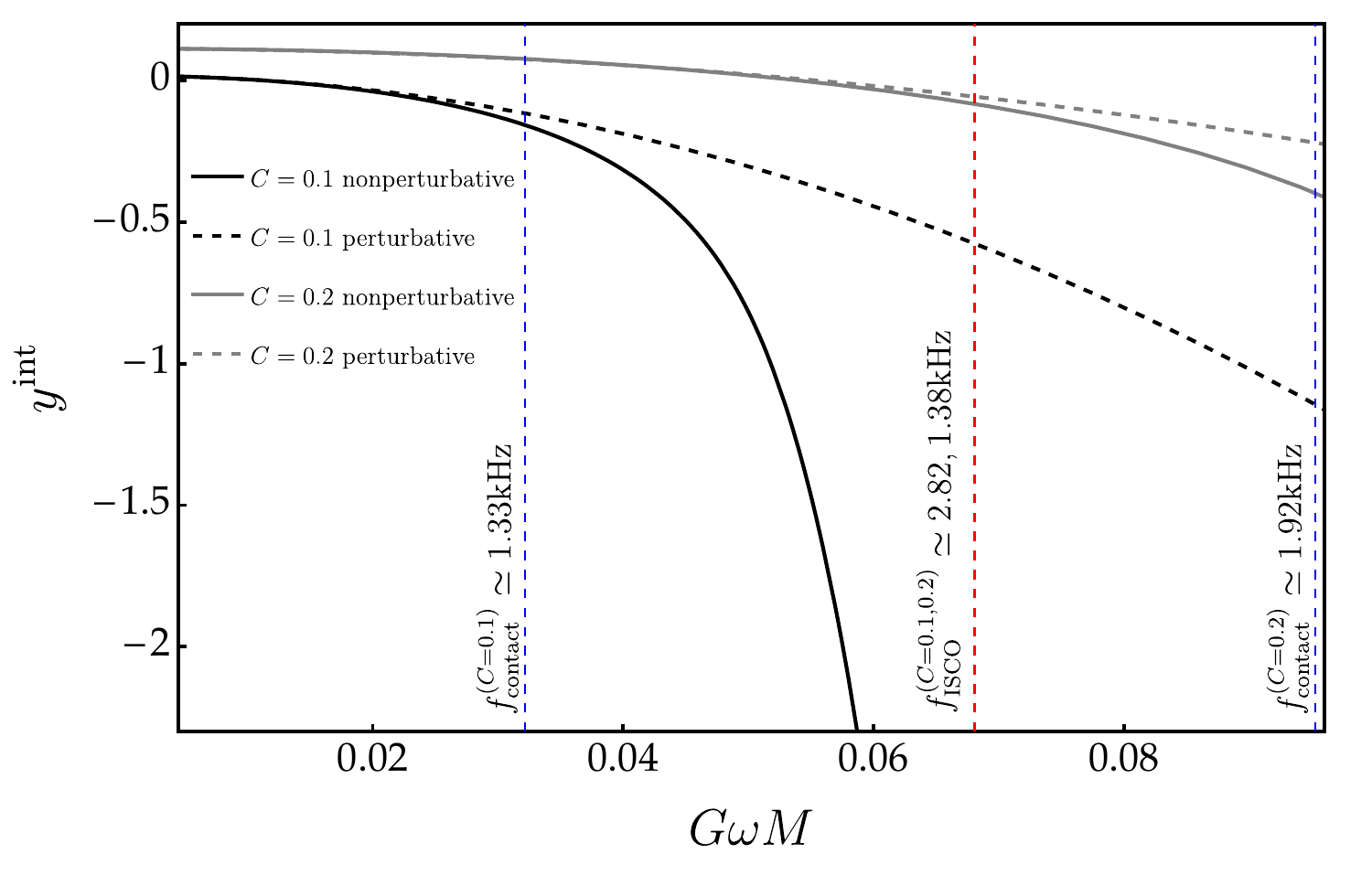}
    \caption{Comparison between the nonperturbative~(solid) and the perturbative results~(dashed) for $y^{\rm int}$ given in Eq.~\eqref{eq:yint}, for a neutron star with AP4 EoS and two representative values of compactness, $C = 0.1$ (black) and $C = 0.2$ (gray), as functions of the frequency. The blue and red dashed vertical lines mark the contact frequency in Eq.~\eqref{eq:contact} and the ISCO frequency in Eq.~\eqref{eq:fISCO}, respectively.}
\label{fig:PerturbativeVSNonperturbative}
\end{figure}

\section{Matching to the vacuum general relativistic solution}\label{app:matchingtoGR}
\noindent
In this Appendix, we derive the most general solution outside the star and match it to Eq.~\eqref{eq:psiZRapp}. This will allow us to establish a relation between the renormalized coefficients $\bar{B}_\text{reg}$ and $\bar{B}_\text{irr}$---which are in turn related to the EFT couplings via Eq.~\eqref{Bratio}---and the integration constants of the Zerilli solution in full general relativity---the latter being determined by the numerical solution in the stellar interior, as discussed in Sec.~\ref{sec:Matching}. For this purpose, we work in $D=4$ throughout this section.

We start from the Zerilli equation in vacuum general relativity~\cite{Zerilli:1970wzz} (see Eq.~\eqref{eq:ddimZeq} with $D=4$):
\begin{equation}
\label{Zerilli}
    f(r)\partial_r\left[f(r) \partial_r \Psi_{\mathrm{Z}}(r)\right]+\left[\omega^2-f(r)\left(\frac{r_s}{r^3}+\frac{2 n}{3 r^2}+\frac{8 n^2(2 n+3)}{3\left(2 n r+3 r_s\right)^2}\right)\right] \Psi_{\mathrm{Z}}(r)=0 \,,
\end{equation}
where $n\equiv (\ell+2)(\ell-1)/2$.
We then solve Eq.~\eqref{Zerilli} perturbatively in $\omega r_s$ in the range $r_s < r \ll 1/\omega$. Note that for a star with radius $R_\star \gtrsim \mathcal{O}(1)r_s$, the perturbative expansion is valid throughout the region $R_\star \leq r \ll 1/\omega$. (In contrast to the black hole case~\cite{Combaluzier--Szteinsznaider:2025eoc}, there is no need to distinguish between a ``near zone'' and an ``intermediate zone''.)

It is useful to introduce the  rescaled field  
\begin{equation}
    u(x(r)) \equiv \left(\frac{r}{r_s}\right)^\ell \Psi_{\mathrm{Z}}(r)\,,
\qquad  x\equiv\frac{r_s}{r} \,.
\end{equation}
Then, the  equation \eqref{Zerilli} for $\ell=2$ reads
\begin{equation}
    x(1-x) u^{\prime\prime}(x)+(6-7 x) u'(x)-\frac{3 x (27 x+58)+32}{(3 x+4)^2}u(x)=\frac{(\omega r_s)^2}{(x-1) x^3}u(x)\,,
\end{equation}
which we solve order by order in $\omega r_s$ for $0 < x < 1$ by expanding $u(x)$ as
\begin{equation}
    u(x)= u_0(x)+\omega r_s u_1(x)+(\omega r_s)^2u_2(x)+\cdots .
\end{equation}
At zeroth and first order in $\omega r_s$ the equation simply reduces to homogeneous one: 
\begin{align}
    x(1-x) u_0^{\prime\prime}(x)+(6-7 x) u'_{0}(x)-\frac{3 x (27 x+58)+32}{(3 x+4)^2}u_0(x)&=0\,,
\\
    x(1-x) u_1^{\prime\prime}(x)+(6-7 x) u'_{1}(x)-\frac{3 x (27 x+58)+32}{(3 x+4)^2}u_1(x)&=0\,,
\end{align}
which admit the following solutions:
\begin{align}
    u_0(x)&=a_0\frac{4+6 x-3 x^3}{x^5 (4+ 3 x)}
    - b_0\frac{ x (12+ (24+13 x)x)+3 \left(4+6 x-3 x^3\right) \log (1-x)}{3x^5 (4+3 x)}\,,
\\
    u_1(x)&=a_1\frac{4+6 x-3 x^3}{x^5 (4+ 3 x)}
    - b_1\frac{ x (12+ (24+13 x)x)+3 \left(4+6 x-3 x^3\right) \log (1-x)}{3x^5 (4+3 x)}\,,
\end{align}
where $a_n$ and $b_n$ are the integration constants of the homogeneous equations at the nth order in $\omega r_s$.

At second order, the equation takes instead the form
\begin{equation}
    x(1-x) u_2^{\prime\prime}(x)+(6-7 x) u_2'(x)-\frac{3 x (27 x+58)+32}{(3 x+4)^2}u_2(x)=\frac{1}{(x-1) x^3}u_0(x)\,.
\label{eq:equ2}
\end{equation}
The most general solution to the inhomogeneous equation \eqref{eq:equ2} is given by a linear superposition of the homogeneous solutions and a particular solution. The latter can be obtained analytically in closed form, although its explicit expression is rather involved and will not be reported here (but see the Mathematica code available at~\cite{gitTA}). Instead, we present its large-$r$ expansion, which is useful for matching to the EFT: 
\begin{equation}
\begin{aligned}
\label{Zer_sol_inf_GR}
     \Psi_{\mathrm{Z}}^{\ell=2}(r) \xrightarrow{r\rightarrow\infty}
      &\,\, a_0\frac{r^3}{r_s^3}-\left(\frac{189 }{1024}a_0-\frac{1 }{5}b_0\right)\frac{r_s^2 }{r^2}
     \\
     &+\omega r_s  \left[a_1\frac{r^3}{r_s^3}-\left(\frac{189 }{1024}a_1-\frac{1 }{5}b_1\right)\frac{r_s^2 }{r^2}\right]\\
     &+\omega^2r_s^2\left[\frac{r^3 }{ r_s^3}\left( a_2+\frac{107}{210}a_0 \log \left(\frac{r_s}{r}\right) \right) +
     \left(b_2 -\left(\frac{107}{1050} b_0 +\frac{3011}{10240}a_0\right)\log \left(\frac{r_s}{r}\right)\right)\frac{r_s^2}{r^2}\right]+\cdots\,,
\end{aligned}
\end{equation}
where we switched back to the Zerilli variable.

Note that in Eq.~\eqref{Zer_sol_inf_GR} we reported only the $r^3$ and $r^{-2}$ falloffs, which will suffice for matching to Eq.~\eqref{eq:psiZRapp}. At each order in $\omega$, additional powers of $r$ generally appear, but are omitted for compactness; these are automatically matched once the coefficients $a_n$ and $b_n$ are fixed using Eq.~\eqref{Zer_sol_inf_GR}. Another aspect worth noting is that the term proportional to $a_0/r^2$ in the first line of Eq.~\eqref{Zer_sol_inf_GR}, despite resembling a genuine induced falloff such as $b_0/r^2$, is in fact an artifact of gravitational nonlinearities and corresponds to a subleading correction to the tidal field $a_0 r^3$. This will become clear later when matching Eq.~\eqref{Zer_sol_inf_GR} to $\Psi_\text{Z}^{\mathrm{R}}$, as this term will cancel exactly against the last term in Eq.~\eqref{eq:psiZRapp}. This is consistent with the black hole case---recovered by setting $b_0 = 0$ (see Ref.~\cite{Combaluzier--Szteinsznaider:2025eoc})---for which the static Love numbers are expected to vanish.

Comparing with Eq.~\eqref{eq:psiZRapp}, we obtain the following matching conditions:
\begin{align}
r_s^3\bar{B}_\text{reg}&  = a_0 + \omega r_s a_1  + \omega^2 r_s^2\left[a_2+  a_0 \left(\frac{2731}{39200}+\frac{107}{210}\log (\mu  r_s)\right) \right]\,,\label{eq:Breg}
\\
r_s^{-2}\bar{B}_\text{irr} & = b_0 +  \omega r_s b_1 + \omega^2r_s^2 \left[\frac{945 }{1024}a_2 +5 b_2 -\frac{1015283 }{1032192}a_0-\frac{111383}{352800}b_0 -  
\left(a_0+\frac{107}{210}b_0\right) \log (\mu   r_s)
\right] \,.\label{eq:Birr}
\end{align}
Equation \eqref{Bratio} then implies the following values for the $\ell = 2$ electric-type response couplings in terms of the integration constants $a_n$ and $b_n$ of the full vacuum solution:
\begin{multline}
\label{Brations}
   \frac{45}{8G^4M^5} K_{+-}^{(E)}(\omega)
    = \frac{b_0}{a_0}
    + \omega r_s \left(\frac{b_1}{a_0} -\frac{b_0a_1}{a_0^2}\right)
    + \omega^2r_s^2\bigg[\!- \left(1+\frac{107}{105}\frac{b_0}{a_0}\right)\log (\mu r_s)
\\
    -\frac{67981 }{176400}\frac{ b_0}{ a_0} +\frac{a_1^2 b_0}{a_0^3}-\frac{a_1b_1}{a_0^2}-\frac{b_0a_2}{a_0^2}+ \frac{945 a_2}{1024 a_0}+\frac{5 b_2}{a_0}-\frac{1015283}{1032192}
    \bigg] \,.
\end{multline}
Using the definition in Eq.~\eqref{eq:KpmcEs}, we obtain the following dictionary: 
\begin{align}
    c_E & = \frac{8G^5M^5}{45 R_\star^5} \frac{b_0}{a_0},
\\
     i  \nu_E & = \frac{16G^5M^5}{45R_\star^5}  \left(\frac{b_1}{a_0} -\frac{b_0a_1}{a_0^2}\right),
\\
    c_{\dot E} & = \frac{32G^8M^8}{45 R_\star^8} \bigg[\!- \left(1+\frac{107}{105}\frac{b_0}{a_0}\right)\log (\mu r_s)
\nonumber \\
&\qquad\qquad\quad\,\,    -\frac{67981 }{176400}\frac{ b_0}{ a_0} +\frac{a_1^2 b_0}{a_0^3}-\frac{a_1b_1}{a_0^2}-\frac{b_0a_2}{a_0^2}+ \frac{945 a_2}{1024 a_0}+\frac{5 b_2}{a_0}-\frac{1015283}{1032192}
    \bigg]\,.
\end{align}
The result \eqref{Brations} can be written more compactly by absorbing the integration constants into two $\omega$-dependent coefficients, defined as
\begin{align}
\label{A,B}
    A(\omega)&\equiv a_0+\omega r_s a_1+\omega^2 r_s^2 a_2 \,, \\
    B(\omega)&\equiv b_0+\omega r_s b_1+\omega^2r_s^2 \left(5 b_2 +\frac{945}{1024}a_2\right)\,.
\end{align}
This yields the following expression for the electric-type Love number with $\ell=2$:
\begin{equation}
    \frac{45}{8G^4M^5} K_{+-}^{(E)}(\omega)
    =\frac{B(\omega)}{A(\omega)}-r_s^2\omega^2 \left[\left(1+\frac{107}{105}\frac{B(\omega)}{A(\omega)}\right) \log (\mu  r_s)+\frac{67981}{176400} \frac{B(\omega)}{A(\omega)} +\frac{1015283}{1032192}\right] + \mathcal{O}(\omega^3)\,,
\label{eq:Kpmapp0}
\end{equation}
where terms of order $\mathcal{O}(\omega^3)$ and higher have been neglected. Note that, in the black hole case, $b_0=0$, and, at the order considered in the small-frequency expansion, the coefficient of $\log(\mu r_s)$ reduces simply to $-r_s^2\omega^2$, in agreement with previous findings, see e.g.~Refs.~\cite{Chakrabarti:2013lua,Chakraborty:2025wvs,Combaluzier--Szteinsznaider:2025eoc}. In particular, this coefficient is found to be related to black hole absorption in a way that is independent of the angular momentum quantum number $\ell$~\cite{Combaluzier--Szteinsznaider:2025eoc}.

\section{Comparison with literature}
\label{app:comparison}
\noindent
Here, we discuss some differences between $c_{\dot E}$ and the dynamical tidal constant introduced in Refs.~\cite{Poisson:2020vap,Pitre:2023xsr,HegadeKR:2024agt}. We stress that such differences merely reflect different conventions. As discussed in the main text, TLNs defined prior to matching to waveform observables are scheme-dependent intermediate quantities. What ultimately matters is the combination of tidal parameters that enters the gravitational waveform after the matching procedure, since observable quantities are independent of the regularization scheme. The only residual physical effect is the renormalization-group running of the couplings.
The comparison presented below may therefore serve as a useful consistency check, both for computations of dynamical tidal responses for a given stellar model prior to matching and for future studies employing alternative matching procedures to connect tidal responses with waveform observables.

We first discuss the tidal constants defined in Refs.~\cite{Poisson:2020vap,Pitre:2023xsr}. Our dynamical TLN~$c_{\dot E}$ can be expressed in terms of the integration constants in the literature as follows:
\begin{align}
    c_{\dot E}=\frac{1}{3}\left[\ddot{k}_2+\left(\frac{140024}{11025}+\frac{4\pi^2}{3}+\ddot{T}_2\right)\left(\frac{GM}{R_\star}\right)^{3}k_2+\frac{6376}{4725}\left(\frac{G M}{R_\star}\right)^8 -\frac{4}{105}\left(\frac{G M}{R_\star}\right)^8\left(56+107k_2\left(\frac{GM}{R_\star}\right)^{-5}\right)\log\left(\mu r_s\right)\right]\,,\label{eq:cEandPP}
\end{align}
where $k_2$, $\ddot{k}_2$, and $\ddot{T}_2$ are the static tidal constant, the dynamical tidal constant, and the tidal moment at second order in the time-derivative expansion, respectively. Equation~\eqref{eq:cEandPP} has been obtained by comparing the gauge-invariant Zerilli functions in the two different approaches. It is worth noting that $c_{\dot E}\simeq \ddot{k}_2/3$ in the Newtonian limit. There appear to be two main differences in the characterization of dynamical tidal responses before matching. First, the dynamical tidal response in Refs.~\cite{Poisson:2020vap,Pitre:2023xsr} is characterized only by the constant~$\ddot{k}_2$ without a scale-dependent logarithmic term. Second, the constant part of $c_{\dot E}$ is also different from $\ddot{k}_2$. More specifically, the former is uniquely determined from the interior problem for a given stellar model, whereas $\ddot{k}_2$ shifts under a redefinition of $\ddot{T}_2$ and a change in the functional form of particular solutions of the tidal perturbation equation~\cite{HegadeKR:2024agt,Katagiri:2024wbg,Katagiri:2024fpn}. In Refs.~\cite{Poisson:2020vap,Pitre:2023xsr}, $\ddot{k}_2$ is fixed by setting $\ddot{T}_2$ to zero. 

To see how to extract $\ddot{k}_2$ more explicitly, let us consider the large-distance asymptotic expansion of the $t$--$t$ component of the perturbed metric in the harmonic coordinates,
\begin{align}
    g_{tt}=&-1+\frac{2GM}{\bar{r}}+\cdots+\left[d^{(0)}+\left(G M \omega\right)^2 d^{(2)}\right]\left(\frac{\bar{r}}{G M}\right)^\ell \left[1+\cdots\right]Y_{\ell m} \e^{-i \omega t}\nonumber\\
    &+\left[I^{(0)}+\left(G M \omega \right)^2 I^{(2)}\right]\left(\frac{G M}{\bar{r}}\right)^{\ell+1}\left[1+\cdots\right] Y_{\ell m} \e^{-i \omega t}\label{eq:largedisgtt}\\
    &+\left(G M \omega \right)^2\left[A^{(2)}\left(\frac{\bar{r}}{G M}\right)^{\ell+2}+\cdots+C_P^+\left(\frac{\bar{r}}{G M}\right)^\ell+\cdots+C_P^-\left(\frac{G M}{\bar{r}}\right)^{\ell+1}+\cdots \right]Y_{\ell m}\e^{-i\omega t}\,,\nonumber
\end{align}
where $\bar{r}$ is the harmonic radial coordinate; $d^{(0)}$, $d^{(2)}$, $I^{(0)}$, and $I^{(2)}$ are the integration constants associated with the homogeneous pieces of the tidal perturbation; the second line corresponds to the contribution of the particular solutions with $A^{(2)}$, $C_P^+$, and $C_P^-$~dimensionless numerical coefficients. It follows that one can shift $d^{(2)}$ and $I^{(2)}$ as $d^{(2)}\to d^{(2)}+C_P^+$ and $I^{(2)}\to I^{(2)}+C_P^-$ by changing the functional form of the particular solutions. It should be noted that, in the exterior, the tidal perturbation is governed by a single second-order differential equation and, hence, the perturbed metric is completely determined by matching with the interior problem, up to an overall normalization. This implies that matching to the interior solution uniquely determines the ratio between the coefficients of the terms proportional to $\bar{r}^\ell$ and $\bar{r}^{-\ell-1}$,
\begin{align}
   \mathfrak{K} \equiv \frac{I^{(0)}+\left(G M \omega \right)^2 \left(I^{(2)}+C_P^-\right)}{d^{(0)}+\left(G  M \omega\right)^2 \left(d^{(2)}+C_P^+\right)} \,.\label{eq:uniqueK}
\end{align}
The scale-independent part of $c_{\dot E}$ is constructed by extracting the quantity analog to $\mathfrak{K}$ from the terms proportional to $r^{\ell+1}$ and $r^{-\ell}$ in the asymptotic expansion of $\Psi_{\rm Z}^{\ell=2}(r)$ at large distances~[see Eq.~\eqref{Zer_sol_inf_GR}].

The tidal response function of Refs.~\cite{Poisson:2020vap,Pitre:2023xsr} is extracted from the homogeneous piece in Eq.~\eqref{eq:largedisgtt} through
\begin{align}
    I^{(0)}+\left(G M \omega \right)^2 I^{(2)} = \left[{d^{(0)}+\left(G  M \omega\right)^2 d^{(2)}}\right]\left[k_2 +\left(G M \omega \right)^2\bar{\ddot{k}}_2\right]\,.\label{eq:linearrelationforPP}
\end{align}
Here, $k_2$ is the static tidal constant identical to that in Eq.~\eqref{eq:cEandPP}, while the constant~$\bar{\ddot{k}}_2$ differs from $\ddot{k}_2$. The left-hand side of Eq.~\eqref{eq:linearrelationforPP} can be viewed as the induced response, while the first and second term on the right-hand side are interpreted as the tidal moment and response function, respectively. Note that the value of $k_2+(G M \omega )^2 \bar{\ddot{k}}_2$ in Eq.~\eqref{eq:linearrelationforPP} depends on the choice of the particular solution because $d^{(2)}$ and $I^{(2)}$ can be shifted, as noted above. By rewriting the right-hand side of Eq.~\eqref{eq:linearrelationforPP} up to quadratic order in $G M \omega$ as
\begin{align}
    I^{(0)}+\left(G M \omega \right)^2 I^{(2)} =  d^{(0)}\left[k_2+\left(G M \omega\right)^2 \left(\bar{\ddot{k}}_2+\frac{d^{(2)}}{d^{(0)}}k_2\right)\right] + \mathcal{O}\left[\left(G  M \omega\right)^4\right]\,,\label{eq:redefinition}
\end{align}
one can appreciate that the constant~$\bar{\ddot{k}}_2+(d^{(2)}/d^{(0)})k_2$ is equivalent to $\ddot{k}_2$ in Eq.~\eqref{eq:cEandPP}, upon setting $\ddot{T}_2$ to zero, as done in Refs.~\cite{Poisson:2020vap,Pitre:2023xsr}. This operation can be viewed as a redefinition of the tidal moment, with a corresponding redefinition of the tidal response function that leaves the left-hand side, i.e.~the induced moment, unchanged. It follows from Eq.~\eqref{eq:redefinition} that: (i) the redefined tidal moment used in Refs.~\cite{Poisson:2020vap,Pitre:2023xsr} corresponds to $d^{(0)}$ and is independent of $G M \omega$; (ii) the extracted value of $\ddot{k}_2 (=\bar{\ddot{k}}_2+(d^{(2)}/d^{(0)})k_2)$ depends on the choice of the particular solution at quadratic order in the $G M \omega$ expansion.   Therefore, $\ddot{k}_2$ differs from the second-order coefficient of the expansion of $\mathfrak{K}$ in Eq.~\eqref{eq:uniqueK}. Indeed,
\begin{align}
\ddot{k}_2=\frac{I^{(2)}}{d^{(0)}}\neq    \left. \frac{1}{2\left(G M\right)^2}\frac{\d^2\mathfrak{K}}{\d\omega^2}\right|_{\omega =0 }= \frac{I^{(2)}+C_P^-}{d^{(0)}}-\frac{I^{(0)} \left(d^{(2)}+C_P^+\right) }{\left(d^{(0)} \right)^2}\,.
\end{align}
The overall approach for computing dynamical TLNs in Ref.~\cite{HegadeKR:2024agt} is similar to that adopted by Refs.~\cite{Poisson:2020vap,Pitre:2023xsr}, with two notable differences discussed in their Appendix~F. 
The first difference is in the choice of the particular solutions of tidal perturbations at quadratic order in the small~$GM\omega$ expansion. Indeed, the particular solutions are normalized by requiring that their large-distance asymptotic expansions, expressed in harmonic coordinates, do not contain terms proportional to $\bar{r}^\ell$ and $\bar{r}^{-\ell-1}$ in Eq.~\eqref{eq:largedisgtt}, at variance with the choice in Refs.~\cite{Poisson:2020vap,Pitre:2023xsr}. This normalization amounts to absorbing the $\bar{r}^\ell$ and $\bar{r}^{-\ell-1}$ contributions of the nonnormalized particular solutions into the homogeneous part. The second difference is in the identification of the integration constants of the metric perturbations with the tidal moment and response function. Specifically, as discussed above, the approach of Refs.~\cite{Poisson:2020vap,Pitre:2023xsr} retains only $d^{(0)}$ on the right-hand side of Eq.~\eqref{eq:redefinition}, while Ref.~\cite{HegadeKR:2024agt} accounts for contributions up to quadratic order in $G M\omega$~[see Eq.~(F6) therein]. As a consequence, the resulting tidal response function in Ref.~\cite{HegadeKR:2024agt} is uniquely fixed by the interior problem, which leads to different dynamical TLNs from those defined in Refs.~\cite{Poisson:2020vap,Pitre:2023xsr}, as explicitly shown in Eq.~(F12) of Ref.~\cite{HegadeKR:2024agt}.

\end{widetext}

\bibliography{biblio}

@article{Bhattacharyya:2026itm,
    author = "Bhattacharyya, Arpan and Kumar, Naman and Kumar, Shailesh",
    title = "{Dynamical Tidal Response of Regular Black Holes: Perturbative Analysis and Shell EFT Interpretation}",
    eprint = "2603.24719",
    archivePrefix = "arXiv",
    primaryClass = "hep-th",
    month = "3",
    year = "2026"
}

@article{HegadeKR:2026kku,
    author = "Hegade K. R., Abhishek and Kwon, K. J. and Venumadhav, Tejaswi and Yu, Hang and Yunes, Nicolas",
    title = "{Relativistic mode sums for neutron-star tidal response}",
    eprint = "2605.08569",
    archivePrefix = "arXiv",
    primaryClass = "gr-qc",
    doi = "10.1103/4jv6-twnk",
    journal = "Phys. Rev. D",
    volume = "114",
    number = "6",
    pages = "064048",
    year = "2026"
}

@article{Saketh:2026trm,
    author = "Saketh, M. V. S. and Ghosh, Suprovo and Andersson, Nils",
    title = "{Dynamical tidal response of neutron stars via scattering amplitudes}",
    eprint = "2606.14405",
    archivePrefix = "arXiv",
    primaryClass = "gr-qc",
    doi = "10.1103/gvv9-cnmx",
    journal = "Phys. Rev. D",
    volume = "114",
    number = "6",
    pages = "064008",
    year = "2026"
}

@article{HegadeKR:2026iou,
    author = "Hegade K. R., Abhishek and Yang, Yumu and Hippert, Mauricio and Noronha-Hostler, Jacquelyn and Noronha, Jorge and Yunes, Nicol{\'a}s",
    title = "{Dynamical tidal response of neutron stars as a probe of dense-matter properties}",
    eprint = "2603.26886",
    archivePrefix = "arXiv",
    primaryClass = "gr-qc",
    doi = "10.1103/92sp-krfc",
    journal = "Phys. Rev. D",
    volume = "114",
    number = "4",
    pages = "044014",
    year = "2026"
}

@article{Kosmopoulos:2025rfj,
    author = "Kosmopoulos, Dimitrios and Perrone, Davide and Solon, Mikhail",
    title = "{Dynamical Love Numbers for Black Holes and Beyond from Shell Effective Field Theory}",
    eprint = "2512.04002",
    archivePrefix = "arXiv",
    primaryClass = "hep-th",
    reportNumber = "CERN-TH-2025-247, CERN-TH-2025-247",
    doi = "10.1103/mtqm-xz2k",
    journal = "Phys. Rev. Lett.",
    volume = "136",
    number = "21",
    pages = "211401",
    year = "2026"
}

@misc{gitTA,
howpublished = {\url{https://github.com/thomasapost/Dynamical_Tidal_Response_of_Neutron_Stars}},
    year = "2026"
}

@article{Andersson:1997eq,
    author = "Andersson, Nils and Kokkotas, Kostas D.",
    title = "{Pulsation modes for increasingly relativistic polytropes}",
    eprint = "gr-qc/9706010",
    archivePrefix = "arXiv",
    doi = "10.1046/j.1365-8711.1998.01541.x",
    journal = "Mon. Not. Roy. Astron. Soc.",
    volume = "297",
    pages = "493",
    year = "1998"
}

@article{Pnigouras:2025muo,
    author = "Pnigouras, P. and Andersson, N. and Gittins, F. and Counsell, A. R.",
    title = "{Dynamical neutron star tides: the signature of a mode resonance}",
    eprint = "2508.06416",
    archivePrefix = "arXiv",
    primaryClass = "gr-qc",
    doi = "10.1093/mnras/staf1285",
    journal = "Mon. Not. Roy. Astron. Soc.",
    volume = "542",
    number = "2",
    pages = "1375--1387",
    year = "2025"
}

@article{Jakobsen:2023pvx,
    author = "Jakobsen, Gustav Uhre and Mogull, Gustav and Plefka, Jan and Sauer, Benjamin",
    title = "{Tidal effects and renormalization at fourth post-Minkowskian order}",
    eprint = "2312.00719",
    archivePrefix = "arXiv",
    primaryClass = "hep-th",
    reportNumber = "HU-EP-23/63-RTG",
    doi = "10.1103/PhysRevD.109.L041504",
    journal = "Phys. Rev. D",
    volume = "109",
    number = "4",
    pages = "L041504",
    year = "2024"
}

@article{Majumder:2015kfa,
    author = "Majumder, Barun and Yagi, Kent and Yunes, Nicolas",
    title = "{Improved Universality in the Neutron Star Three-Hair Relations}",
    eprint = "1504.02506",
    archivePrefix = "arXiv",
    primaryClass = "gr-qc",
    doi = "10.1103/PhysRevD.92.024020",
    journal = "Phys. Rev. D",
    volume = "92",
    number = "2",
    pages = "024020",
    year = "2015"
}

@article{Vines:2011ud,
      author         = "Vines, Justin and Flanagan, Eanna E. and Hinderer, Tanja",
      title          = "{Post-1-Newtonian tidal effects in the gravitational
                        waveform from binary inspirals}",
      journal        = "Phys. Rev.",
      volume         = "D83",
      year           = "2011",
      pages          = "084051",
      doi            = "10.1103/PhysRevD.83.084051",
      eprint         = "1101.1673",
      archivePrefix  = "arXiv",
      primaryClass   = "gr-qc",
      SLACcitation   = "%%CITATION = ARXIV:1101.1673;%%"
}

@article{Vines:2010ca,
    author = "Vines, Justin E. and Flanagan, Eanna E.",
    title = "{Post-1-Newtonian quadrupole tidal interactions in binary systems}",
    eprint = "1009.4919",
    archivePrefix = "arXiv",
    primaryClass = "gr-qc",
    doi = "10.1103/PhysRevD.88.024046",
    journal = "Phys. Rev. D",
    volume = "88",
    pages = "024046",
    year = "2013"
}

@article{Barbosa:2025uau,
    author = "Barbosa, Sergio and Brax, Philippe and Fichet, Sylvain and de Souza, Lucas",
    title = "{Running Love numbers and the Effective Field Theory of gravity}",
    eprint = "2501.18684",
    archivePrefix = "arXiv",
    primaryClass = "hep-th",
    doi = "10.1088/1475-7516/2025/07/071",
    journal = "JCAP",
    volume = "07",
    pages = "071",
    year = "2025"
}

@article{DeLuca:2022tkm,
    author = "De Luca, Valerio and Khoury, Justin and Wong, Sam S. C.",
    title = "{Implications of the weak gravity conjecture for tidal Love numbers of black holes}",
    eprint = "2211.14325",
    archivePrefix = "arXiv",
    primaryClass = "hep-th",
    doi = "10.1103/PhysRevD.108.044066",
    journal = "Phys. Rev. D",
    volume = "108",
    number = "4",
    pages = "044066",
    year = "2023"
}

@article{Branchesi:2023mws,
    author = "Branchesi, Marica and others",
    title = "{Science with the Einstein Telescope: a comparison of different designs}",
    eprint = "2303.15923",
    archivePrefix = "arXiv",
    primaryClass = "gr-qc",
    reportNumber = "ET-0084A-23",
    doi = "10.1088/1475-7516/2023/07/068",
    journal = "JCAP",
    volume = "07",
    pages = "068",
    year = "2023"
}

@article{ET:2025xjr,
    author = "Abac, Adrian and others",
    collaboration = "ET",
    title = "{The Science of the Einstein Telescope}",
    eprint = "2503.12263",
    archivePrefix = "arXiv",
    primaryClass = "gr-qc",
    reportNumber = "ET-0036C-25",
    doi = "10.1088/1475-7516/2026/03/081",
    journal = "JCAP",
    volume = "03",
    pages = "081",
    year = "2026"
}

@article{Vallisneri:2007ev,
    author = "Vallisneri, Michele",
    title = "{Use and abuse of the Fisher information matrix in the assessment of gravitational-wave parameter-estimation prospects}",
    eprint = "gr-qc/0703086",
    archivePrefix = "arXiv",
    reportNumber = "LIGO-P070009-00-Z",
    doi = "10.1103/PhysRevD.77.042001",
    journal = "Phys. Rev. D",
    volume = "77",
    pages = "042001",
    year = "2008"
}

@article{Khan:2015jqa,
	title        = {{Frequency-domain gravitational waves from nonprecessing black-hole binaries. II. A phenomenological model for the advanced detector era}},
	author       = {Khan, Sebastian and Husa, Sascha and Hannam, Mark and Ohme, Frank and P\"urrer, Michael and Jim\'enez Forteza, Xisco and Boh\'e, Alejandro},
	year         = 2016,
	journal      = {Phys. Rev. D},
	volume       = 93,
	number       = 4,
	pages        = {044007},
	doi          = {10.1103/PhysRevD.93.044007},
	eprint       = {1508.07253},
	archiveprefix = {arXiv},
	primaryclass = {gr-qc}
}

@article{Husa:2015iqa,
	title        = {{Frequency-domain gravitational waves from nonprecessing black-hole binaries. I. New numerical waveforms and anatomy of the signal}},
	author       = {Husa, Sascha and Khan, Sebastian and Hannam, Mark and P\"urrer, Michael and Ohme, Frank and Jim\'enez Forteza, Xisco and Boh\'e, Alejandro},
	year         = 2016,
	journal      = {Phys. Rev. D},
	volume       = 93,
	number       = 4,
	pages        = {044006},
	doi          = {10.1103/PhysRevD.93.044006},
	eprint       = {1508.07250},
	archiveprefix = {arXiv},
	primaryclass = {gr-qc}
}

@article{Baiotti:2019sew,
    author = "Baiotti, Luca",
    title = "{Gravitational waves from neutron star mergers and their relation to the nuclear equation of state}",
    eprint = "1907.08534",
    archivePrefix = "arXiv",
    primaryClass = "astro-ph.HE",
    doi = "10.1016/j.ppnp.2019.103714",
    journal = "Prog. Part. Nucl. Phys.",
    volume = "109",
    pages = "103714",
    year = "2019"
}

@article{Jarequi:2026cyp,
    author = "Jarequi, Gregory and Mitra, Soumodeep and Vaidya, Varun",
    title = "{Dynamical Tidal response of compact stars -- An EFT approach}",
    eprint = "2603.12331",
    archivePrefix = "arXiv",
    primaryClass = "gr-qc",
    month = "3",
    year = "2026"
}

@article{HegadeKR:2025qwj,
    author = "Hegade K. R., Abhishek and Kwon, K. J. and Venumadhav, Tejaswi and Yu, Hang and Yunes, Nicol{\'a}s",
    title = "{Relativistic and Dynamical Love Numbers}",
    eprint = "2507.10693",
    archivePrefix = "arXiv",
    primaryClass = "gr-qc",
    doi = "10.1103/1wdp-6x27",
    journal = "Phys. Rev. Lett.",
    volume = "136",
    number = "7",
    pages = "071401",
    year = "2026"
}

@article{Counsell:2024pua,
    author = "Counsell, Rhys and Gittins, Fabian and Andersson, Nils and Pnigouras, Pantelis",
    title = "{Neutron star g modes in the relativistic Cowling approximation}",
    eprint = "2409.20178",
    archivePrefix = "arXiv",
    primaryClass = "gr-qc",
    doi = "10.1093/mnras/stae2721",
    journal = "Mon. Not. Roy. Astron. Soc.",
    volume = "536",
    number = "2",
    pages = "1967--1979",
    year = "2024"
}

@article{Yu:2024uxt,
    author = "Yu, Hang and Arras, Phil and Weinberg, Nevin N.",
    title = "{Dynamical tides during the inspiral of rapidly spinning neutron stars: Solutions beyond mode resonance}",
    eprint = "2404.00147",
    archivePrefix = "arXiv",
    primaryClass = "gr-qc",
    doi = "10.1103/PhysRevD.110.024039",
    journal = "Phys. Rev. D",
    volume = "110",
    number = "2",
    pages = "024039",
    year = "2024"
}

@article{Pnigouras:2022zpx,
    author = "Pnigouras, Pantelis and Gittins, Fabian and Nanda, Amlan and Andersson, Nils and Jones, David Ian",
    title = "{The dynamical tides of spinning Newtonian stars}",
    eprint = "2205.07577",
    archivePrefix = "arXiv",
    primaryClass = "gr-qc",
    doi = "10.1093/mnras/stad3593",
    journal = "Mon. Not. Roy. Astron. Soc.",
    volume = "527",
    number = "3",
    pages = "8409--8428",
    year = "2024"
}

@article{Passamonti:2022yqp,
    author = "Passamonti, Andrea and Andersson, Nils and Pnigouras, Pantelis",
    title = "{Dynamical tides in superfluid neutron stars}",
    eprint = "2202.05161",
    archivePrefix = "arXiv",
    primaryClass = "astro-ph.HE",
    doi = "10.1093/mnras/stac1380",
    journal = "Mon. Not. Roy. Astron. Soc.",
    volume = "514",
    number = "1",
    pages = "1494--1510",
    year = "2022"
}

@article{Passamonti:2020fur,
    author = "Passamonti, A. and Andersson, N. and Pnigouras, P.",
    title = "{Dynamical tides in neutron stars: The impact of the crust}",
    eprint = "2012.09637",
    archivePrefix = "arXiv",
    primaryClass = "astro-ph.HE",
    doi = "10.1093/mnras/stab870",
    journal = "Mon. Not. Roy. Astron. Soc.",
    volume = "504",
    number = "1",
    pages = "1273--1293",
    year = "2021"
}

@article{Andersson:2019ahb,
    author = "Andersson, N. and Pnigouras, P.",
    title = "{Exploring the effective tidal deformability of neutron stars}",
    eprint = "1906.08982",
    archivePrefix = "arXiv",
    primaryClass = "astro-ph.HE",
    doi = "10.1103/PhysRevD.101.083001",
    journal = "Phys. Rev. D",
    volume = "101",
    number = "8",
    pages = "083001",
    year = "2020"
}

@article{Andersson:2019dwg,
    author = "Andersson, N. and Pnigouras, P.",
    title = "{The phenomenology of dynamical neutron star tides}",
    eprint = "1905.00012",
    archivePrefix = "arXiv",
    primaryClass = "gr-qc",
    doi = "10.1093/mnras/stab371",
    journal = "Mon. Not. Roy. Astron. Soc.",
    volume = "503",
    number = "1",
    pages = "533--539",
    year = "2021"
}

@article{Chakrabarti:2013xza,
    author = "Chakrabarti, Sayan and Delsate, T{\'e}rence and Steinhoff, Jan",
    title = "{Effective action and linear response of compact objects in Newtonian gravity}",
    eprint = "1306.5820",
    archivePrefix = "arXiv",
    primaryClass = "gr-qc",
    doi = "10.1103/PhysRevD.88.084038",
    journal = "Phys. Rev. D",
    volume = "88",
    pages = "084038",
    year = "2013"
}

@article{Ho:1998hq,
    author = "Ho, Wynn C. G. and Lai, Dong",
    title = "{Resonant tidal excitations of rotating neutron stars in coalescing binaries}",
    eprint = "astro-ph/9812116",
    archivePrefix = "arXiv",
    doi = "10.1046/j.1365-8711.1999.02703.x",
    journal = "Mon. Not. Roy. Astron. Soc.",
    volume = "308",
    pages = "153",
    year = "1999"
}

@article{Lai:1997wh,
    author = "Lai, Dong",
    title = "{Dynamical tides in rotating binary stars}",
    eprint = "astro-ph/9704132",
    archivePrefix = "arXiv",
    doi = "10.1086/304899",
    journal = "Astrophys. J.",
    volume = "490",
    pages = "847",
    year = "1997"
}

@article{Lai:1993di,
    author = "Lai, Dong",
    title = "{Resonant oscillations and tidal heating in coalescing binary neutron stars}",
    eprint = "astro-ph/9404062",
    archivePrefix = "arXiv",
    reportNumber = "CRSR-1064",
    doi = "10.1093/mnras/270.3.611",
    journal = "Mon. Not. Roy. Astron. Soc.",
    volume = "270",
    pages = "611",
    year = "1994"
}

@article{Postnikov:2010yn,
    author = "Postnikov, Sergey and Prakash, Madappa and Lattimer, James M.",
    title = "{Tidal Love Numbers of Neutron and Self-Bound Quark Stars}",
    eprint = "1004.5098",
    archivePrefix = "arXiv",
    primaryClass = "astro-ph.SR",
    doi = "10.1103/PhysRevD.82.024016",
    journal = "Phys. Rev. D",
    volume = "82",
    pages = "024016",
    year = "2010"
}

@article{Hinderer:2009ca,
    author = "Hinderer, Tanja and Lackey, Benjamin D. and Lang, Ryan N. and Read, Jocelyn S.",
    title = "{Tidal deformability of neutron stars with realistic equations of state and their gravitational wave signatures in binary inspiral}",
    eprint = "0911.3535",
    archivePrefix = "arXiv",
    primaryClass = "astro-ph.HE",
    doi = "10.1103/PhysRevD.81.123016",
    journal = "Phys. Rev. D",
    volume = "81",
    pages = "123016",
    year = "2010"
}

@article{Engvik:1994tj,
    author = "Engvik, L. and Hjorth-Jensen, M. and Osnes, E. and Bao, G. and Ostgaard, E.",
    title = "{Asymmetric nuclear matter and neutron star properties}",
    eprint = "nucl-th/9406028",
    archivePrefix = "arXiv",
    doi = "10.1103/PhysRevLett.73.2650",
    journal = "Phys. Rev. Lett.",
    volume = "73",
    pages = "2650--2653",
    year = "1994"
}

@misc{lalsuite,
       author         = "{LIGO Scientific Collaboration} and {Virgo Collaboration} and {KAGRA Collaboration}",
       title          = "{LVK} {A}lgorithm {L}ibrary - {LALS}uite",
       howpublished   = "Free software (GPL)",
       doi            = "10.7935/GT1W-FZ16",
       year           = "2018"
 }

@ARTICLE{1981NuPhA.361..502F,
       author = {{Friedman}, B. and {Pandharipande}, V.~R.},
        title = "{Hot and cold, nuclear and neutron matter}",
      journal = {Nucl. Phys. A},
         year = 1981,
        month = may,
       volume = {361},
       number = {2},
        pages = {502-520},
          doi = {10.1016/0375-9474(81)90649-7},
       adsurl = {https://ui.adsabs.harvard.edu/abs/1981NuPhA.361..502F}
}

@article{Yagi:2013awa,
    author = "Yagi, Kent and Yunes, Nicolas",
    title = "{I-Love-Q Relations in Neutron Stars and their Applications to Astrophysics, Gravitational Waves and Fundamental Physics}",
    eprint = "1303.1528",
    archivePrefix = "arXiv",
    primaryClass = "gr-qc",
    doi = "10.1103/PhysRevD.88.023009",
    journal = "Phys. Rev. D",
    volume = "88",
    number = "2",
    pages = "023009",
    year = "2013"
}

@article{Rodriguez:2026iot,
    author = "Rodr{\'\i}guez, Mar{\'\i}a J. and Santoni, Luca and Solomon, Adam R.",
    title = "{Love numbers of black holes and compact objects}",
    eprint = "2604.08653",
    archivePrefix = "arXiv",
    primaryClass = "gr-qc",
    month = "4",
    year = "2026"
}

@article{Chakraborty:2026qru,
    author = "Chakraborty, Sumanta and Pani, Paolo",
    title = "{Tidal Response of Compact Objects}",
    eprint = "2604.08679",
    archivePrefix = "arXiv",
    primaryClass = "gr-qc",
    month = "4",
    year = "2026"
}

@article{Katagiri:2025qze,
    author = "Katagiri, Takuya and Mukkamala, Gowtham Rishi and Yagi, Kent",
    title = "{Theoretical modeling of approximate universality of tidally deformed neutron stars}",
    eprint = "2505.05429",
    archivePrefix = "arXiv",
    primaryClass = "gr-qc",
    doi = "10.1103/8yfc-v76l",
    journal = "Phys. Rev. D",
    volume = "112",
    number = "2",
    pages = "023030",
    year = "2025"
}

@article{Goldberger:2022rqf,
    author = "Goldberger, Walter D.",
    title = "{Effective Field Theory for Compact Binary Dynamics}",
    eprint = "2212.06677",
    archivePrefix = "arXiv",
    primaryClass = "hep-th",
    month = "12",
    year = "2022"
}

@article{Sotani:2001bb,
    author = "Sotani, Hajime and Tominaga, Kazuhiro and Maeda, Kei-ichi",
    title = "{Density discontinuity of a neutron star and gravitational waves}",
    eprint = "gr-qc/0108060",
    archivePrefix = "arXiv",
    reportNumber = "WU-AP-135-01",
    doi = "10.1103/PhysRevD.65.024010",
    journal = "Phys. Rev. D",
    volume = "65",
    pages = "024010",
    year = "2002"
}

@misc{gitTK,
howpublished = {\url{https://github.com/TakuyaKatagiri/Theory_universal_Love}},
author = " ",
    year = "2025"
}

@article{Katagiri:2024fpn,
    author = "Katagiri, Takuya and Cardoso, Vitor and Ikeda, Tact and Yagi, Kent",
    title = "{Tidal response beyond vacuum general relativity with a canonical definition}",
    eprint = "2410.02531",
    archivePrefix = "arXiv",
    primaryClass = "gr-qc",
    reportNumber = "RUP-24-19",
    doi = "10.1103/PhysRevD.111.084081",
    journal = "Phys. Rev. D",
    volume = "111",
    number = "8",
    pages = "084081",
    year = "2025"
}

@article{Martinez-Rodriguez:2026omk,
    author = "Mart{\'\i}nez-Rodr{\'\i}guez, Irvin",
    title = "{Neutron Stars as Perfect Fluids: Extracting the Linearized Response Function}",
    eprint = "2602.07115",
    archivePrefix = "arXiv",
    primaryClass = "gr-qc",
    month = "2",
    year = "2026"
}

@article{Mandal:2023hqa,
    author = "Mandal, Manoj K. and Mastrolia, Pierpaolo and Silva, Hector O. and Patil, Raj and Steinhoff, Jan",
    title = "{Renormalizing Love: tidal effects at the third post-Newtonian order}",
    eprint = "2308.01865",
    archivePrefix = "arXiv",
    primaryClass = "hep-th",
    reportNumber = "HU-EP-23/43-RTG",
    doi = "10.1007/JHEP02(2024)188",
    journal = "JHEP",
    volume = "02",
    pages = "188",
    year = "2024"
}

@article{Flanagan:1997sx,
    author = "Flanagan, Eanna E. and Hughes, Scott A.",
    title = "{Measuring gravitational waves from binary black hole coalescences: 1. Signal-to-noise for inspiral, merger, and ringdown}",
    eprint = "gr-qc/9701039",
    archivePrefix = "arXiv",
    reportNumber = "GRP-456",
    doi = "10.1103/PhysRevD.57.4535",
    journal = "Phys. Rev. D",
    volume = "57",
    pages = "4535--4565",
    year = "1998"
}

@article{Mandal:2023lgy,
    author = "Mandal, Manoj K. and Mastrolia, Pierpaolo and Silva, Hector O. and Patil, Raj and Steinhoff, Jan",
    title = "{Gravitoelectric dynamical tides at second post-Newtonian order}",
    eprint = "2304.02030",
    archivePrefix = "arXiv",
    primaryClass = "hep-th",
    reportNumber = "HU-EP-23/09-RTG",
    doi = "10.1007/JHEP11(2023)067",
    journal = "JHEP",
    volume = "11",
    pages = "067",
    year = "2023"
}

@article{Pitre:2025qdf,
    author = "Pitre, Tristan and Poisson, Eric",
    title = "{Impact of nonlinearities on relativistic dynamical tides in compact binary inspirals}",
    eprint = "2506.08722",
    archivePrefix = "arXiv",
    primaryClass = "gr-qc",
    doi = "10.1103/1wwz-6jyc",
    journal = "Phys. Rev. D",
    volume = "112",
    number = "8",
    pages = "084017",
    year = "2025"
}

@article{Porto:2005ac,
    author = "Porto, Rafael A.",
    title = "{Post-Newtonian corrections to the motion of spinning bodies in NRGR}",
    eprint = "gr-qc/0511061",
    archivePrefix = "arXiv",
    reportNumber = "CMU-TH-05-10",
    doi = "10.1103/PhysRevD.73.104031",
    journal = "Phys. Rev. D",
    volume = "73",
    pages = "104031",
    year = "2006"
}

@article{Rothstein:2014sra,
    author = "Rothstein, Ira Z.",
    title = "{Progress in effective field theory approach to the binary inspiral problem}",
    doi = "10.1007/s10714-014-1726-y",
    journal = "Gen. Rel. Grav.",
    volume = "46",
    pages = "1726",
    year = "2014"
}

@article{Saes:2025jvr,
    author = "Saes, Jayana A. and Hegade K. R., Abhishek and Yunes, Nicol{\'a}s",
    title = "{Universal relations with dynamical tides}",
    eprint = "2511.19626",
    archivePrefix = "arXiv",
    primaryClass = "gr-qc",
    doi = "10.1103/5hcp-szdg",
    journal = "Phys. Rev. D",
    volume = "113",
    number = "10",
    pages = "104045",
    year = "2026"
}

@article{Saketh:2024juq,
    author = "Saketh, M. V. S. and Zhou, Zihan and Ghosh, Suprovo and Steinhoff, Jan and Chatterjee, Debarati",
    title = "{Investigating tidal heating in neutron stars via gravitational Raman scattering}",
    eprint = "2407.08327",
    archivePrefix = "arXiv",
    primaryClass = "gr-qc",
    doi = "10.1103/PhysRevD.110.103001",
    journal = "Phys. Rev. D",
    volume = "110",
    pages = "103001",
    year = "2024"
}

@article{Riva:2023rcm,
    author = "Riva, Massimiliano Maria and Santoni, Luca and Savi{\'c}, Nikola and Vernizzi, Filippo",
    title = "{Vanishing of nonlinear tidal Love numbers of Schwarzschild black holes}",
    eprint = "2312.05065",
    archivePrefix = "arXiv",
    primaryClass = "gr-qc",
    doi = "10.1016/j.physletb.2024.138710",
    journal = "Phys. Lett. B",
    volume = "854",
    pages = "138710",
    year = "2024"
}

@article{Iteanu:2024dvx,
    author = "Iteanu, Simon and Riva, Massimiliano Maria and Santoni, Luca and Savi{\'c}, Nikola and Vernizzi, Filippo",
    title = "{Vanishing of quadratic Love numbers of Schwarzschild black holes}",
    eprint = "2410.03542",
    archivePrefix = "arXiv",
    primaryClass = "gr-qc",
    reportNumber = "DESY 24-141",
    doi = "10.1007/JHEP02(2025)174",
    journal = "JHEP",
    volume = "02",
    pages = "174",
    year = "2025"
}

@article{Perry:2023wmm,
    author = "Perry, Malcolm and Rodriguez, Maria J.",
    title = "{Dynamical Love Numbers for Kerr Black Holes}",
    eprint = "2310.03660",
    archivePrefix = "arXiv",
    primaryClass = "gr-qc",
    month = "10",
    year = "2023"
}

@article{LIGOScientific:2017vwq,
	archiveprefix = {arXiv},
	author = {Abbott, B. P. and others},
	collaboration = {LIGO Scientific, Virgo},
	doi = {10.1103/PhysRevLett.119.161101},
	eprint = {1710.05832},
	journal = {Phys. Rev. Lett.},
	number = {16},
	pages = {161101},
	primaryclass = {gr-qc},
	reportnumber = {LIGO-P170817},
	title = {{GW170817: Observation of Gravitational Waves from a Binary Neutron Star Inspiral}},
	volume = {119},
	year = {2017}}

@article{Kokkotas:1999bd,
    author = "Kokkotas, Kostas D. and Schmidt, Bernd G.",
    title = "{Quasinormal modes of stars and black holes}",
    eprint = "gr-qc/9909058",
    archivePrefix = "arXiv",
    doi = "10.12942/lrr-1999-2",
    journal = "Living Rev. Rel.",
    volume = "2",
    pages = "2",
    year = "1999"
}

@article{Lindblom:1983ps,
    author = "Lindblom, L and Detweiler, Steven L.",
    title = "{The quadrupole oscillations of neutron stars}",
    doi = "10.1086/190884",
    journal = "Astrophys. J. Suppl.",
    volume = "53",
    pages = "73--92",
    year = "1983"
}

@article{Detweiler:1985zz,
    author = "Detweiler, Steven L. and Lindblom, L.",
    title = "{On the nonradial pulsations of general relativistic stellar models}",
    doi = "10.1086/163127",
    journal = "Astrophys. J.",
    volume = "292",
    pages = "12--15",
    year = "1985"
}

@article{Regge:1957td,
	author = {Regge, Tullio and Wheeler, John A.},
	doi = {10.1103/PhysRev.108.1063},
	journal = {Phys. Rev.},
	pages = {1063-1069},
	slaccitation = {%%CITATION = PHRVA,108,1063;%%},
	title = {{Stability of a Schwarzschild singularity}},
	volume = {108},
	year = {1957}}

@article{Flanagan:2007ix,
    author = "Flanagan, Eanna E. and Hinderer, Tanja",
    title = "{Constraining neutron star tidal Love numbers with gravitational wave detectors}",
    eprint = "0709.1915",
    archivePrefix = "arXiv",
    primaryClass = "astro-ph",
    doi = "10.1103/PhysRevD.77.021502",
    journal = "Phys. Rev. D",
    volume = "77",
    pages = "021502",
    year = "2008"
}

@book{Poisson_Will_2014, place={Cambridge}, title={Gravity: Newtonian, Post-Newtonian, Relativistic}, publisher={Cambridge University Press}, author={Poisson, Eric and Will, Clifford M.}, year={2014}}

@article{Burgio:2021vgk,
    author = "Burgio, G. F. and Schulze, H. -J. and Vidana, I. and Wei, J. -B.",
    title = "{Neutron stars and the nuclear equation of state}",
    eprint = "2105.03747",
    archivePrefix = "arXiv",
    primaryClass = "nucl-th",
    doi = "10.1016/j.ppnp.2021.103879",
    journal = "Prog. Part. Nucl. Phys.",
    volume = "120",
    pages = "103879",
    year = "2021"
}

@article{Lattimer:2021emm,
    author = "Lattimer, J. M.",
    title = "{Neutron Stars and the Nuclear Matter Equation of State}",
    doi = "10.1146/annurev-nucl-102419-124827",
    journal = "Ann. Rev. Nucl. Part. Sci.",
    volume = "71",
    pages = "433--464",
    year = "2021"
}

@article{Combaluzier-Szteinsznaider:2024sgb,
    author = "Combaluzier-Szteinsznaider, Oscar and Hui, Lam and Santoni, Luca and Solomon, Adam R. and Wong, Sam S. C.",
    title = "{Symmetries of vanishing nonlinear Love numbers of Schwarzschild black holes}",
    eprint = "2410.10952",
    archivePrefix = "arXiv",
    primaryClass = "gr-qc",
    doi = "10.1007/JHEP03(2025)124",
    journal = "JHEP",
    volume = "03",
    pages = "124",
    year = "2025"
}

@inproceedings{Goldberger:2007hy,
	archiveprefix = {arXiv},
	author = {Goldberger, Walter D.},
	booktitle = {{Les Houches Summer School - Session 86: Particle Physics and Cosmology: The Fabric of Spacetime}},
	eprint = {hep-ph/0701129},
	month = {1},
	title = {{Les Houches lectures on effective field theories and gravitational radiation}},
	year = {2007}}

@article{Zerilli:1970wzz,
    author = "Zerilli, F. J.",
    title = "{Gravitational field of a particle falling in a schwarzschild geometry analyzed in tensor harmonics}",
    doi = "10.1103/PhysRevD.2.2141",
    journal = "Phys. Rev. D",
    volume = "2",
    pages = "2141--2160",
    year = "1970"
}

@article{Glazer:2024eyi,
    author = "Glazer, Daniel and Joyce, Austin and Rodriguez, Maria J. and Santoni, Luca and Solomon, Adam R. and Temoche, Luis Fernando",
    title = "{Higher-dimensional spinning black holes and effective field theory}",
    eprint = "2412.21090",
    archivePrefix = "arXiv",
    primaryClass = "hep-th",
    doi = "10.1007/JHEP03(2026)036",
    journal = "JHEP",
    volume = "03",
    pages = "036",
    year = "2026"
}

@article{Saketh:2022xjb,
    author = "Saketh, M. V. S. and Steinhoff, Jan and Vines, Justin and Buonanno, Alessandra",
    title = "{Modeling horizon absorption in spinning binary black holes using effective worldline theory}",
    eprint = "2212.13095",
    archivePrefix = "arXiv",
    primaryClass = "gr-qc",
    doi = "10.1103/PhysRevD.107.084006",
    journal = "Phys. Rev. D",
    volume = "107",
    number = "8",
    pages = "084006",
    year = "2023"
}

@article{Correia:2024jgr,
    author = "Correia, Miguel and Isabella, Giulia",
    title = "{The Born regime of gravitational amplitudes}",
    eprint = "2406.13737",
    archivePrefix = "arXiv",
    primaryClass = "hep-th",
    doi = "10.1007/JHEP03(2025)144",
    journal = "JHEP",
    volume = "03",
    pages = "144",
    year = "2025"
}

@article{Caron-Huot:2025tlq,
    author = "Caron-Huot, Simon and Correia, Miguel and Isabella, Giulia and Solon, Mikhail",
    title = "{Gravitational Wave Scattering via the Born Series: Scalar Tidal Matching to O(G7) and Beyond}",
    eprint = "2503.13593",
    archivePrefix = "arXiv",
    primaryClass = "hep-th",
    doi = "10.1103/qd3c-nfz6",
    journal = "Phys. Rev. Lett.",
    volume = "135",
    number = "19",
    pages = "191601",
    year = "2025"
}

@article{Pani:2025qxs,
    author = "Pani, Paolo and Riva, Massimiliano Maria and Santoni, Luca and Savi{\'c}, Nikola and Vernizzi, Filippo",
    title = "{Nonlinear relativistic tidal response of neutron stars}",
    eprint = "2512.14663",
    archivePrefix = "arXiv",
    primaryClass = "gr-qc",
    doi = "10.1007/JHEP05(2026)074",
    journal = "JHEP",
    volume = "05",
    pages = "074",
    year = "2026"
}

@article{Saketh:2023bul,
    author = "Saketh, M. V. S. and Zhou, Zihan and Ivanov, Mikhail M.",
    title = "{Dynamical tidal response of Kerr black holes from scattering amplitudes}",
    eprint = "2307.10391",
    archivePrefix = "arXiv",
    primaryClass = "hep-th",
    doi = "10.1103/PhysRevD.109.064058",
    journal = "Phys. Rev. D",
    volume = "109",
    number = "6",
    pages = "064058",
    year = "2024"
}

@article{Levi:2018nxp,
    author = "Levi, Mich{\`e}le",
    title = "{Effective Field Theories of Post-Newtonian Gravity: A comprehensive review}",
    eprint = "1807.01699",
    archivePrefix = "arXiv",
    primaryClass = "hep-th",
    doi = "10.1088/1361-6633/ab12bc",
    journal = "Rept. Prog. Phys.",
    volume = "83",
    number = "7",
    pages = "075901",
    year = "2020"
}

@article{Porto:2016pyg,
	archiveprefix = {arXiv},
	author = {Porto, Rafael A.},
	doi = {10.1016/j.physrep.2016.04.003},
	eprint = {1601.04914},
	journal = {Phys. Rept.},
	pages = {1--104},
	primaryclass = {hep-th},
	title = {{The effective field theorist\textquoteright{}s approach to gravitational dynamics}},
	volume = {633},
	year = {2016}}

@article{Charalambous:2021mea,
	archiveprefix = {arXiv},
	author = {Charalambous, Panagiotis and Dubovsky, Sergei and Ivanov, Mikhail M.},
	doi = {10.1007/JHEP05(2021)038},
	eprint = {2102.08917},
	journal = {JHEP},
	pages = {038},
	primaryclass = {hep-th},
	reportnumber = {INR-TH-2021-001},
	title = {{On the Vanishing of Love Numbers for Kerr Black Holes}},
	volume = {05},
	year = {2021}}

@article{Charalambous:2023jgq,
	archiveprefix = {arXiv},
	author = {Charalambous, Panagiotis and Ivanov, Mikhail M.},
	doi = {10.1007/JHEP07(2023)222},
	eprint = {2303.16036},
	journal = {JHEP},
	pages = {222},
	primaryclass = {hep-th},
	title = {{Scalar Love numbers and Love symmetries of 5-dimensional Myers-Perry black holes}},
	volume = {07},
	year = {2023}}

@article{Chakrabarti:2013lua,
	archiveprefix = {arXiv},
	author = {Chakrabarti, Sayan and Delsate, T\'erence and Steinhoff, Jan},
	eprint = {1304.2228},
	primaryclass = {gr-qc},
	title = {{New perspectives on neutron star and black hole spectroscopy and dynamic tides}},
	year = {2013}}

@article{Poisson:2020vap,
	archiveprefix = {arXiv},
	author = {Poisson, Eric},
	doi = {10.1103/PhysRevD.103.064023},
	eprint = {2012.10184},
	journal = {Phys. Rev. D},
	number = {6},
	pages = {064023},
	primaryclass = {gr-qc},
	title = {{Compact body in a tidal environment: New types of relativistic Love numbers, and a post-Newtonian operational definition for tidally induced multipole moments}},
	volume = {103},
	year = {2021}}

@article{Bern:2020uwk,
	archiveprefix = {arXiv},
	author = {Bern, Zvi and Parra-Martinez, Julio and Roiban, Radu and Sawyer, Eric and Shen, Chia-Hsien},
	doi = {10.1007/JHEP05(2021)188},
	eprint = {2010.08559},
	journal = {JHEP},
	pages = {188},
	primaryclass = {hep-th},
	reportnumber = {CALT-TH/2020-041},
	title = {{Leading Nonlinear Tidal Effects and Scattering Amplitudes}},
	volume = {05},
	year = {2021}}

@article{DeLuca:2023mio,
	archiveprefix = {arXiv},
	author = {De Luca, Valerio and Khoury, Justin and Wong, Sam S. C.},
	doi = {10.1103/PhysRevD.108.024048},
	eprint = {2305.14444},
	journal = {Phys. Rev. D},
	number = {2},
	pages = {024048},
	primaryclass = {gr-qc},
	title = {{Nonlinearities in the tidal Love numbers of black holes}},
	volume = {108},
	year = {2023}}

@article{JimenezForteza:2018rwr,
    author = "Jim{\'e}nez Forteza, Xisco and Abdelsalhin, Tiziano and Pani, Paolo and Gualtieri, Leonardo",
    title = "{Impact of high-order tidal terms on binary neutron-star waveforms}",
    eprint = "1807.08016",
    archivePrefix = "arXiv",
    primaryClass = "gr-qc",
    doi = "10.1103/PhysRevD.98.124014",
    journal = "Phys. Rev. D",
    volume = "98",
    number = "12",
    pages = "124014",
    year = "2018"
}

@article{Kobayashi:2025vgl,
    author = "Kobayashi, Hajime and Mukohyama, Shinji and Oshita, Naritaka and Takahashi, Kazufumi and Yingcharoenrat, Vicharit",
    title = "{Dynamical tidal response of nonrotating black holes: Connecting the Mano-Suzuki-Takasugi formalism and worldline EFT}",
    eprint = "2511.12580",
    archivePrefix = "arXiv",
    primaryClass = "gr-qc",
    reportNumber = "YITP-25-176, RESCEU-25/25, IPMU25-0052, RIKEN-iTHEMS-Report-25",
    doi = "10.1103/4ql9-mgj5",
    journal = "Phys. Rev. D",
    volume = "113",
    number = "8",
    pages = "084011",
    year = "2026"
}

@article{Gamba:2020wgg,
    author = "Gamba, Rossella and Breschi, Matteo and Bernuzzi, Sebastiano and Agathos, Michalis and Nagar, Alessandro",
    title = "{Waveform systematics in the gravitational-wave inference of tidal parameters and equation of state from binary neutron star signals}",
    eprint = "2009.08467",
    archivePrefix = "arXiv",
    primaryClass = "gr-qc",
    doi = "10.1103/PhysRevD.103.124015",
    journal = "Phys. Rev. D",
    volume = "103",
    number = "12",
    pages = "124015",
    year = "2021"
}

@article{Gupta:2024gun,
    author = "Gupta, Anuradha and others",
    title = "{Possible causes of false general relativity violations in gravitational wave observations}",
    eprint = "2405.02197",
    archivePrefix = "arXiv",
    primaryClass = "gr-qc",
    doi = "10.21468/SciPostPhysCommRep.5",
    month = "5",
    year = "2024"
}

@article{Akmal:1998cf,
    author = "Akmal, A. and Pandharipande, V. R. and Ravenhall, D. G.",
    title = "{The Equation of state of nucleon matter and neutron star structure}",
    eprint = "nucl-th/9804027",
    archivePrefix = "arXiv",
    doi = "10.1103/PhysRevC.58.1804",
    journal = "Phys. Rev. C",
    volume = "58",
    pages = "1804--1828",
    year = "1998"
}

@article{Douchin:2001sv,
    author = "Douchin, F. and Haensel, P.",
    title = "{A unified equation of state of dense matter and neutron star structure}",
    eprint = "astro-ph/0111092",
    archivePrefix = "arXiv",
    doi = "10.1051/0004-6361:20011402",
    journal = "Astron. Astrophys.",
    volume = "380",
    pages = "151",
    year = "2001"
}

@article{Mueller:1996pm,
    author = "Mueller, Horst and Serot, Brian D.",
    title = "{Relativistic mean field theory and the high density nuclear equation of state}",
    eprint = "nucl-th/9603037",
    archivePrefix = "arXiv",
    reportNumber = "IU-NTC-96-03",
    doi = "10.1016/0375-9474(96)00187-X",
    journal = "Nucl. Phys. A",
    volume = "606",
    pages = "508--537",
    year = "1996"
}

@article{Ivanov:2026icp,
    author = "Ivanov, Mikhail M. and Li, Yue-Zhou and Parra-Martinez, Julio and Zhou, Zihan",
    title = "{Gravitational Raman Scattering: a Systematic Toolkit for Tidal Effects in General Relativity}",
    eprint = "2602.06951",
    archivePrefix = "arXiv",
    primaryClass = "hep-th",
    reportNumber = "MIT-CTP/6001",
    month = "2",
    year = "2026"
}

@article{Ivanov:2024sds,
    author = "Ivanov, Mikhail M. and Li, Yue-Zhou and Parra-Martinez, Julio and Zhou, Zihan",
    title = "{Gravitational Raman Scattering in Effective Field Theory: A Scalar Tidal Matching at O(G3)}",
    eprint = "2401.08752",
    archivePrefix = "arXiv",
    primaryClass = "hep-th",
    reportNumber = "MIT-CTP/5664",
    doi = "10.1103/PhysRevLett.132.131401",
    journal = "Phys. Rev. Lett.",
    volume = "132",
    number = "13",
    pages = "131401",
    year = "2024",
    note = "[Erratum: Phys.Rev.Lett. 134, 159901 (2025)]"
}

@article{Rodriguez:2023xjd,
	archiveprefix = {arXiv},
	author = {Rodriguez, Maria J. and Santoni, Luca and Solomon, Adam R. and Temoche, Luis Fernando},
	doi = {10.1103/PhysRevD.108.084011},
	eprint = {2304.03743},
	journal = {Phys. Rev. D},
	number = {8},
	pages = {084011},
	primaryclass = {hep-th},
	title = {{Love numbers for rotating black holes in higher dimensions}},
	volume = {108},
	year = {2023}}

@article{Hui:2020xxx,
	archiveprefix = {arXiv},
	author = {Hui, Lam and Joyce, Austin and Penco, Riccardo and Santoni, Luca and Solomon, Adam R.},
	doi = {10.1088/1475-7516/2021/04/052},
	eprint = {2010.00593},
	journal = {JCAP},
	pages = {052},
	primaryclass = {hep-th},
	title = {{Static response and Love numbers of Schwarzschild black holes}},
	volume = {04},
	year = {2021}}

@article{Kodama:2003jz,
	archiveprefix = {arXiv},
	author = {Kodama, Hideo and Ishibashi, Akihiro},
	doi = {10.1143/PTP.110.701},
	eprint = {hep-th/0305147},
	journal = {Prog. Theor. Phys.},
	pages = {701-722},
	primaryclass = {hep-th},
	slaccitation = {%%CITATION = HEP-TH/0305147;%%},
	title = {{A Master equation for gravitational perturbations of maximally symmetric black holes in higher dimensions}},
	volume = {110},
	year = {2003}}

@article{Kol:2011vg,
	archiveprefix = {arXiv},
	author = {Kol, Barak and Smolkin, Michael},
	doi = {10.1007/JHEP02(2012)010},
	eprint = {1110.3764},
	journal = {JHEP},
	pages = {010},
	primaryclass = {hep-th},
	slaccitation = {%%CITATION = ARXIV:1110.3764;%%},
	title = {{Black hole stereotyping: Induced gravito-static polarization}},
	volume = {02},
	year = {2012}}

@article{Cardoso:2017cfl,
	archiveprefix = {arXiv},
	author = {Cardoso, Vitor and Franzin, Edgardo and Maselli, Andrea and Pani, Paolo and Raposo, Guilherme},
	doi = {10.1103/PhysRevD.95.089901, 10.1103/PhysRevD.95.084014},
	eprint = {1701.01116},
	journal = {Phys. Rev.},
	note = {[Addendum: Phys. Rev.D95,no.8,089901(2017)]},
	number = {8},
	pages = {084014},
	primaryclass = {gr-qc},
	slaccitation = {%%CITATION = ARXIV:1701.01116;%%},
	title = {{Testing strong-field gravity with tidal Love numbers}},
	volume = {D95},
	year = {2017}}

@article{Binnington:2009bb,
	archiveprefix = {arXiv},
	author = {Binnington, Taylor and Poisson, Eric},
	doi = {10.1103/PhysRevD.80.084018},
	eprint = {0906.1366},
	journal = {Phys. Rev.},
	pages = {084018},
	primaryclass = {gr-qc},
	slaccitation = {%%CITATION = ARXIV:0906.1366;%%},
	title = {{Relativistic theory of tidal Love numbers}},
	volume = {D80},
	year = {2009}}

@article{Hinderer:2007mb,
	archiveprefix = {arXiv},
	author = {Hinderer, Tanja},
	doi = {10.1086/533487},
	eprint = {0711.2420},
	journal = {Astrophys. J.},
	pages = {1216-1220},
	primaryclass = {astro-ph},
	slaccitation = {%%CITATION = ARXIV:0711.2420;%%},
	title = {{Tidal Love numbers of neutron stars}},
	volume = {677},
	year = {2008}}

@article{Damour:2009vw,
	archiveprefix = {arXiv},
	author = {Damour, Thibault and Nagar, Alessandro},
	doi = {10.1103/PhysRevD.80.084035},
	eprint = {0906.0096},
	journal = {Phys. Rev. D},
	pages = {084035},
	primaryclass = {gr-qc},
	title = {{Relativistic tidal properties of neutron stars}},
	volume = {80},
	year = {2009}}

@article{Hinderer:2016eia,
    author = "Hinderer, Tanja and others",
    title = "{Effects of neutron-star dynamic tides on gravitational waveforms within the effective-one-body approach}",
    eprint = "1602.00599",
    archivePrefix = "arXiv",
    primaryClass = "gr-qc",
    doi = "10.1103/PhysRevLett.116.181101",
    journal = "Phys. Rev. Lett.",
    volume = "116",
    number = "18",
    pages = "181101",
    year = "2016"
}

@article{Katagiri:2024wbg,
    author = "Katagiri, Takuya and Yagi, Kent and Cardoso, Vitor",
    title = "{Relativistic dynamical tides: Subtleties and calibration}",
    eprint = "2409.18034",
    archivePrefix = "arXiv",
    primaryClass = "gr-qc",
    doi = "10.1103/PhysRevD.111.084080",
    journal = "Phys. Rev. D",
    volume = "111",
    number = "8",
    pages = "084080",
    year = "2025"
}

@article{Yagi:2016bkt,
    author = "Yagi, Kent and Yunes, Nicol{\'a}s",
    title = "{Approximate Universal Relations for Neutron Stars and Quark Stars}",
    eprint = "1608.02582",
    archivePrefix = "arXiv",
    primaryClass = "gr-qc",
    doi = "10.1016/j.physrep.2017.03.002",
    journal = "Phys. Rept.",
    volume = "681",
    pages = "1--72",
    year = "2017"
}

@article{Yagi:2013bca,
    author = "Yagi, Kent and Yunes, Nicolas",
    title = "{I-Love-Q}",
    eprint = "1302.4499",
    archivePrefix = "arXiv",
    primaryClass = "gr-qc",
    doi = "10.1126/science.1236462",
    journal = "Science",
    volume = "341",
    pages = "365--368",
    year = "2013"
}

@article{Combaluzier--Szteinsznaider:2025eoc,
    author = "Combaluzier--Szteinsznaider, Oscar and Glazer, Daniel and Joyce, Austin and Rodriguez, Maria J. and Santoni, Luca",
    title = "{Dynamical tidal response of Schwarzschild Black Holes}",
    eprint = "2511.02372",
    archivePrefix = "arXiv",
    primaryClass = "gr-qc",
    doi = "10.1007/JHEP06(2026)032",
    journal = "JHEP",
    volume = "06",
    pages = "032",
    year = "2026"
}

@article{Chakraborty:2025wvs,
    author = "Chakraborty, Sumanta and De Luca, Valerio and Gualtieri, Leonardo and Pani, Paolo",
    title = "{Dynamical Love numbers of black holes: Theory and gravitational waveforms}",
    eprint = "2507.22994",
    archivePrefix = "arXiv",
    primaryClass = "gr-qc",
    doi = "10.1103/fr3y-s1sz",
    journal = "Phys. Rev. D",
    volume = "112",
    number = "10",
    pages = "104015",
    year = "2025"
}

@article{HegadeKR:2024agt,
    author = "Hegade K. R., Abhishek and Ripley, Justin L. and Yunes, Nicol{\'a}s",
    title = "{Dynamical tidal response of nonrotating relativistic stars}",
    eprint = "2403.03254",
    archivePrefix = "arXiv",
    primaryClass = "gr-qc",
    doi = "10.1103/PhysRevD.109.104064",
    journal = "Phys. Rev. D",
    volume = "109",
    number = "10",
    pages = "104064",
    year = "2024"
}

@article{Pitre:2023xsr,
    author = "Pitre, Tristan and Poisson, Eric",
    title = "{General relativistic dynamical tides in binary inspirals without modes}",
    eprint = "2311.04075",
    archivePrefix = "arXiv",
    primaryClass = "gr-qc",
    doi = "10.1103/PhysRevD.109.064004",
    journal = "Phys. Rev. D",
    volume = "109",
    number = "6",
    pages = "064004",
    year = "2024"
}

@article{Steinhoff:2016rfi,
    author = "Steinhoff, Jan and Hinderer, Tanja and Buonanno, Alessandra and Taracchini, Andrea",
    title = "{Dynamical Tides in General Relativity: Effective Action and Effective-One-Body Hamiltonian}",
    eprint = "1608.01907",
    archivePrefix = "arXiv",
    primaryClass = "gr-qc",
    doi = "10.1103/PhysRevD.94.104028",
    journal = "Phys. Rev. D",
    volume = "94",
    number = "10",
    pages = "104028",
    year = "2016"
}

@article{Chatziioannou:2020pqz,
	archiveprefix = {arXiv},
	author = {Chatziioannou, Katerina},
	doi = {10.1007/s10714-020-02754-3},
	eprint = {2006.03168},
	journal = {Gen. Rel. Grav.},
	number = {11},
	pages = {109},
	primaryclass = {gr-qc},
	title = {{Neutron star tidal deformability and equation of state constraints}},
	volume = {52},
	year = {2020}}

@article{Goldberger:2004jt,
	archiveprefix = {arXiv},
	author = {Goldberger, Walter D. and Rothstein, Ira Z.},
	doi = {10.1103/PhysRevD.73.104029},
	eprint = {hep-th/0409156},
	journal = {Phys. Rev. D},
	pages = {104029},
	reportnumber = {UCSD-PTH-04-17, CMU-HEP-04-06},
	title = {{An Effective field theory of gravity for extended objects}},
	volume = {73},
	year = {2006}}

@article{Goldberger:2020fot,
	archiveprefix = {arXiv},
	author = {Goldberger, Walter D. and Li, Jingping and Rothstein, Ira Z.},
	doi = {10.1007/JHEP06(2021)053},
	eprint = {2012.14869},
	journal = {JHEP},
	pages = {053},
	primaryclass = {hep-th},
	title = {{Non-conservative effects on spinning black holes from world-line effective field theory}},
	volume = {06},
	year = {2021}}

@article{Goldberger:2005cd,
	archiveprefix = {arXiv},
	author = {Goldberger, Walter D. and Rothstein, Ira Z.},
	doi = {10.1103/PhysRevD.73.104030},
	eprint = {hep-th/0511133},
	journal = {Phys. Rev. D},
	pages = {104030},
	title = {{Dissipative effects in the worldline approach to black hole dynamics}},
	volume = {73},
	year = {2006}}

\end{document}